\documentclass{article}

\usepackage{microtype}
\usepackage{graphicx}
\usepackage{booktabs}
\usepackage{hyperref}
\usepackage{xcolor}
\usepackage{bm}
\usepackage[preprint]{icml2027}
\usepackage{longtable}
\usepackage{amsmath}
\usepackage{amssymb}
\usepackage{mathtools}
\usepackage{amsthm}
\usepackage{array}
\usepackage{float}

\theoremstyle{plain}
\newtheorem{theorem}{Theorem}[section]
\newtheorem{corollary}[theorem]{Corollary}
\newtheorem{lemma}[theorem]{Lemma}
\newtheorem{proposition}[theorem]{Proposition}
\theoremstyle{definition}
\newtheorem{definition}{Definition}[section]
\newtheorem{remark}[theorem]{Remark}

\newcommand{\pending}[1]{\textcolor{orange}{[PENDING: #1]}}
\newcommand{\figbox}[2]{\fbox{\parbox[c][3.2cm][c]{\dimexpr#1-2\fboxsep-2\fboxrule\relax}{\centering\small\textbf{FIGURE PLACEHOLDER}\\[2pt] #2}}}
\newcommand{\maybefig}[3]{\IfFileExists{#1}{\includegraphics[width=#2]{#1}}{\figbox{#2}{#3}}}
\newcommand{\maybetab}[2]{\IfFileExists{#1}{\input{#1}}{\pending{#2}}}

\newcommand{\pf}{\rho}                 %

\newcommand{\Jtop}{J_{\rm top}}
\newcommand{\Jlow}{J_{\rm low}}
\newcommand{\Kl}{\mathcal K}
\newcommand{\Pp}{P_{\Kl^\perp}}

\newcommand{\repourl}{\url{https://github.com/mpoppel/sameSpec_diffFate}}
\icmltitlerunning{Same Spectrum, Different Fate}

\begin{document}

\twocolumn[
  \icmltitle{Same Spectrum, Different Fate: Trainability of Single-Qubit Re-Uploading Circuits through the Lens of Quantum Signal Processing}

  \icmlsetsymbol{equal}{*}
  \begin{icmlauthorlist}
    \icmlauthor{Michael Poppel}{lmu,aq}
    \icmlauthor{David Bucher}{aq}
    \icmlauthor{Maximilian Zorn}{lmu}
    \icmlauthor{Markus Baumann}{lmu}
    \icmlauthor{Sebastian W\"olckert}{lmu}
    \icmlauthor{Claudia Linnhoff-Popien}{lmu}
    \icmlauthor{Philipp Altmann}{lmu}
    \icmlauthor{Jonas Stein}{lmu}
  \end{icmlauthorlist}
  \icmlaffiliation{lmu}{Department of Computer Science, LMU Munich, Germany}
  \icmlaffiliation{aq}{Aqarios GmbH, Munich, Germany}
  \icmlcorrespondingauthor{Michael Poppel}{michael.poppel@aqarios.com}
  \icmlkeywords{Quantum Machine Learning, Variational Quantum Circuits, Trainability, Initialisation, Quantum Signal Processing}
  \vskip 0.3in
]

\printAffiliationsAndNotice{}

\begin{abstract}
Re-uploading models with angle encoding compute truncated Fourier series of their input. For one-dimensional inputs, a single qubit with
$E$ re-uploading layers represents exactly the same functions as a two-qubit circuit with the same encoding: all real trigonometric
polynomials of degree $E$ bounded by one. Yet from Haar-random gates and for $E\ge24$, gradient descent rarely fits these functions with the
single qubit and almost always with two. We explain the gap with quantum signal processing (QSP). Up to gauge rotations of the gates, a
function has finitely many realisations as a single-qubit circuit, one for each choice of roots in a Fej\'er--Riesz factorisation, its \emph{QSP branch}. We prove that the
single-qubit circuit is rigid: it has no spare tangent direction, so its conditioning is, locally, a function of the realised function and
its branch. An exact identity ties the branch to the circuit's distance from the no-flip point, at which the highest Fourier coefficient has a
private direction: the extreme branches, whose roots all lie on one side of the unit circle, are close to it for every $E$, balanced
branches are far, and in our measurements closeness comes with a private direction. Haar starts
land on balanced branches and stay balanced. Realising the same stalled function on the minimum-phase branch removes the stall at every
stalled point we tested, and so does starting near an extreme branch: on three real-world series and two synthetic families, gradient
descent from such starts reaches the least-squares optimum in 28 to 30 of 30 runs within a median of a few hundred steps, with 7 to 70 times
less work than L-BFGS, against no run from Haar starts.
\end{abstract}

\section{Introduction}
\label{sec:intro}
Variational quantum circuits with angle encoding implement truncated Fourier series~\citep{schuld2021effect}. For one-dimensional inputs,
a single qubit with $E$ re-uploading layers (the \emph{serial} circuit) realises every real trigonometric polynomial of degree $\le E$ with
values in $[-1,1]$~\citep{yu2022power,motlagh2024generalized}: the same model class, and the same spectrum, as a two-qubit circuit with the
same encoding budget. Width can be exchanged for depth without losing expressivity. In training it cannot. We draw every gate from the Haar measure, the
uniform distribution on $\mathrm{SU}(2)$, which unlike uniform rotation angles does not depend on how the gate is
parameterised~\citep{mezzadri2007generate}; uniform rotation angles, the common choice in practice, start on the same kind of branch
and converge about as rarely (Table~\ref{tab:starts}). From such starts, gradient descent with a budget of 50{,}000 steps fits random flat-spectrum targets of
degree 32 with the serial circuit in 3 of 100 runs and with the two-qubit control in 99. More steps help only slowly: at $E=24$, 45 of 80
serial runs converge within 200{,}000 steps, each doubling of the budget beyond 25{,}000 adding 9 to 19 of them. On three real-world series, Adam with a tuned learning rate brings the serial
circuit to the least-squares optimum in at most 9 of 100 runs and the control in at least 98, and with four times the steps the serial
circuit still reaches it in at most 18 of 30 runs (Section~\ref{sec:cost}).

We explain this gap and remove it, using the theory of quantum signal processing (QSP)~\citep{haah2019product,motlagh2024generalized}. A
QSP sequence alternates fixed single-qubit rotations with a rotation by the signal, here the encoding gate $R_X(x)$; the serial circuit is
therefore a QSP sequence. Up to gauge rotations of the gates and under a mild condition (Lemma~\ref{lem:strip}), a function has finitely many realisations as such a sequence, one for each choice of roots in the Fej\'er--Riesz
factorisation of $(1\pm f)/2$; we call the choice the \emph{QSP branch}. For symmetric QSP, \citet{wang2022energy} characterise the
realisations of a polynomial in this way and show that the structured start of phase finding lies next to the one with all roots inside
the unit circle. We extend this picture to general gates trained from random starts and ask why most starts fail. Our contributions are three.
\begin{enumerate}
  \item We show that the conditioning of the serial circuit is fixed by the function it computes and by its QSP branch
  (Theorem~\ref{thm:rigid}, Lemma~\ref{lem:strip}), and that the branch fixes how far the circuit is from the no-flip point (Proposition~\ref{thm:identity}). The
  two-qubit control, with the same model class, has spare directions; in our measurements, its conditioning changes along them.
  \item Haar starts begin far from the no-flip point, with $\Pi\approx e^{-(E-1)}$ (Remark~\ref{rem:clt}). We measure that they begin
  and end on balanced branches, and runs stall there. The branch can change only where the function touches $\pm1$ (Proposition~\ref{prop:branchcount}).
  \item Changing the branch removes the stall. In our tests, the same stalled function realised on the minimum-phase branch converges, and
  so do starts near an extreme branch (all roots inside, or all outside, the unit circle). On two synthetic families and three real-world
  series, nearly every run from such starts converges, in a median of a few hundred steps (Section~\ref{sec:cures}).
\end{enumerate}

\section{Related work}
\label{sec:related}
\emph{Fourier view.} The spectrum of angle-encoded circuits depends only on the encoding~\citep{schuld2021effect,holzer2024spectral}; which
coefficients are realised depends on the gates~\citep{Wiedmann_2024,oh2026fourier}, and their statistics at random parameters are well
studied~\citep{mhiri2025coefficients,Barthe_2024,strobl2026fingerprints}. \citet{duffy2026spectralbiasvariationalquantum} bound each
frequency's gradient by its redundancy. Redundancy is fixed by the encoding and the same for every start; it cannot explain why one circuit
converges from one start and stalls from another.

\emph{Quantum signal processing.} Single-qubit re-uploading circuits are generalised QSP
sequences~\citep{yu2022power,motlagh2024generalized,haah2019product}. Finding the rotation angles of a QSP sequence (its ``phases'' in QSP
terminology) by optimisation fails from random angles and succeeds from a structured start with trivial middle rotations~\citep{Dong_2021}.
For symmetric QSP, \citet{wang2022energy} characterise all global minima by the choice of roots of a factorisation, show that the choice with
all roots inside the unit circle maximises the leading coefficient and lies next to the structured start, where the landscape is strongly
convex for targets of norm $O(1/d)$, and observe that optimisation towards other root choices is slower.
\citet{wang2026representation} train QSP with random rotation angles as a learning model and prove convergence for sparse samples. We carry
the root-choice picture over to general $\mathrm{SU}(2)$ gates trained on data, where the realisations of a branch form an orbit of gauge
rotations rather than a single point; we read the leading-coefficient identity in the gates, as the no-flip product of the middle gates; and
we ask why random starts fail, which the private fraction and the statistics of $\Pi$ on random branches answer. We work in the dense regime
($2E+1$ samples for $2E+1$ coefficients).

\emph{Trainability.} Barren plateaus~\citep{mcclean2018barren,cerezo2021cost} and their Lie-algebraic theory~\citep{ragone2024lie,larocca2023theory}
concern gradients that concentrate as the number of qubits grows; \citet{casado2026controllability} argue that intermediate controllability
trades expressivity against such concentration. Our circuits have one or two qubits, where concentration in the number of qubits plays no role; the larger
circuit trains better, and the obstruction is the rigidity of the smaller one. Escape from small products takes time logarithmic in the start
scale~\citep{saxe2014exact,pesme2023saddle,woodworth2020kernel}; we find such an escape, but it does not decide the outcome. Near-identity initialisations, by identity blocks~\citep{grant2019initialization} or small Gaussian
angles~\citep{zhang2022escaping}, were proposed against barren plateaus; in our setting such a start lies near an extreme QSP branch,
and it trains the single qubit as well (Section~\ref{sec:cures}).

\section{Setting}
\label{sec:structure}
\begin{definition}[VQC with angle encoding]
\label{def:vqc}
An $L$-layer VQC on $N$ qubits applies encoding layers $S(x)$ interleaved with blocks of trainable gates,
$U(x)=W^{(L)}S(x)W^{(L-1)}\cdots S(x)W^{(0)}$, and outputs $f(x)=\langle0|U(x)^\dagger MU(x)|0\rangle$. Each block $W^{(\ell)}$ is a product of
general $\mathrm{SU}(2^N)$ gates; $S(x)$ applies a single-qubit rotation $e^{-ixH/2}$ to each encoded qubit, with a fixed generator $H$ of
eigenvalues $\pm1$ (for instance a Pauli matrix); and $M$ is a non-identity Pauli word.
\end{definition}
We use general gates so that the model class is the largest one the encoding allows (Lemma~\ref{lem:rowrank}): a failure to train cannot be
attributed to a restricted ansatz, and Haar-random starts are defined without reference to a parameterisation.
Without loss of generality we take $H=X$, i.e.\ $R_X(x)=e^{-ixX/2}$, $M=Z$ on one qubit and the initial state $|0\rangle$: any other generator with
eigenvalues $\pm1$, observable or initial state differs from these by fixed unitaries that the neighbouring gates absorb
(Appendix~\ref{sec:background}). Generators with other eigenvalues, such as scaled or ternary encodings, change the spectrum and are not
covered.

\paragraph{Parameterisation.} Every trainable gate is a general $\mathrm{SU}(2^N)$ matrix $G$. Its parameters are the coordinates
$\theta_Q$ of an update $G\mapsto Ge^{i\theta_QQ}$, one for each non-identity Pauli word $Q$ (three per single-qubit gate); a training step
updates these coordinates and multiplies the update into $G$. Unlike Euler angles, this parameterisation has no singular points, and the
counts below do not depend on how a gate is written. $P$ denotes the total number of parameters.

\paragraph{Circuits.} The \emph{serial} circuit has one qubit ($N=1$): $E$ encoding gates $R_X(x)$ interleaved with $E+1$ trainable
single-qubit gates $W_0,\dots,W_E$, with readout $Z$; $W_0$ and $W_E$ are its \emph{end gates}, $W_1,\dots,W_{E-1}$ its
\emph{middle gates}. The \emph{ancilla} circuit, our control, has two qubits: the same $E$ encoding gates,
all on qubit~0, interleaved with $E+1$ general two-qubit gates, with readout $Z$ on qubit~1. Both circuits have the encoding budget $E$ and the same model class (Lemma~\ref{lem:rowrank}); the ancilla
circuit has a second qubit and 15 instead of 3 parameters per gate.

\begin{theorem}[Fourier representation~\citep{schuld2021effect,holzer2024spectral}]
\label{thm:fourier}
With $E$ encoding gates whose generators have eigenvalues $\pm1$, $f_{\boldsymbol\theta}(x)=\sum_{\omega=-E}^{E}c_\omega(\boldsymbol\theta)e^{i\omega x}$
with $c_{-\omega}=\overline{c_\omega}$: no other frequencies occur, whatever the trainable gates; the gates determine the coefficients.
\end{theorem}
\begin{lemma}[Model class]
\label{lem:rowrank}
\textup{(i)}~\citep{yu2022power,motlagh2024generalized} The serial and the ancilla circuit both realise exactly the real trigonometric
polynomials of degree $\le E$ bounded by one, for any number of gates per block (a product of general gates is again a general gate). \textup{(ii)} Their coefficient Jacobian (below) has full rank
$2E+1$ for almost every choice of gates.
\end{lemma}
\begin{proof}[Proof sketch]
(i) For the serial circuit this is the Fej\'er--Riesz construction of QSP; the ancilla circuit contains the serial one as a special case. (ii) The rank is full at
one point, by (i) and Sard's theorem, and each Fourier coefficient is a polynomial in the gate entries, so the gates where the rank drops form the zero set of a nonzero polynomial, a closed null set. Full rank
does not mean good conditioning: at Haar starts with $E=24$, $J$ has full rank, with $\sigma_{\min}/\sigma_{\max}$ as small as $10^{-33}$
(Appendix~\ref{app:nb37}).
Full proof: Appendix~\ref{sec:background}.
\end{proof}

\paragraph{Reference optimum.} By Lemma~\ref{lem:rowrank}(i), the least-squares trigonometric polynomial of degree $E$ is the optimum
of the loss whenever its supremum is at most one, and QSP constructs gates that realise it. We use it as the reference for training and
ask why gradient-based training, the standard way to train re-uploading circuits, does not reach it.

\paragraph{Coefficient Jacobian.} The Fourier coefficients form the real vector
\[\mathbf c=(c_0,\sqrt2\,\mathrm{Re}\,c_1,\sqrt2\,\mathrm{Im}\,c_1,\dots,\sqrt2\,\mathrm{Re}\,c_E,\sqrt2\,\mathrm{Im}\,c_E)\]
in $\mathbb R^{2E+1}$. The factor
$\sqrt2$ (Parseval weighting) makes $\|\mathbf c\|^2=\int f^2\,\frac{dx}{2\pi}$, so distances between coefficient vectors are distances between
functions. The \emph{coefficient Jacobian} $J=\partial\mathbf c/\partial\boldsymbol\theta\in\mathbb R^{(2E+1)\times P}$ depends on the gates, but not on the data or the target. Under gradient descent the coefficients move by
$-\eta JJ^\top\nabla_{\mathbf c}L$. The step size $\eta$ is limited by the largest singular value of $J$, so for linearised steps the error
along the smallest singular direction shrinks at best by a factor $1-2\sigma_{\min}^2/\sigma_{\max}^2$ per step: a badly conditioned $J$
stalls training.

\paragraph{Data and loss.} We train on the $n=2E+1$ equispaced inputs of $[-\pi,\pi)$, the smallest grid that determines a polynomial of degree $E$. On this grid the
values of a degree-$E$ polynomial are $\sqrt n$ times an orthogonal transform of its weighted coefficients (discrete Fourier transform), so
the Jacobian of the outputs is $\sqrt n$ times an orthogonal transform of $J$ and has the same conditioning (Lemma~\ref{lem:bridge}). The loss
is the mean squared error divided by the target's variance, $L=\|f-f^*\|^2/\mathrm{Var}(f^*)$, which equals one for the best constant fit
whatever the target's scale, so that one threshold applies to all targets and degrees: a run \emph{converges} if $L\le10^{-6}$.

\section{Rigidity of the serial circuit}
\label{sec:rigid}
The serial circuit can change its conditioning only by changing its function or its branch; the two-qubit control can also change it at a
fixed function.

\paragraph{Phase-locking and the tangent count.} On one qubit every parameter moves the state, at each input, along the same vector
$|\psi^\perp(x)\rangle$ with its own complex weight (Lemma~\ref{lem:phaselock}); the weights, as functions of $x$, span the circuit's \emph{tangent directions}. Over the $E+1$ blocks this gives the serial circuit at most
$2E+2$ tangent directions for $2E+1$ coefficients, and the ancilla circuit at least $6E+6$ (Proposition~\ref{prop:tangentmain}). Phase-locking
holds at every gate configuration of the serial circuit, at starts that converge and at starts that stall alike. It therefore accounts for
the absence of a spare tangent direction, and through Theorem~\ref{thm:rigid} for the rigidity of the serial circuit, but not for which
starts fail. In our experiments the latter is predicted by the QSP branch (Section~\ref{sec:branch}).

\paragraph{Gauge rotations.} Write $\mathbf W=(W_0,\dots,W_E)$ for the gates of the serial circuit, with $3(E+1)$ parameters. Not every
change of the parameter values changes the function: $E+2$ directions never do, at any $\mathbf W$. They move a rotation $e^{isX}$ about the
encoding axis from one gate to the next (it commutes with $R_X$), rotate the input state $|0\rangle$ by $e^{itZ}$ (a global phase), or rotate by
$e^{itZ}$ before the readout (it commutes with $Z$). These \emph{gauge rotations} form a torus $G=\mathrm U(1)^{E+2}$; they change how the
circuit is written, not what it computes. Where $J$ has full rank, every other direction changes the function. The \emph{fibre} of a
function $f$ is the set of all parameter values that realise it. It contains the gauge orbit of each of its points, and
Theorem~\ref{thm:rigid} shows that, near a point of full rank, it contains nothing else.
\begin{theorem}[Rigidity of the serial circuit]
\label{thm:rigid}
Let the gates $\mathbf W$ have $\mathrm{rank}\,J=2E+1$, and let not both the input state $W_0|0\rangle$ and the readout axis lie on the
encoding axis. Then all gates in a neighbourhood of $\mathbf W$ that realise the same function are gauge rotations of $\mathbf W$, and a gauge
rotation multiplies $J$ from the right by an orthogonal matrix. Hence, near $\mathbf W$, the matrix $JJ^\top$ and the singular values of $J$
depend on the gates only through the function they realise.
\end{theorem}
\begin{proof}[Proof sketch]
At full rank the fibre is locally a manifold of dimension $3(E+1)-(2E+1)=E+2$, and the $G$-orbit inside it has the same dimension; so near
$\mathbf W$ the fibre is the orbit. A gauge rotation rotates the parameter coordinates of each gate, which multiplies $J$ by an orthogonal
matrix. Full proof: Appendix~\ref{app:pf-rigid}.
\end{proof}
Any change of the gates near $\mathbf W$ that keeps the function is therefore a gauge rotation and leaves the singular values of $J$
unchanged: the serial circuit cannot improve its conditioning without changing what it computes. A gradient step moves the coefficients by
$-\eta JJ^\top\nabla_{\mathbf c}L$, which near $\mathbf W$ depends on the gates only through the current function. The ancilla control is
different: at full rank its fibres are manifolds of dimension $13E+14$, while its symmetry group has dimension at most $7E+16$, so at
least $6E-2$ parameter directions keep the function without being gauge; along them we measure changes of the conditioning
(Appendix~\ref{app:numcomp}). Theorem~\ref{thm:rigid} is local. Globally, the fibre of $f$ consists of one gauge orbit for each QSP branch
(Section~\ref{sec:branch}; Lemma~\ref{lem:strip} gives the conditions), so the conditioning depends on the function and its branch.

\section{The QSP branch decides the conditioning}
\label{sec:branch}
\subsection{The private fraction}
The \emph{top coefficient} $c_E$, the coefficient of the highest frequency, is the one gradient descent struggles to fit from Haar
starts (Section~\ref{sec:cost}).
Each row of $J$ is the gradient, with respect to all parameters, of one entry of $\mathbf c$; the two rows $\Jtop$ belong to
$\mathrm{Re}\,c_E$ and $\mathrm{Im}\,c_E$, the other rows $\Jlow$ to the lower coefficients. Let $\Kl$ be the span of the lower rows and $\Pp$ the
projection onto its orthogonal complement: the parameter directions that change no lower coefficient to first order.
\begin{definition}[Private fraction]
\label{def:private}
$\pf=\|\Pp\Jtop^\top\|_F/\|\Jtop\|_F\in[0,1]$.
\end{definition}
\begin{lemma}[Private fraction and smallest singular value]
\label{lem:private}
$\sigma_{\min}(J)\le\|\Pp\Jtop^\top\|_2\le\pf\|\Jtop\|_F$.
\end{lemma}
\begin{proof}
For a unit $w\in\mathbb R^2$ write the
projection of $\Jtop^\top w$ onto $\Kl$ as $\Jlow^\top u$ and set $a=(-u,w)$; then $J^\top a=\Pp\Jtop^\top w$ and $\|a\|\ge1$.
\end{proof}
Like $JJ^\top$, the private fraction is unchanged by gauge rotations (Lemma~\ref{lem:cov}), so by Theorem~\ref{thm:rigid} (Corollary~\ref{cor:rigidflow}) it, too, depends
locally only on the function. In words: $\pf$ is the share of the top coefficient's gradient that lies in directions of its own, which move $c_E$ without disturbing
any lower coefficient. By Lemma~\ref{lem:private}, a small private fraction forces a small singular value of $J$, i.e.\ an ill-conditioned problem on which
gradient descent is slow. A small singular value can also have other causes; where runs stall, i.e.\ stay above the convergence threshold
for tens of thousands of steps (Section~\ref{sec:cost}), we measure that it comes from the top coefficient: there the rank correlation of $\pf$ and $\sigma_{\min}(J)$ is $0.99$, and the
smallest left singular vector of $J$ puts a median $86\%$ of its weight on the two top rows (Appendix~\ref{app:baselines}).

\begin{figure}[t]
  \centering
  \includegraphics[width=\columnwidth]{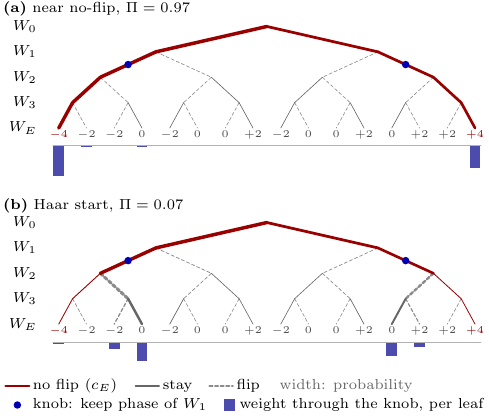}
  \caption{\textbf{Why the top coefficient has a direction of its own near no-flip} ($E=4$). Paths through the serial circuit: $W_0$
  splits the state between the eigenstates $|{\pm}\rangle$ of the encoding, each encoding gate multiplies $|{+}\rangle$ by $z^{-1}$ and
  $|{-}\rangle$ by $z$, and each middle gate keeps the eigenstate (solid) or flips it (dashed). Leaves show the power of $z$; only the two
  paths that never flip (red) reach $\pm E$, and $c_E$ rests on them alone. Edge width is the probability flowing along the edge. Both panels
  share the same Haar-random first and last gates $W_0$ and $W_E$; they differ only in the middle gates. Blue dots mark one of the gates' knobs, the phase
  of $W_1$'s keep amplitude, on the two edges it acts on. It changes the amplitude of every path through the dots, and the bars show how much of their weight ends at each leaf.
  \textbf{(a)} Middle gates with flip amplitude $0.1$: $98\%$ of that weight ends at the two $c_E$ leaves, so the knob moves
  essentially only $c_E$. At $\Pi=1$ this is exact (Proposition~\ref{prop:noflip}), and the private fraction is $0.99$ at $E=24$
  (Appendix~\ref{app:numcomp}).
  \textbf{(b)} A Haar start with $\Pi$ at the Haar median: $93\%$ of that weight ends at lower leaves, so moving $c_E$ moves lower
  coefficients too. The panel illustrates the small private fractions we measure along Haar paths (Section~\ref{sec:cost}).}
  \label{fig:tree}
\end{figure}
\subsection{An exact identity for the distance from no-flip}
The private fraction is hard to control directly; the no-flip product $\Pi$ defined below is not. At $\Pi=1$ the top coefficient's
directions are orthogonal to those of $c_1,\dots,c_{E-1}$. Away from $\Pi=1$, $\Pi$ alone does not determine $\pf$; on the realisations
of our targets, however, large $\Pi$ comes with large $\pf$ and small $\Pi$ with small $\pf$ (end of this section). This subsection shows
that the branch fixes $\Pi$.

In the eigenbasis $|{\pm}\rangle$ of the encoding, each encoding gate multiplies $|{+}\rangle$ by $z^{-1}$ and $|{-}\rangle$ by $z$, with
$z=e^{ix/2}$. A middle gate $W_l$ keeps the eigenstate with amplitude $\alpha_l$ or \emph{flips} it into the other with amplitude $\beta_l$;
$q_l=|\beta_l|^2$ is its \emph{flip probability} and $\Pi=\prod_{l=1}^{E-1}(1-q_l)$ the \emph{no-flip product}. A path through the circuit
therefore carries a power $z^k$ with $-E\le k\le E$, and a pair of paths contributes to the frequency $(k-k')/2$ (Figure~\ref{fig:tree}).
The top frequency $E$ needs the path that stays in $|{+}\rangle$ and the path that stays in $|{-}\rangle$: the two paths that never flip.
At the \emph{no-flip point} ($\Pi=1$) only these two carry weight. A flip at middle gate $l$ then changes only the frequencies $l$ and
$E-l$, a phase change at a middle gate only $c_E$, and an end gate only $c_0$ and $c_E$. So $c_E$ and $c_1,\dots,c_{E-1}$ depend on
disjoint sets of parameters, and their rows of $J$ are orthogonal (Proposition~\ref{prop:noflip}).

Each amplitude of the output state in the readout basis is a sum of powers $z^k$ with $k=-E,-E+2,\dots,E$; written as $z^{-E}a(w)$ and
$z^{-E}b(w)$ with $w=z^2=e^{ix}$, $a$ and $b$ are polynomials of degree $\le E$, with $|a|^2=(1+f)/2$ and $|b|^2=(1-f)/2$ on the unit
circle. Their extreme coefficients $a_0,a_E,b_0,b_E$ come only from the two paths that never flip. By Fej\'er--Riesz, for $c_E\neq0$ the roots of
$w^E(1+f)/2$ come in $E$ pairs $(r_j,1/\bar r_j)$ with $|r_j|\le1$, and $a$ takes exactly one root of each pair; likewise $b$ for $(1-f)/2$.
These choices are the \emph{QSP branch}; for symmetric QSP, \citet[Theorem~4]{wang2022energy} construct all realisations this way. Let
$\lambda^a_j=\log(1/|r_j|)$, and $\varepsilon^a_j=+1$ if $a$ takes the inner root of pair $j$ and $-1$ otherwise; likewise
$\lambda^b_j,\varepsilon^b_j$. Through these numbers the branch enters Proposition~\ref{thm:identity}. On the \emph{extreme} branches all
roots of $a$, and all of $b$, lie on one side of the circle; on \emph{balanced} branches about half lie inside.

\begin{proposition}[No-flip identity]
\label{thm:identity}
Let $c_E\neq0$, $\gamma=|c_E|/2$, $t_a=|a_E|/|a_0|$, $t_b=|b_E|/|b_0|$, and let $\alpha$ and $\beta$ be the angles, on the Bloch
sphere, of the input state $W_0|0\rangle$ and of the readout axis $W_E^\dagger ZW_E$ to the encoding axis. Then
\begin{gather*}
|c_E|=\tfrac12\,\Pi\,\sin\alpha\,\sin\beta,\quad \Pi=\gamma\,\bigl(t_a+t_a^{-1}+t_b+t_b^{-1}\bigr),\\
\log t_a=\textstyle\sum_j\varepsilon^a_j\lambda^a_j,\qquad \log t_b=\sum_j\varepsilon^b_j\lambda^b_j .
\end{gather*}
For every circuit, also with $c_E=0$, $\Pi=|a_0|^2+|a_E|^2+|b_0|^2+|b_E|^2$.
\end{proposition}
\begin{proof}[Proof sketch]
The extreme powers $z^{\pm E}$ come only from the two paths that never flip; their weights add up to $\Pi$, which gives
$\Pi=|a_0|^2+|a_E|^2+|b_0|^2+|b_E|^2$. The end gates decide how this weight splits between the four coefficients, which gives the
end-gate form. The $\gamma$-form uses $|a|^2+|b|^2\equiv1$, and the form of $\log t_a$ uses $a=a_E\prod_j(w-\zeta_j)$ over the chosen
roots. Full proof: Appendix~\ref{app:branch}.
\end{proof}
The $\gamma$-form is the general-gate analogue of the leading-coefficient identity of \citet[Lemma~18]{wang2022energy}; the end-gate form
extends Proposition~\ref{prop:noflip}(i) from $\Pi=1$ to every circuit and reads the identity in the gates. The angles $\alpha$ and $\beta$
depend only on the end gates: $\alpha=0$ means the input state is an eigenstate of the encoding, $\alpha=\pi/2$ an equal superposition of
both, and likewise for $\beta$. For a fixed function $|c_E|$ is fixed, so the branch fixes $\Pi$, and $\Pi=2|c_E|/(\sin\alpha\sin\beta)$:
since $c_E$ needs both no-flip paths, the more unevenly the end gates split the weight between the two eigenstates, the less weight the
middle gates may lose to flips. Every realisation of $f$ has $\Pi\le\Pi_{\max}$, a bound fixed by $f$ alone (Corollary~\ref{cor:extreme}). It is below one
as soon as $f$ has lower frequencies, since only paths with flips produce them, and the extreme branches attain it; on random branches
$\Pi$ stays near the floor $2|c_E|$ of Corollary~\ref{cor:random} (measured, below).
\begin{corollary}[Extreme branches are nearly no-flip]
\label{cor:extreme}
\textup{(i)} If $c_E\neq0$, every realisation of $f$ has $\Pi\le\Pi_{\max}=G_++G_-+\gamma^2(G_+^{-1}+G_-^{-1})$ with $G_\pm=\exp\int\log\tfrac{1\pm f}{2}\tfrac{dx}{2\pi}$,
with equality exactly when $a$ and $b$ each have all roots in the closed unit disc or all roots outside the open disc (for both in the disc: the
\emph{minimum-phase} realisation, a term from signal processing).
\textup{(ii)} If $|f|<1$, such a realisation has total flip
$\sum_l\log\frac1{1-q_l}\le\frac12\int\log\frac1{1-f^2}\frac{dx}{2\pi}\le\frac{\langle f^2\rangle}{2(1-\|f\|_\infty^2)}$, independent of $E$, with $\langle f^2\rangle=\int f^2\frac{dx}{2\pi}$.
\end{corollary}
\begin{proof}[Proof sketch]
Jensen's formula holds on every branch: $G_+=|a_E|^2P_{\rm out}^2$ and $|a_0|=|a_E|P_{\rm in}P_{\rm out}$, with $P_{\rm in}$ and
$P_{\rm out}$ the products of $\min(1,|\zeta_j|)$ and $\max(1,|\zeta_j|)$ over the roots of $a$. With $\gamma=|a_E||a_0|$,
$G_++\gamma^2/G_+-|a_0|^2-|a_E|^2=|a_E|^2(1-P_{\rm in}^2)(P_{\rm out}^2-1)\ge0$, zero exactly when no root lies on one side of the
circle; likewise for $b$, and add. Full proof: Appendix~\ref{app:branch}.
\end{proof}
Part (i) is the general-gate form of \citet[Lemma~18]{wang2022energy}, who call the inside-disc realisation the maximal solution; part (ii)
parallels their Theorem~5, which bounds its distance from the structured start independently of the degree when $\|f\|_\infty\le\frac12$.
\begin{corollary}[The floor]
\label{cor:random}
Every realisation has $\Pi\ge2|c_E|$; if $c_E\neq0$, with equality iff $t_a=t_b=1$.
\end{corollary}
\begin{proof}[Proof sketch]
$t+t^{-1}\ge2$ in Proposition~\ref{thm:identity}, with equality iff $t=1$. Full proof: Appendix~\ref{app:branch}.
\end{proof}
On random branches $\Pi$ stays near the floor $2|c_E|$ and far from one: at $E=24$ and amplitude $0.9$ the median of $\Pi/(2|c_E|)$ over
2{,}000 random branches is $1.26$, against $\Pi_{\max}\approx0.9$ on the extreme branches (check T3). Heuristically, with about $E/2$ inner
roots the signed sums $\sum_j\varepsilon_j\lambda_j$ largely cancel.
\begin{proposition}[The branch count is invariant]
\label{prop:branchcount}
Along a continuous path of gates on which $f>-1$ (resp.\ $f<1$) everywhere, the number $k_a$ (resp.\ $k_b$) of zeros of $a$ (resp.\ $b$) in
the open unit disc $|w|<1$ is constant.
\end{proposition}
\begin{proof}[Proof sketch]
On the unit circle $|a|^2=(1+f)/2>0$, so no zero of $a$ crosses the circle, and $k_a=(2\pi i)^{-1}\oint_{|w|=1}a'(w)/a(w)\,dw$ is continuous and integer-valued along the path. Full proof: Appendix~\ref{app:branch}.
\end{proof}
Along a continuous path of gates the branch count therefore changes only where the function touches $\pm1$. A run therefore keeps
the branch it starts on unless its function touches $\pm1$, and a start that begins balanced stays near the floor. Gradient descent moves in
discrete steps, and along 40 runs all 32 changes we observed occurred where $|f|\ge1-10^{-2}$ (check T4). Branches and the gauge rotations of
Section~\ref{sec:rigid} together describe all realisations: two realisations of $f\not\equiv\pm1$ with the same root choices are gauge rotations of
each other if one of them has no middle gate that is a full flip (Lemma~\ref{lem:strip}, by layer stripping; for symmetric phases the corresponding uniqueness is
Theorem~1 of \citealp{wang2022energy}). Theorem~\ref{thm:rigid} is the local form of this
statement. Each branch carries its own conditioning.

\paragraph{From $\Pi$ to $\pf$.} The identity speaks about $\Pi$; gradient descent cares about $\pf$. At $\Pi=1$ the top rows of $J$ are orthogonal
to the rows of $c_1,\dots,c_{E-1}$ (Proposition~\ref{prop:noflip}), so only the row of $c_0$, which the end gates also move, can lower $\pf$;
with Haar end gates we measure $\pf=0.99$ at $E=24$. Away from $\Pi=1$ no bound in terms of $\Pi$ alone can hold: with random gate
phases, $\pf$ drops to $0.08$ at $\Pi=0.998$, and special phases make $J$ singular to machine precision even at $\Pi=0.99$ (check C6,
Appendix~\ref{app:nb37}). What does bound $\pf$ is a factorisation through the state. Since $f=2|a|^2-1$ on the unit
circle, $J=RD$, with $D$ the Jacobian from the gates to the coefficients of $(a,b)$ and $R$ the readout $\delta a\mapsto4\,\mathrm{Re}(\bar a\,\delta a)$;
hence
\begin{equation*}
  \kappa_D\,\pf_R\;\le\;\pf\;\le\;\pf_R/\kappa_D
\end{equation*}
(Proposition~\ref{prop:readout}), where $\kappa_D=\sigma_r(D)/\sigma_1(D)$, $r=\operatorname{rank}D$, measures how evenly the gates move the
state, and the \emph{readout privacy} $\pf_R$ depends, for generic gates, only on $(a,b)$ up to unit factors, that is, on the function and
its branch. On the extreme branches, the branches of largest $\Pi$, $\pf_R$ is bounded below by an explicit function of $f$ that does not
depend on $E$, for instance $\pf_R\ge(1-F)\sqrt{(1+F)/(1-F+(1+F)^2)}$ with $F=\|f\|_\infty$ (Theorem~\ref{thm:privacy-extreme}); at amplitude
$0.9$ the full bound of the theorem is about half the measured $\pf_R$. That gate control stays bounded there we have measured, not proved:
in the coordinates of Theorem~\ref{thm:nnf}, $\kappa_{D_S}\ge0.72$ at the minimum-phase realisations of fresh amplitude-$0.9$ targets from $E=8$
to $128$, without trend in $E$, and Proposition~\ref{prop:gate-control} writes it exactly in terms of the partial products of the circuit.
Together, the private fraction in these coordinates is at least $0.22$ at all 240 such realisations from $E=16$ to $128$, with only the
measured gate control unproved; in the coordinates of Section~\ref{sec:structure}, $\pf$ is within a factor $1.25$ of $\pf_R$ at every
minimum-phase realisation of Appendices~\ref{app:readoutexp} and~\ref{app:d13}. Where Haar runs stall, $\pf_R$ carries the branch: at the stalled points of the restart experiment of
Section~\ref{sec:cost} and of a repetition at $E=16$, and at their realisations on other branches, $\pf_R\le10^{-2}$ implies $\pf\le10^{-2}$
at all 113 such points, against 69 of 78 for $\Pi\le10^{-2}$ (Appendices~\ref{app:readoutexp}--\ref{app:d14}).

\section{Why random starts fail}
\label{sec:cost}\label{sec:exact}\label{sec:gd}\label{sec:lm}\label{sec:adam}\label{sec:mechanism}
This section and the next test the theory of Sections~\ref{sec:rigid}--\ref{sec:branch} in training; the numerical checks of the individual
statements are listed in Table~\ref{tab:status}. Haar starts begin far from the no-flip point: for Haar gates the flip probabilities of the
middle gates are independent and uniform on $[0,1]$, so $-\ln\Pi$ is a sum of $E-1$ independent standard exponentials,
$-\ln\Pi\sim\mathrm{Gamma}(E-1,1)$, and $\Pi\approx e^{-(E-1)}$ at the start (Remark~\ref{rem:clt}; check T1).

\paragraph{Experimental setup.} Targets are random real trigonometric polynomials of degree $E$, scaled to $\sup|f^*|=0.9$ so that both
circuits can represent them exactly. \emph{Flat} targets draw all coefficients from the same distribution; \emph{$1/k$} targets let the size of
$c_k$ decay like $1/k$, as smooth data do, so their top coefficient is small. They are harder for every circuit, since the gradient on $c_E$
is proportional to its residual. We compare the serial circuit with the ancilla control (one gate per block; at matched parameter counts the
gap persists, Appendix~\ref{app:additional}) from Haar-random gates, the uniform distribution on
the gates; uniform rotation angles, the common practical start, behave the same (Table~\ref{tab:starts}). Optimisers: gradient descent with
Armijo backtracking~\citep{armijo1966minimization}, Adam~\citep{kingma2014adam} with a tuned learning rate, and Levenberg--Marquardt
(LM)~\citep{levenberg1944method,marquardt1963algorithm}. Every target and start has its own seed, shared by all circuits and optimisers.
The experiments behind Tables~\ref{tab:branch}--\ref{tab:data} are confirmatory: we fixed their predictions before running them, and
Appendix~\ref{app:valid} lists all predictions with their outcomes, including those that failed. Full protocol:
Appendix~\ref{app:experiments}.

\begin{figure*}[t]
  \centering
  \includegraphics[width=\textwidth]{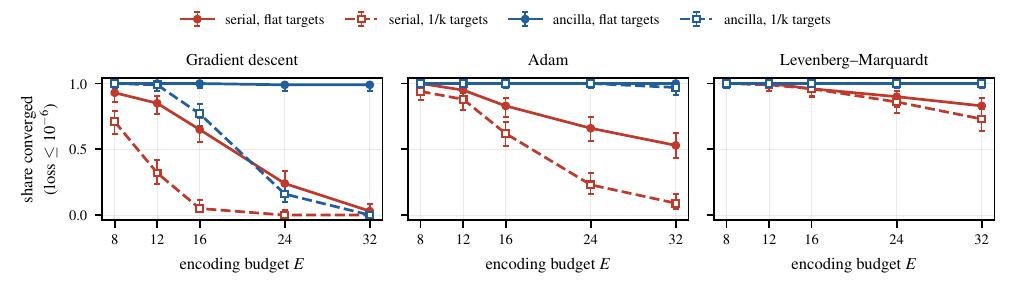}
  \caption{Share of runs that converge (final loss $\le10^{-6}$) against the encoding budget $E$, serial circuit and ancilla control from
  Haar-random gates, flat and $1/k$ targets; 100 runs per point, 95\% Wilson intervals. Budgets: gradient descent and Adam 50{,}000 steps,
  LM 2{,}000 iterations; protocol in Appendix~\ref{app:experiments}.}
  \label{fig:convergence}
\end{figure*}

\paragraph{The gap.} Figure~\ref{fig:convergence} shows the share of runs that converge. Under gradient descent the serial circuit fits flat targets
in 65 of 100 runs at $E=16$ and in 3 at $E=32$, the control in 100 and 99; Adam and LM narrow the gap without closing it at $E\ge24$, and on
real-world series the gap persists (Section~\ref{sec:cures}). The failure is not one of step size: gradient descent uses a line search and
Adam a tuned learning rate. More steps help the serial circuit only partly, while the control converges throughout (Table~\ref{tab:budget}).

\paragraph{Escape, then stall.} An obvious explanation for the failure is the small start: from Haar starts $|c_E|\approx10^{-10}$ at $E=24$
(Proposition~\ref{thm:identity} with $\Pi\approx e^{-23}$), and for products of many small factors the time to escape grows with the logarithm of
the start scale~\citep{saxe2014exact}. This explains the delay but not the outcome. Every run first grows $|c_E|$, by about a decade per 2,000
steps, and none converges during this escape. All Haar starts begin at the same scale, so the start cannot tell runs apart (AUC $0.53$).
Which runs stall is decided later: at 25k steps the private fraction separates runs that converge later from runs that never do
(cross-validated AUC $0.80$ over Haar and uniform-angle starts pooled, $0.77$ on Haar starts alone, against $0.60$ for the gradient norm
and $0.41$ for the current loss, so it does not just track progress), as does the conditioning $\kappa_J=\sigma_{\min}(J)/\sigma_{\max}(J)$
($0.77$, pooled), which by Lemma~\ref{lem:private} measures the same direction.

\paragraph{The stall belongs to the branch.} Haar starts are balanced: at $E=24$ their branch counts lie in $[9,15]$, never near $0$ or $E$.
Their counts change by $\pm1$ or $\pm2$, mostly in the first 100 steps, each time where the function comes within $10^{-3}$ of $\pm1$ (check T4);
between 12.5k and 25k steps they are unchanged in $95\%$ of unconverged runs, between 25k and 50k in $97\%$ (Proposition~\ref{prop:branchcount}; Appendix~\ref{app:starts}); along the paths
$\Pi/(2|c_E|)\approx1.1$--$1.4$, near the floor of Corollary~\ref{cor:random}. In the end-gate form, $\sin\alpha\sin\beta$ has risen from a median $0.65$ at
the start to $0.7$--$0.9$: the end gates turn towards the equator, which can raise $|c_E|=\frac12\Pi\sin\alpha\sin\beta$ by at most a factor
$1.5$ at the median, while the remaining eight to nine decades would have to come from the middle gates. By Section~\ref{sec:rigid}, the conditioning of a stalled run is a property of its function and its branch. To separate the two, we take
every run that has not converged at 25k or 50k steps, compute from its current function $f$ (not the target) gates that realise $f$ exactly
on the minimum-phase branch, and restart gradient descent from them towards the same target (Table~\ref{tab:branch}). Both starting
points compute the same function and have the same loss; they differ only in the branch. The private fraction jumps from $10^{-6}$--$10^{-3}$ to above $0.5$, and the restart converges at every point within 5k steps, while
continuing the run converges at $1$--$13\%$ and a random branch at $4$--$38\%$. The same function is a stall on its balanced branch and not a
stall on the extreme one. For symmetric QSP phase finding, \citet[Section~7.3]{wang2022energy} observed that optimisation towards
root choices other than the inside-disc one is markedly slower; the private fraction is our account of why.

\begin{table}[t]
\centering\small
\caption{Restarting stalled runs on another branch of the same function (flat targets, Haar starts; fresh seeds, predictions fixed in
advance; Appendix~\ref{app:branchE}). Points: runs not converged at 25k or 50k steps. Restarts: 5k gradient-descent steps. The points pool the 25k and 50k snapshots of 40, 60 and 40 runs (28+17, 59+49 and 40+39 unconverged runs), so a run that stalls throughout counts twice. At $E=24$ the realisations of one point miss the $10^{-8}$ reconstruction check; the rows on realisations use the other 107.}
\label{tab:branch}
\setlength{\tabcolsep}{3pt}
\begin{tabular}{lccc}
\toprule
 & $E=16$ & $E=24$ & $E=32$ \\
\midrule
stalled points & 45 & 108 & 79 \\
run's $\pf$ (median) & $3\times10^{-3}$ & $6\times10^{-4}$ & $1\times10^{-6}$ \\
min.-phase: $\pf\ge0.5$ & $100\%$ & $100\%$ & $100\%$ \\
restart converges: min.\ phase & $100\%$ & $100\%$ & $100\%$ \\
\quad random branch & $38\%$ & $12\%$ & $4\%$ \\
\quad same gates & $13\%$ & $3\%$ & $1\%$ \\
\bottomrule
\end{tabular}
\end{table}

The effect grows with $E$: the run's private fraction falls from $3\times10^{-3}$ to $10^{-6}$ between $E=16$ and $32$, and at $E=16$ a random
branch is already noticeably better than the run's own (two of the predictions at $E=16$ failed for this reason; Appendix~\ref{app:branchE}).

\paragraph{Curvature-aware optimisers.} Methods that are not slowed by a small singular value avoid the stall on these targets: from 80 Haar
starts at $E=24$, natural gradient converges in 80, L-BFGS in 75, LM in 74, damped Gauss--Newton in 72, Adam in 48 and gradient descent in 17
(Appendix~\ref{app:nb31}). L-BFGS is the cheap one: median cost $7.6\times10^3$ forward-pass equivalents, against $2.4\times10^5$ for gradient descent on its
15 easiest runs. They do not undo every stall: LM restarted at gradient descent's stalled points converges in 23 of 35 runs that gradient
descent never finishes (Appendix~\ref{app:nb31}). They pay with a Jacobian or curvature estimate per step, and on data they do not cure
every stall (Section~\ref{sec:cures}).

\section{Starting near no-flip}
\label{sec:cures}\label{sec:remedies}
By Lemma~\ref{lem:strip} and Theorem~\ref{thm:rigid}, the conditioning of a serial circuit with $\Pi>0$ is fixed by the function it
computes and by its QSP branch; gradient descent cannot improve it while it stays on the same function and branch. We therefore choose the branch instead, at initialisation, where it costs nothing. None of the
starts below uses the target.

\paragraph{Starts.} Table~\ref{tab:starts} compares three starts with large $\Pi$ against Haar starts, on fresh seeds ($E=24$;
Appendix~\ref{app:starts}). The \emph{minimum-phase} start realises an unrelated random function exactly on the minimum-phase branch, a
computation done once and offline that depends only on $E$. The \emph{no-flip} start makes every middle gate a rotation by a random angle
about the encoding axis, tilted off that axis by a flip amplitude of $0.1$; such a gate commutes with the encoding up to the tilt, so it
leaves the encoding eigenstates nearly in place. The \emph{small-middle} start is the special case of a rotation angle near zero, middle
gates near the identity, with Haar end gates. It is the near-identity start proposed against barren plateaus~\citep{grant2019initialization,zhang2022escaping}
and the structured start of QSP phase finding~\citep{Dong_2021}. All three start with $\Pi\ge0.79$, against $10^{-9.6}$ for Haar, and all
three converge where Haar starts do not: minimum phase and small middle in every run within 500 steps, no-flip in 54 of 60 runs, more slowly
(Figure~\ref{fig:starts} in Appendix~\ref{app:starts}). Uniform rotation angles start on balanced branches like Haar gates ($\Pi\approx10^{-10}$,
counts $(12,12)$) and converge in 9 and 0 of 30 runs, against 4 and 0 for Haar on the same targets (Appendix~\ref{app:nb35}).
In every arm, angles only sample the start: training updates the gate matrices in Lie-algebra coordinates. Trained in Euler angles, the
small-middle start would sit at their coordinate singularity at the identity, which we did not study.

\begin{table}[t]
\centering\small
\caption{Starts for the serial circuit, $E=24$, 30 runs per target family, gradient descent for 50k steps (Appendix~\ref{app:starts}; fresh
seeds, the same start seeds in every arm). Steps: median first step at loss $\le10^{-6}$ over converged runs, logged every 500 steps. Branch
counts $(k_a,k_b)$ are medians at the start; $\Pi$ at the end is the median over converged runs. Uniform angles: every gate a
\texttt{qml.Rot} with angles uniform in $[0,2\pi)$, on the targets and start seeds of the Haar row (NB35). Angles only sample the start: for
every start the optimiser acts on the gate matrices; training in Euler angles, which are singular at the identity, was not studied.}
\label{tab:starts}
\setlength{\tabcolsep}{1.6pt}
\begin{tabular}{lccccccc}
\toprule
 & \multicolumn{2}{c}{converged} & \multicolumn{2}{c}{steps} & & & end \\
start & flat & $1/k$ & flat & $1/k$ & $\Pi(0)$ & $(k_a,k_b)$ & $\Pi$ \\
\midrule
minimum phase & 30 & 30 & $\le$500 & $\le$500 & 0.93 & (24,24) & 0.9 \\
no-flip & 30 & 24 & 1.3k & 9.8k & 0.79 & (16,7) & 0.3--0.5 \\
small middle & 30 & 30 & $\le$500 & $\le$500 & 0.98 & (20,0) & 0.8--0.9 \\
Haar & 4 & 0 & 30k & -- & $10^{-9.6}$ & (12,12) & 0.13 \\
uniform angles & 9 & 0 & 25k & -- & $10^{-10.3}$ & (12,12) & 0.11 \\
\bottomrule
\end{tabular}
\end{table}

\paragraph{Large $\Pi$, not small angles.} A natural reading is that these starts work because they are close to the identity, as the
near-identity start proposed against barren plateaus is. They are not: the middle gates of the
minimum-phase start are rotations by a median angle of $\pi$, those of the no-flip start by $1.6$, and those of the small-middle start by
$0.05$; Haar gates lie at $2.3$. What the three share is a large no-flip product. The flip amplitude $\varepsilon=|\beta_l|$ of the no-flip start's middle gates
(Section~\ref{sec:branch}), with $\Pi=(1-\varepsilon^2)^{E-1}$, acts as a dose: convergence falls monotonically with it, in 30, 25, 19, 11, 3 and 0 of 30 runs for $\varepsilon=0.1,0.3,0.5,0.7,0.9,0.995$ on flat
targets, and 24, 2, 0, 0, 0, 0 on $1/k$ targets (Appendix~\ref{app:sweep}). The no-flip start also shows that the branch count is only a coarse
summary. Its counts are mixed and change in about two thirds of its runs, yet it converges: its middle gates keep $\Pi$ large, and it fits the
target's small top coefficient by tilting its end gates, ending at $\Pi/(2|c_E|)=1/(\sin\alpha\sin\beta)$ with median $6.3$ (flat) and $21$
($1/k$), far above the floor. Along its converged paths the private fraction stays above $10^{-3}$, while along Haar paths it falls to a median
minimum of $10^{-9}$, at the resolution of float64 (Appendix~\ref{app:noflip}). In our runs, what the converging starts share is $\Pi$ well above
$2|c_E|$, and convergence falls as the flip amplitude grows; a near-identity start is one way to get large $\Pi$, since a rotation by a small
angle barely flips.

\paragraph{On data.}\label{sec:data} Table~\ref{tab:data} repeats the comparison on data: three real-world series and two synthetic families,
serial circuit with five gates per block, $E=24$, nottem chosen before the screen of Appendix~\ref{app:experiments}, the extreme-branch gates in the first gate of each block (with $t$ gates per block, $JJ^\top$ is $t$ times that of the circuit
of block products, so rank and conditioning are unchanged; Appendix~\ref{sec:background}; under Adam, which this identity does not cover, the stacking does not change the outcome either, Appendix~\ref{app:nb38}). Gradient descent from Haar starts
reaches the least-squares optimum in no run; from small middle angles or from a minimum-phase start in 28--30 of 30 on every
data set, and Adam from the same starts in 22--30. The serial circuit then matches the two-qubit control on all three real series, and exceeds it on the bump
target (14 of 30 for the control). The start
is also cheap: gradient descent from a minimum-phase start reaches the optimum in a median of 100--700 steps,
or $0.6$--$4.2\times10^3$ forward-pass equivalents, 7 to 70 times less than L-BFGS from Haar starts and two to three orders of magnitude less
than natural gradient. The curvature-aware optimisers, which cure the stall on exactly representable targets, do so only partly here.
Natural gradient is slow rather than stuck: each of its steps costs a full Jacobian, and with 20k instead of 5k iterations it reaches the optimum on four of five data sets (30 of 30 runs),
but not on the bump target (9). L-BFGS stalls on Nottingham (10 of 30 after 50k iterations, 9 after 20k). On held-out points (the 30 runs of Table~\ref{tab:data}), the held-out error $1-R^2$ of Adam from Haar
starts is a median $0.4\%$, $8\%$ and $9\%$ above that of the optimum on the three real series (at most $17\%$), that of gradient descent
from Haar starts $62$--$1200\%$ above (Appendix~\ref{app:data}).

\begin{table}[t]
\centering\small
\caption{Runs out of 30 that reach the least-squares optimum (excess training loss $\le10^{-6}$ of the variance) and, in brackets, the median
work of those runs in thousands of forward-pass equivalents (a gradient costs 3, a Jacobian $P=375$). Serial circuit with five gates per block
unless noted, $E=24$ (Appendix~\ref{app:data}); budgets 20k iterations unless noted. GD: gradient descent; NG: natural gradient; MP: minimum phase; SM: small middle; bump: Poisson-kernel targets; nottem, gas(price), ibm(close):
real series. Work counts every function evaluation, including line searches, plus 3 per gradient and $P$ per Jacobian; it measures classical
statevector simulation of this parameterisation (with one gate per block $P$ would be $75$; on hardware a parameter-shift gradient costs $2P$
circuit evaluations). Fresh targets, start seeds and learning-rate selection (Appendix~\ref{app:data}), so the Haar rows differ from the first
30 runs of the main experiment in Table~\ref{tab:budget}.}
\label{tab:data}\label{tab:cost}
\setlength{\tabcolsep}{1.5pt}\footnotesize
\begin{tabular}{lccccc}
\toprule
 & flat & bump & nottem & gas & ibm \\
\midrule
\multicolumn{6}{l}{\emph{Haar starts}}\\
GD & 0 & 0 & 0 & 0 & 0 \\
Adam & 16 (17) & 0 & 2 (37) & 3 (54) & 1 (48) \\
L-BFGS (50k) & 26 (8.1) & 25 (44) & 10 (16) & 21 (32) & 30 (29) \\
NG & 30 (420) & 9 (2900) & 30 (590) & 30 (2500) & 30 (3000) \\
\midrule
\multicolumn{6}{l}{\emph{Starts of this section}}\\
GD, MP & 30 (0.6) & 30 (0.6) & 30 (1.2) & 30 (4.2) & 30 (0.6) \\
GD, SM & 30 (0.6) & 28 (0.6) & 30 (1.2) & 30 (4.8) & 28 (0.6) \\
Adam, MP & 30 (0.6) & 30 (0.6) & 30 (0.6) & 30 (0.6) & 30 (0.6) \\
Adam, SM & 30 (1.6) & 22 (7.5) & 30 (13) & 30 (4.7) & 30 (5.7) \\
\midrule
\multicolumn{6}{l}{\emph{Two-qubit control, Haar starts}}\\
Adam & 30 (3.1) & 14 (44) & 30 (16) & 30 (20) & 30 (18) \\
\bottomrule
\end{tabular}
\end{table}

\section{Discussion and limitations}
\label{sec:conclusion}
The serial circuit and the two-qubit control represent the same functions. What separates them in training is that the serial circuit is
rigid: it cannot change how it realises a function without changing the function, so it inherits the conditioning of whichever QSP branch
it is on. Haar starts put it on a balanced branch, far from no-flip, where in our measurements the top coefficient has almost no private direction; the extreme
branches are close to no-flip for every $E$. In the gates, $|c_E|=\frac12\Pi\sin\alpha\sin\beta$: from a Haar start the end gates can supply
only a small factor, and the middle product is many decades too small. Starting there, or moving there, removes the stall for plain gradient descent and for Adam, on
synthetic targets and on real data. This agrees with the account of structured starts in QSP phase finding by \citet{wang2022energy}, and extends it to random starts and
general gates.

\emph{Only the top coefficient?} The stall concerns one coefficient, so what it costs in loss is the target's energy at the highest
frequency: a few percent of the variance for flat targets, much less for decaying spectra, and on our real series a held-out error a median $0.4$--$9\%$ above that of
the optimum for Adam from Haar starts. We do not read our results as a statement about generalisation; they concern
optimisation. The degree $E$ is the modeller's choice. Where the target has content at frequency $E$, as our synthetic targets do, missing it
is underfitting, not regularisation; and a circuit that cannot reach its own top frequency behaves like a model of lower degree while paying
for degree $E$ in parameters and steps. Raising $E$ to compensate makes the stall worse (Figure~\ref{fig:convergence}).

\emph{Limitations.} On the extreme branches the link from $\Pi$ to $\pf$ is proved up to one measured constant, and only for the private
fraction in the coordinates of Theorem~\ref{thm:nnf}: Proposition~\ref{prop:readout} bounds it on both sides by $\pf_R$ and
$\kappa_{D_S}$, Theorem~\ref{thm:privacy-extreme} bounds $\pf_R$ below, and that $\kappa_{D_S}$ is bounded below independently of $E$ we
have measured up to $E=128$ and reduced to two frame constants of the circuit's partial products (Corollary~\ref{cor:frames}), but not
proved; the stability bounds of \citet{ni2025inlfft} for inverting the nonlinear Fourier transform also grow with the length.
Theorem~\ref{thm:nnf} bounds gate control only for $(E-1)\ln(1/\Pi)<0.31$; the singular point of check C6 lies in that regime, so what fails
there is the readout of the state, not its control. That $\pf_R$ is small on the balanced branches where Haar runs stall is measured ($E=16$
and $24$), not proved, and there gate control matters too: the median of $\log\pf_R$ is $58\%$ and $68\%$ of that of $\log\pf$. Which slow no-flip runs converge we
cannot predict. Rigidity is local and assumes full rank. Realising a function from its coefficients alone is ill-conditioned when $|c_E|$ is small: at stalled Haar points with
$|c_E|\lesssim10^{-3}$ the realisation from $f$ in float64 coefficients need not reproduce the run's branch, although the gates stripped from
the run's own $(a,b)$ always do (Appendix~\ref{app:unique}). Inputs are one-dimensional, simulation is noise-free, and the confirmatory experiments use the dense regime of $2E+1$ samples or
three real-world series. The
step sizes of our gradient descent exceed the edge of stability along most of the path ($\eta\lambda_{\max}$ median $2.1$), so bounds that
assume small steps describe the mechanism, not the runs. Our Lean check of the proofs (Appendix~\ref{app:lean}) covers the
algebraic statements, with cited analytic results as hypotheses.

\section{Conclusion}
\label{sec:final}
Single-qubit re-uploading circuits represent the same functions as their two-qubit control, yet from Haar starts they fail ever more often, as $E$ grows,
to fit the top Fourier coefficient. We traced this failure to the way the function is realised, not to the model class. The serial circuit is rigid:
near almost every choice of gates, every realisation of the same function is a gauge rotation of it, so the conditioning depends only on the
function and on its QSP branch. An exact identity, $|c_E|=\frac12\Pi\sin\alpha\sin\beta$, ties the top coefficient to the no-flip product
$\Pi$ of the middle gates: the extreme branches are close to no-flip for every $E$, the balanced ones sit near the floor $\Pi=2|c_E|$. Haar
starts land on balanced branches; along a continuous path the branch count changes only where the function touches $\pm1$, and the
gradient-descent runs we measured stay on balanced branches, with a top coefficient that has almost no direction of its own. Starting on or near an extreme branch, by a minimum-phase, a
no-flip or a near-identity start, removes the stall for gradient descent and Adam, on synthetic targets and on real data. For
first-order training of serial re-uploading circuits, what matters is the branch a start lands on. The common random starts, Haar
gates as well as uniform rotation angles, land on balanced branches, and it is the branch that fails, not the circuit.

The stall is not confined to the top coefficient. The frequencies just below it are carried by few paths as well ($\binom Ej$ paths have the power $z^{E-2j}$, and a frequency
needs a pair of them; Figure~\ref{fig:tree}), and at the points where Haar runs stall they share its conditioning: the private fractions of $E-1$, $E-2$
and $E-3$ are at most three times that of the top (medians from $10^{-2}$ at $E=16$ to $10^{-16}$ at $E=32$, at 80 digits), on flat targets 90--100\% of the
residual sits in the four highest frequencies, and on flat targets at $E=24$ the two smallest singular values of $J$ lie three orders of magnitude below the third. The minimum-phase realisations of the same functions make each of the four private
($\pf\ge0.5$ at every point; Appendix~\ref{app:nb35}). On flat targets the stall thus lacks about one complex direction, shared by a band near the top; on $1/k$ targets more directions are
missing, over a wider band (Appendix~\ref{app:nb35}). The extreme branch restores the top four at every point.

What remains open is higher dimensions. In one dimension adding a single qubit is a cheap cure. In higher
dimensions one ancilla can still serve all inputs, but noise limits the depth of near-term circuits~\citep{preskill2018quantum}, which
favours spreading the encoding over more qubits, and barren plateaus~\citep{mcclean2018barren,larocca2023theory} limit how far that can go.
Our analysis does not carry over as it stands: a circuit of fixed width whose depth grows linearly in $E$ has far fewer tangent directions
than the $(2E+1)^d$ Fourier coefficients, so serial and ancilla circuits no longer share a model class, and QSP branches have no general
analogue in several variables~\citep{rossi2022multivariable}. The path structure does carry over. With every input encoded on one qubit by
the same generator, the coefficient at a corner $(s_1E,\dots,s_dE)$ of the spectrum is the top coefficient of the circuit restricted to
$x_r=s_rx$, so it is reached by a single pair of paths, which flip wherever the encoding switches between inputs of opposite sign in $s$
(Corollary~\ref{cor:corner}; checked at 1{,}680 corners, $d=2,3$). With alternating inputs, a start near no-flip therefore leaves the coefficient at $(E,-E)$ (for $d=2$) out of reach
(median $|c|=3\times10^{-14}$ at $E=3$, flip amplitude $0.05$, against $0.28$ at $(E,E)$); with the inputs in blocks the paths flip only at the
block boundaries, where a start then needs some flip amplitude, and it stays within reach ($8\times10^{-4}$). Which start, and which encoding order, serves all corners at once is the question this raises.

\section*{Acknowledgements}
This work was supported by the LMU Sustainability Fund (EfOiE), the German Federal Ministry of Research, Technology and Space (BMFTR)
through QuCUN, QuaRDS, CAQAO, IgnisQNC, Q-NextGrid and the Bavarian Ministry of Economic Affairs project 6GQT.

\section*{AI Use Statement}
This work was carried out with the assistance of a large language model (Claude, Anthropic), which supported the development of the
proofs, the code of the experiments and the writing of the paper. To check the written proofs, we formalised them in Lean~4 with the help of
the language model and, in early rounds, of the automated prover Aristotle (Harmonic), and compiled them with \texttt{lake};
Appendix~\ref{app:lean} matches each written proof to its Lean statement and lists the steps argued by hand, and the Lean files are in the
repository (\repourl) for readers who want to retrace the proofs. The authors take full
responsibility for the content of the paper.

\section*{Impact Statement}
This paper presents work whose goal is to advance the field of machine learning, specifically the understanding of the trainability of
variational quantum circuits. There are many potential societal consequences of our work, none which we feel must be specifically
highlighted here.

\bibliography{phaseLocking}
\bibliographystyle{icml2027}

\appendix
\onecolumn
\section*{Overview of the appendices and the repository}
The appendices are ordered from statements to proofs to experiments.
\begin{itemize}
  \item Appendix~\ref{app:proofs}: all statements with their proofs, in the order of the main text: circuits, Jacobians and data
  (\ref{sec:background}); polynomial structure and phase-locking (\ref{sec:phaselock}); tangent count (\ref{sec:tangent}); rigidity
  (\ref{app:pf-rigid}); private fraction, single path and no-flip point (\ref{app:new}); the no-flip identity and the QSP branches
  (\ref{app:branch}); steering the state near the no-flip point (\ref{app:nnf}); the readout factorisation (\ref{app:readout}); readout privacy on the extreme branches (\ref{app:theorem-a}); gate control (\ref{app:gate-control}).
  \item Appendix~\ref{app:lean}: the Lean formalisation, and one table with the proof and the numerical check of every statement.
  \item Appendix~\ref{app:experiments}: experimental protocol (circuits, optimisers, data, seeds, notebooks).
  \item Appendix~\ref{app:valid}: validation and confirmatory experiments (NB31--NB42), each with its pre-registered predictions, and the
  numerical companions of Appendix~\ref{app:proofs} (\ref{app:numcomp}).
  \item Appendix~\ref{app:additional}: further results of the exploratory experiments.
\end{itemize}
Table~\ref{tab:notation} collects the notation. The repository (\repourl) contains the simulation code, the result files behind every figure and
table, the notebooks that regenerate them, and the Lean~4/Mathlib project; results cited from the literature enter there as explicit
hypotheses (Appendix~\ref{app:lean}).

\begin{table}[h]
\centering\small
\caption{Notation.}
\label{tab:notation}
\begin{tabular}{p{0.16\linewidth}p{0.74\linewidth}}
\toprule
$N$, $L$ & number of qubits; number of encoding layers \\
$E$ & encoding budget, the number of encoding gates ($NL$ if all qubits are encoded; $L$ for the ancilla circuit) \\
$\boldsymbol\theta$, $P$ & Lie-algebra coordinates of all gates; their number \\
$S(x)$, $W^{(\ell)}$ & encoding layer ($R_X(x)$ on the encoded qubits); block of trainable gates, $\ell=0,\dots,L$ \\
$M$ & observable, $Z$ on the last qubit \\
$J$, $\Jtop$, $\Jlow$ & coefficient Jacobian $\partial\tilde{\mathbf c}/\partial\boldsymbol\theta$; its two rows for $c_E$; its other rows \\
$\kappa_J$ & $\sigma_{2E+1}/\sigma_1$ of $J$ \\
$\pf$, $q$ & private fraction of the top rows (Definition~\ref{def:private}); private pull (Lemma~\ref{lem:pull}) \\
$D$, $R$, $\mathcal A(a,b)$ & Jacobian from the gates to the coefficients of $(a,b)$; readout $\delta a\mapsto4\,\mathrm{Re}(\bar a\,\delta a)$; allowed variations (App.~\ref{app:readout}) \\
$\pf_R$, $\kappa_D$ & readout privacy; gate control $\sigma_r(D)/\sigma_1(D)$, $r=\operatorname{rank}D$ (Proposition~\ref{prop:readout}) \\
$\pf_S$, $\kappa_{D_S}$ & the same in the coordinates of Theorem~\ref{thm:nnf} \\
$s_k$, $f_k$, $A_S$, $\sigma_F$ & stay and flip amplitudes of the middle circuit before $W_k$; frame constants (Proposition~\ref{prop:gate-control}, Corollary~\ref{cor:frames}) \\
$\mu(x)$ & serial envelope $\langle\psi^\perp|M|\psi\rangle$ \\
$z$, $w$, $D(z)$ & $z=e^{ix/2}$, $w=z^2=e^{ix}$; the encoding gate in its eigenbasis, $\mathrm{diag}(z^{-1},z)$ \\
$q_l$, $\Pi$ & flip probability of middle gate $l$ in the encoding basis; no-flip product $\prod_{l=1}^{E-1}(1-q_l)$ \\
$a$, $b$ & polynomials of degree $\le E$ with output state $z^{-E}(a(w),b(w))$; $|a|^2=(1+f)/2$, $|b|^2=(1-f)/2$ on $|w|=1$ \\
$t_a$, $t_b$, $\gamma$ & $|a_E|/|a_0|$, $|b_E|/|b_0|$; $\gamma=|c_E|/2$ \\
$\lambda_j$, $\varepsilon_j$ & $\log(1/|r_j|)$ for the inner root $r_j$ of a root pair of $w^E(1\pm f)/2$; $+1$ if the realisation uses the inner root \\
$k_a$, $k_b$ & branch counts: zeros of $a$, $b$ in the open unit disc \\
$\alpha$, $\beta$ & angles of the input state and of the readout axis to the encoding axis \\
\bottomrule
\end{tabular}
\end{table}
\clearpage

\section{Proofs}
\label{app:proofs}
This appendix contains every statement of the paper with its proof, in the order of the main text. Statements of the main text are
restated before their proofs; statements needed only here are stated in full. Each numbered result of the main text that we prove also
has a numerical check; the checks are collected in Appendix~\ref{app:valid}, the Lean counterparts of the proofs in
Appendix~\ref{app:lean}, and Table~\ref{tab:status} summarises both. Conventions: the serial circuit has one $\mathrm{SU}(2)$ gate
per block unless stated (for several gates per block, read every statement in block coordinates); gates are varied in right-trivialised
coordinates $W_l\mapsto W_le^{i\theta_l\cdot\sigma}$, $P=3(E+1)$; Table~\ref{tab:notation} collects the
notation.

\subsection{Circuits, Jacobians and data}
\label{sec:background}\label{sec:theory}
\paragraph{Generality of Definition~\ref{def:vqc}.} Let $d=2^N$ and let a circuit have the encoding layer $S'(x)=VS(x)V^\dagger$, the
observable $M'=V_MMV_M^\dagger$ and the initial state $V_0|0\rangle$, with fixed $V,V_M,V_0\in\mathrm{SU}(d)$; this covers every
encoding generator and every observable that has the eigenvalues, including multiplicities, of ours, and every initial state. With blocks
$W^{(\ell)}$ its output is the output of the circuit of Definition~\ref{def:vqc} with the blocks $\tilde W^{(L)}=V_M^\dagger W^{(L)}V$,
$\tilde W^{(\ell)}=V^\dagger W^{(\ell)}V$ for $0<\ell<L$ and $\tilde W^{(0)}=V^\dagger W^{(0)}V_0$ ($L\ge1$). The map $W\mapsto\tilde W$ is a
bijection of $\mathrm{SU}(d)^{L+1}$ composed of left and right translations, which preserve Haar measure; with several gates per block it
acts on the first and the last gate of each block. The model classes therefore agree. A block $\tilde W=AWB$ satisfies $\tilde W\exp(i\boldsymbol\theta\cdot\mathbf P)=AW\exp(i(R\boldsymbol\theta)\cdot\mathbf P)B$,
with $R$ the orthogonal map of conjugation by $B$ (as in the proof of Lemma~\ref{lem:stack}), so the Lie-algebra coordinates of the two
circuits differ by orthogonal maps. Hence every statement that holds for all or for almost all gates carries over: the Fourier
coefficients agree as functions of the gates, so that $J$ is multiplied on the right by an orthogonal matrix and keeps its rank, and the
state of the circuit is $V_M$ times that of the circuit with blocks $\tilde W$, so that the tangent vectors are moved by the fixed
unitary $V_M$ and the dimension of their span is unchanged. A statement that names $Z$, $|0\rangle$ or $R_X$ carries over in translated
form: the rotation $e^{isZ}$ of the last block $\tilde W^{(L)}$ in Corollary~\ref{cor:noredundancy}(i) is the rotation $e^{isM'}$ of
$W^{(L)}$, and for $N=1$ Lemma~\ref{lem:phaselock}(iii) holds because $T$ commutes with $V_M\in\mathrm{SU}(2)$, so that $\mu$ is that of the
circuit with blocks $\tilde W$ and each $h_j$ is a real combination of the functions of the coordinates of the same gate there. Gradient-descent and Levenberg--Marquardt steps are equivariant under such maps, so that the runs of
the two circuits correspond. Adam normalises coordinate by coordinate: its runs correspond if every $R$ is a signed permutation, which
holds when $V$ and $V_0$ map Pauli words to Pauli words up to sign (for instance between $R_X$, $R_Y$ and $R_Z$ encodings); otherwise
they correspond to Adam in a rotated basis of the Lie algebra. \qed

\begin{definition}[Coefficient and data Jacobians]
\label{def:jacobian}
Let $\tilde{\mathbf c}=(\mathrm{Re}\,c_0,\mathrm{Re}\,c_1,\mathrm{Im}\,c_1,\dots,\mathrm{Re}\,c_E,\mathrm{Im}\,c_E)^\top\in\mathbb R^{2E+1}$.
The \emph{coefficient Jacobian} is $J=\partial\tilde{\mathbf c}/\partial\boldsymbol\theta\in\mathbb R^{(2E+1)\times P}$.
The \emph{data Jacobian} at inputs $x_1,\dots,x_n$ is $(\mathbf J_{\rm data})_{ij}=\partial f_{\boldsymbol\theta}(x_i)/\partial\theta_j$.
\end{definition}
In this appendix $J$ denotes the unweighted $\partial\tilde{\mathbf c}/\partial\boldsymbol\theta$; the main text uses the Parseval-weighted
$\partial\mathbf c/\partial\boldsymbol\theta=DJ$. The two have the same rank and row space, hence the same private fraction and the same directions
that leave $f$ unchanged; their singular values differ by at most the factor $\sqrt2$.

The theory concerns $J$, which depends only on the circuit; a loss on data sees $\mathbf J_{\rm data}$. The two differ by a factor that
depends only on the inputs:
\begin{lemma}[From $J$ to data]
\label{lem:bridge}
$\mathbf J_{\rm data}=BJ$ with the real Fourier design matrix $B\in\mathbb R^{n\times(2E+1)}$ of the inputs (columns $1$, $2\cos kx$,
$-2\sin kx$, $k=1,\dots,E$). On the $n=2E+1$ equispaced points of $[-\pi,\pi)$, $B^\top B=\mathrm{diag}(n,2n,\dots,2n)$, so the
singular values of $B$ lie between $\sqrt n$ and $\sqrt{2n}$, and the largest and smallest singular values of $\mathbf J_{\rm data}$ are
those of $J$ up to factors in $[\sqrt n,\sqrt{2n}]$. With the Parseval-weighted vector $\mathbf c=D\tilde{\mathbf c}$ of
Section~\ref{sec:structure}, $D=\mathrm{diag}(1,\sqrt2,\dots,\sqrt2)$, the matrix $BD^{-1}/\sqrt n$ is orthogonal on these points, so
$\mathbf J_{\rm data}$ is $\sqrt n$ times an orthogonal transform of $\partial\mathbf c/\partial\boldsymbol\theta$, and their singular values agree
up to the factor $\sqrt n$ exactly.
\end{lemma}
\begin{proof} $f=c_0+\sum_{k=1}^E2(\mathrm{Re}\,c_k\cos kx-\mathrm{Im}\,c_k\sin kx)$, so the vector of
outputs at the inputs is $B\tilde{\mathbf c}$, with $B$ the matrix of the functions $1,2\cos kx,-2\sin kx$ at the inputs. $B$ does not
depend on $\boldsymbol\theta$, hence $\mathbf J_{\rm data}=BJ$. On the $n=2E+1$ equispaced points $x_1,\dots,x_n$ the columns are
orthogonal: the product of two different columns is a combination of $\cos mx$ and $\sin mx$ with $0<m\le2E$, and
$\sum_ie^{imx_i}=0$ for $0<|m|<n$. The squared norms are $n$ for the constant column and $2n$ for the others. $B$ is then square with
$B^\top B=\mathrm{diag}(n,2n,\dots,2n)$, hence invertible, and $BB^\top$ has the same eigenvalues, so that
$n\|y\|^2\le\|B^\top y\|^2\le2n\|y\|^2$ for every $y\in\mathbb R^n$. Since $\mathbf J_{\rm data}^\top y=J^\top(B^\top y)$ and $B^\top$ is
onto, the largest and the smallest value of $\|\mathbf J_{\rm data}^\top y\|$ over unit $y$ are those of $\|J^\top z\|$ over unit
$z\in\mathbb R^{2E+1}$ up to factors in $[\sqrt n,\sqrt{2n}]$. These values are the largest and the $(2E+1)$-th singular value of
$\mathbf J_{\rm data}$ and of $J$.
\end{proof}

Section~\ref{sec:exact} trains on these points, so the conditioning measured there is that
of $J$ up to a factor $\sqrt2$; on other inputs it is that of $J$ only up to the conditioning of $B$.

\begin{lemma}[Genericity]
\label{lem:genericity}
Let a circuit of Definition~\ref{def:vqc} have $m$ gates, and let $J(G)$ be its coefficient Jacobian at the gates
$G=(G_1,\dots,G_m)\in\mathcal G=\mathrm{SU}(d)^m$. If $\mathrm{rank}\,J=2E+1$ at one point of $\mathcal G$, then it holds outside a set of
measure zero for the Haar measure of $\mathcal G$, the product of the Haar measures of the gates.
\end{lemma}
\begin{proof}
By Lemma~\ref{lem:poly} every $c_\omega$, and hence every entry of $\tilde{\mathbf c}$, is a polynomial in the real and imaginary parts of
the gate entries. The entry of $J$ for the coordinate of $G_i$ with Pauli word $P$ is the derivative at $s=0$ along $s\mapsto G_ie^{isP}$;
this curve has the velocity $iG_iP$, as has $s\mapsto G_i(I+isP)$, so the entry is the derivative of that polynomial along the second
curve, again a polynomial in the gate entries. $\mathrm{rank}\,J<2E+1$ if and only if $\Delta=\det(JJ^\top)=0$, and by assumption $\Delta$ is
nonzero at one point. A polynomial is real-analytic on the connected real-analytic manifold $\mathcal G$. By the identity
theorem~\citep{krantz2002primer}, applied in charts and extended by connectedness, $\Delta$ does not vanish identically on any open subset of
$\mathcal G$. So in every real-analytic chart on a connected domain of $\mathbb R^D$ (exponential charts, for instance) it is a
real-analytic function that is not identically zero, and its zero set there has Lebesgue measure zero~\citep{mityagin2020zero}. Finitely
many such charts cover the compact $\mathcal G$, and Haar measure is given by a nowhere-vanishing smooth form, so its null sets are the
sets that are null in every chart~\citep[Thm.~8.21 and~(8.25)]{knapp2023lie}; hence the zero set of $\Delta$ has Haar measure zero. The
argument uses of $\Delta$ only that it is a polynomial in the real and imaginary parts of the gate entries that is nonzero at one point, and
the proof of Proposition~\ref{prop:tangent} applies it to another such polynomial. In the Lean project this null-set property of
polynomials enters as a hypothesis (Appendix~\ref{app:lean}).
\end{proof}

\par\smallskip\noindent\textbf{Lemma~\ref{lem:rowrank} (restated).} \emph{\textup{(i)} The serial and the ancilla circuit both realise exactly the real trigonometric
polynomials of degree $\le E$ bounded by one, for any number of gates per block (a product of general gates is again a general gate). \textup{(ii)} Their coefficient Jacobian has full rank
$2E+1$ for almost every choice of gates.}\par\smallskip
\begin{proof} \emph{(i), serial.} Every real trigonometric polynomial $f$ of degree $\le L$
with $\|f\|_\infty\le1$ is realised by $L+1$ elements of $\mathrm{SU}(2)$. By the Fejér--Riesz
theorem~\citep{fejer1916trigonometrische,riesz1916problem}, $(1+f)/2=|p|^2$ and $(1-f)/2=|q|^2$ on the circle for polynomials $p,q$ of
degree $\le L$ in $e^{ix}$, so that $|p|^2+|q|^2=1$. By Theorem~3 of~\citet{motlagh2024generalized}, such a pair is the first column of
a circuit of $L+1$ rotations in $\mathrm{U}(2)$ interleaved with $L$ copies of their signal $\mathrm{diag}(e^{ix},1)$. Each rotation is a
constant phase times an element of $\mathrm{SU}(2)$, and, with the Hadamard matrix $H$,
\[
\mathrm{diag}(e^{ix},1)=-(iX)\,\mathrm{diag}(1,e^{ix})\,(iX),\qquad \mathrm{diag}(1,e^{ix})=-e^{ix/2}\,(iH)\,R_X(x)\,(iH),
\]
with $iX,iH\in\mathrm{SU}(2)$ ($\det H=-1$). The factors $iX$ and $iH$ are absorbed into the neighbouring gates, so the circuit is a serial
circuit times a global phase $e^{i(\alpha+Lx/2)}$ with constant $\alpha$. A global phase cancels in $f$, even where it depends on $x$.
The serial circuit therefore prepares $p(e^{ix})|0\rangle+q(e^{ix})|1\rangle$ up to that phase, and its output is $|p|^2-|q|^2=f$.
Conversely every serial output has degree $\le L$ and $|f|\le\|Z\|=1$. A block of several gates realises exactly $\mathrm{SU}(2)$ (set all but one gate to the identity), so the class
does not depend on the number of gates per block. \emph{(i), ancilla.} Choosing $W^{(\ell)}=V_\ell\otimes I$ for $\ell<L$ and
$W^{(L)}=e^{i\pi/4}\,\mathrm{SWAP}\,(V_L\otimes I)\in\mathrm{SU}(4)$ keeps qubit~1 in $|0\rangle$ until the last block, which moves the
serial state onto qubit~1; the output is the serial output. Conversely, the ancilla output has degree $\le E$ (Theorem~\ref{thm:fourier})
and modulus $\le\|M\|=1$. Blocks of several gates realise exactly $\mathrm{SU}(4)$.
\emph{(ii).} Parameterise every serial gate by Euler angles, $R_Z R_Y R_Z$, which is surjective onto $\mathrm{SU}(2)$. The map $F$ from
the angles to $\tilde{\mathbf c}$ is smooth, and by (i) its image is $\mathcal F_E$, which contains the open set
$\{\,|c_0|+2\sum_{k\ge1}|c_k|<1\,\}$. If $\mathrm{rank}\,dF<2E+1$ everywhere, the image would consist of critical values, which have
measure zero by Sard's theorem~\citep{sard1942measure}; a contradiction. Every Euler-angle direction moves a gate $G_i$ with velocity $G_i\,(i\,\mathbf t\cdot\mathbf P)$ for some real vector $\mathbf t$, since
$G_i^{-1}\partial G_i$ is anti-Hermitian and traceless. So $dF$ at the angles is $J$ in Lie-algebra coordinates at the corresponding gates
times a real matrix, and $\mathrm{rank}\,J\ge\mathrm{rank}\,dF=2E+1$ there; $J$ has only $2E+1$ rows, so $\mathrm{rank}\,J=2E+1$. For the ancilla circuit, the
columns of $J$ for Pauli words $P\otimes I$ on the embedded blocks of (i), with every other gate of a block the identity, are the columns
of the serial $J$, so the embedding of a serial
full-rank point is a full-rank point. With several gates per block, setting all but one gate of each block to the identity gives a
full-rank point in the same way. For both circuits Lemma~\ref{lem:genericity} extends the full-rank point to almost every
choice of gates.
\end{proof}

\begin{lemma}[Stationary points]
\label{lem:stationary}
Let $\mathcal L(\boldsymbol\theta)=\frac1n\sum_{i=1}^n(f_{\boldsymbol\theta}(x_i)-y_i)^2$ and let $B$ be the design matrix of the inputs
(Lemma~\ref{lem:bridge}). If $J(\boldsymbol\theta)$ has full row rank and $\nabla\mathcal L(\boldsymbol\theta)=0$, then $f_{\boldsymbol\theta}$ is a
least-squares fit of the data among all real trigonometric polynomials of degree $\le E$.
\end{lemma}
\begin{proof}
$\nabla\mathcal L=\frac2n\mathbf J_{\rm data}^\top\mathbf r=\frac2nJ^\top B^\top\mathbf r$ with the residual
$\mathbf r=B\tilde{\mathbf c}-\mathbf y$. Full row rank of $J$ gives $B^\top\mathbf r=0$, the normal equations of least squares over the
column space of $B$, which consists of the values at the inputs of the real trigonometric polynomials of degree $\le E$.
\end{proof}
If the least-squares fit lies in the model class and the loss is above its optimum, a stationary point is therefore a point where $J$ is
rank-deficient, with $B^\top\mathbf r$ in its left null space. We use the lemma in Section~\ref{sec:lm}.

\begin{lemma}[Stacked gates rescale the Riemannian step]
\label{lem:stack}
Let a block be a product $W=G_t\cdots G_1$ of $t$ gates in $\mathrm{SU}(d)$, and update every gate by a right-trivialised gradient
step $G_i\leftarrow G_i\exp(-i\eta\,\mathbf g_i\cdot\mathbf P)$ (or any update with the same first-order term), where $\mathbf g_i$ is the gradient of the loss with respect to the
coordinates $\boldsymbol\theta_i$ of the perturbation $G_i\exp(i\boldsymbol\theta_i\cdot\mathbf P)$ and $\mathbf P$ is an orthogonal basis of the traceless Hermitian matrices, each of squared norm $d$,
$\mathrm{tr}(P_aP_b)=d\,\delta_{ab}$ (for example the non-identity Pauli words). Then the block moves as $W\exp(-it\eta\,\mathbf g\cdot\mathbf P)$ to first order in $\eta$, where $\mathbf g$ is the gradient with respect to the
perturbation $W\exp(i\boldsymbol\theta\cdot\mathbf P)$ of the block itself.
\end{lemma}
\begin{proof}
Write $B_i=G_{i-1}\cdots G_1$ ($B_1=I$). Perturbing gate $i$ gives
$G_t\cdots G_ie^{i\Lambda_i}G_{i-1}\cdots G_1=W\exp\bigl(iB_i^\dagger\Lambda_iB_i\bigr)$. With $\Lambda_i=\boldsymbol\theta_i\cdot\mathbf P$ define
$R_i\boldsymbol\theta_i$ by $B_i^\dagger(\boldsymbol\theta_i\cdot\mathbf P)B_i=(R_i\boldsymbol\theta_i)\cdot\mathbf P$. Conjugation by a unitary
preserves the traceless Hermitian matrices and the inner product $\mathrm{tr}(A^\dagger B)$, in which $\mathbf P$ is orthogonal with equal norms, so
$R_i$ is orthogonal.
By the chain rule $\mathbf g_i=R_i^\top\mathbf g$. After the update of every gate by $-\eta\mathbf g_i$, the block changes, to first order in $\eta$,
by $W\bigl(i\sum_k(R_k(-\eta\mathbf g_k))\cdot\mathbf P\bigr)=W\bigl(-it\eta\,\mathbf g\cdot\mathbf P\bigr)$ (product rule), since
$R_iR_i^\top=I$.
\end{proof}
The same maps give the Jacobian with respect to the gates of a block as $(J_WR_1,\dots,J_WR_t)$, where $J_W$ is the Jacobian with
respect to the block, so $J_{\rm gates}J_{\rm gates}^\top=t\,J_{\rm blocks}J_{\rm blocks}^\top$: rank and condition number are those of the
circuit whose gates are the block products, and the singular values are larger by $\sqrt t$ (Lean: \texttt{MultiGate.gram\_stack}).
To first order in the step size, the number of gates in a block therefore acts under Riemannian gradient descent as a factor on the
step, which a line search can absorb. The argument does not extend to Adam, whose coordinate-wise normalisation is not invariant under
the orthogonal maps $R_i$.

\subsection{Polynomial structure and phase-locking}
\label{sec:chain}\label{sec:phaselock}
\begin{lemma}[Polynomial structure]
\label{lem:poly}
For any circuit of Definition~\ref{def:vqc}, each Fourier coefficient is a sum over pairs of index paths,
\[c_\omega=\sum_{p,p'}\gamma_\omega(p,p')\prod_{\ell=0}^{L}W^{(\ell)}_{p_\ell}\,\overline{W^{(\ell)}_{p'_\ell}},\]
where a path $p=(p_0,\dots,p_L)$ selects one matrix entry $W^{(\ell)}_{p_\ell}$ of every block, and the numbers $\gamma_\omega(p,p')$
depend on the encoding layer $S$, the observable $M$, the initial state and $\omega$, but not on the blocks. Every monomial of $c_\omega$ therefore
contains exactly one entry and one conjugated entry of each of the $L+1$ blocks.
\end{lemma}
We state the lemma because it makes every $c_\omega$ a polynomial in the real and imaginary parts of the gate entries, on which
the ``almost every choice of gates'' statement of Lemma~\ref{lem:rowrank}(ii) rests.

\begin{proof} Write $p_\ell=(a_\ell,b_\ell)$ for the row and column of the entry that a path selects
in block $\ell$, and $\beta,\beta'\in\{0,\dots,d-1\}$ for computational basis states. Multiplying out $U|0\rangle$,
\[
\langle\beta|U|0\rangle=\sum_{p}s_\beta(p;x)\prod_{\ell=0}^{L}W^{(\ell)}_{p_\ell},\qquad
s_\beta(p;x)=\delta_{a_L\beta}\,\delta_{b_00}\prod_{\ell=1}^{L}S(x)_{b_\ell a_{\ell-1}},
\]
so every term contains exactly one entry of each block, and $s_\beta$ does not depend on the blocks. Since
$f=\sum_{\beta,\beta'}\overline{\langle\beta|U|0\rangle}M_{\beta\beta'}\langle\beta'|U|0\rangle$,
\[
f(x)=\sum_{p,p'}\Gamma(p,p';x)\prod_{\ell=0}^{L}W^{(\ell)}_{p_\ell}\,\overline{W^{(\ell)}_{p'_\ell}},\qquad
\Gamma(p,p';x)=\sum_{\beta,\beta'}\overline{s_\beta(p';x)}\,M_{\beta\beta'}\,s_{\beta'}(p;x).
\]
Every entry of a Pauli rotation by the angle $x$ is a combination of $e^{ix/2}$ and $e^{-ix/2}$, so every entry of $S(x)$ is a
combination of exponentials whose exponent is a sum of one term $\pm\frac12$ per encoded qubit. $\Gamma(p,p';\cdot)$ multiplies $L$
entries of $S(x)$ and $L$ conjugated entries, which together contain $2E$ such terms. It is therefore a finite sum of exponentials
$e^{i\nu x}$, and each $\nu$ is a sum of $2E$ terms $\pm\frac12$, hence an integer with $|\nu|\le E$. Let $\gamma_\omega(p,p')$ be the
coefficient of $e^{i\omega x}$ in $\Gamma(p,p';\cdot)$. It depends on $S$, $M$, the initial state (through $\delta_{b_00}$) and $\omega$
only. The blocks do not depend on $x$, so comparing coefficients of $e^{i\omega x}$ with Theorem~\ref{thm:fourier} gives the formula for
$c_\omega$. Each summand contains exactly one entry and one conjugated entry of each block; collecting equal monomials changes their
coefficients, not this form. With several gates per block, each block entry is a polynomial in the entries of its gates, so every
$c_\omega$ is a polynomial in the real and imaginary parts of the gate entries.
\end{proof}

\begin{figure}[t]
  \centering
  \includegraphics[width=\textwidth]{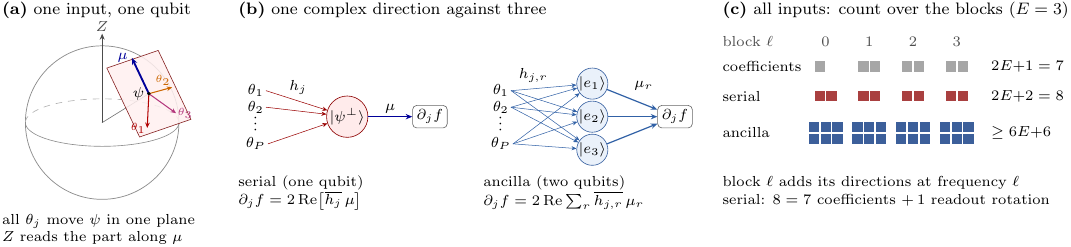}
  \caption{\textbf{Phase-locking.} (a) At one input, every parameter $\theta_j$ of the serial circuit moves the state within the same
  tangent plane of the Bloch sphere, i.e.\ along the one complex direction $|\psi^\perp\rangle$ with its own complex weight $h_j$, and the
  readout $Z$ sees only the component along $\mu$ (Lemma~\ref{lem:phaselock}). (b) Hence every gradient $\partial_jf$ is read off the same
  factor $\mu$; on the two qubits of the ancilla circuit the parameters move the state along three orthogonal vectors. (c) Accumulated over
  the $E+1$ blocks, each adding its directions at a new frequency, this gives the serial circuit $2E+2$ tangent directions for $2E+1$
  Fourier coefficients, the extra one being the readout rotation, and the ancilla circuit at least $6E+6$ for almost every choice of gates
  (Proposition~\ref{prop:tangentmain}). The serial circuit thus has no direction to spare, the first-order counterpart of its rigidity
  (Theorem~\ref{thm:rigid}).}
  \label{fig:coupling}
\end{figure}
\begin{lemma}[Serial phase-locking]
\label{lem:phaselock}
Let $N=1$, $M=Z$ and $|\psi(x,\boldsymbol\theta)\rangle=U|0\rangle=\psi_0|0\rangle+\psi_1|1\rangle$, and let
$|\psi^\perp\rangle=\overline{\psi_1}\,|0\rangle-\overline{\psi_0}\,|1\rangle$, a unit vector orthogonal to $|\psi\rangle$. Then for
every parameter $\theta_j$:

\textup{(i)} $(I-|\psi\rangle\langle\psi|)|\partial_j\psi\rangle=h_j|\psi^\perp\rangle$ with
$h_j(x)=\langle\psi^\perp|\partial_j\psi\rangle$. All tangent vectors, the components $(I-|\psi\rangle\langle\psi|)|\partial_j\psi\rangle$ of
the derivatives orthogonal to the state, are complex multiples of the one vector $|\psi^\perp\rangle$.

\textup{(ii)} With $\mu(x)=\langle\psi^\perp|M|\psi\rangle$, which is the same for all $j$,
\begin{equation}
  \partial_jf(x)=2\,\mathrm{Re}\bigl[\overline{h_j(x)}\,\mu(x)\bigr].
  \label{eq:envelope}
\end{equation}
We call $\mu$ the \emph{envelope}.

\textup{(iii)} $h_j$ is determined by the sub-circuit up to the gate containing $\theta_j$ and is a trigonometric polynomial of degree
at most the number of encoding layers preceding that gate; $\mu$ is a trigonometric polynomial of degree at most $L$.
\end{lemma}
\begin{proof} Let $T$ be the map $(v_0,v_1)\mapsto(\overline{v_1},-\overline{v_0})$ on $\mathbb C^2$,
that is, $T=i\sigma_y\mathcal C$ with $\mathcal C$ the complex conjugation in the computational basis; then $|\psi^\perp\rangle=T|\psi\rangle$. $T$ is
anti-unitary with $\langle v|Tv\rangle=0$ and $\|Tv\|=\|v\|$.
\emph{(i)} $\{|\psi\rangle,T|\psi\rangle\}$ is an orthonormal basis of $\mathbb C^2$, so
$|\partial_j\psi\rangle=\lambda_j|\psi\rangle+h_j\,T|\psi\rangle$ with $\lambda_j=\langle\psi|\partial_j\psi\rangle$ and
$h_j=\langle T\psi|\partial_j\psi\rangle$; applying $I-|\psi\rangle\langle\psi|$ gives (i). Differentiating $\langle\psi|\psi\rangle=1$
gives $\mathrm{Re}\,\lambda_j=0$.
\emph{(ii)} $\partial_jf=2\,\mathrm{Re}\langle\partial_j\psi|M|\psi\rangle=2\,\mathrm{Re}\bigl[\bar\lambda_j\langle\psi|M|\psi\rangle
+\bar h_j\langle T\psi|M|\psi\rangle\bigr]$. The first term is imaginary times real, so its real part vanishes; this gives
Eq.~\eqref{eq:envelope} with $\mu=\langle T\psi|M|\psi\rangle$, independent of $j$.
\emph{(iii), locality.} Let $G_j$ be the gate that carries $\theta_j$, $|\phi_j\rangle$ the state right after it and $D_j(x)$ the
downstream circuit (later gates and encoding layers), so $|\psi\rangle=D_j|\phi_j\rangle$ and
$|\partial_j\psi\rangle=D_j\Omega_j|\phi_j\rangle$ with $\Omega_j=(\partial_jG_j)G_j^{-1}$, which does not depend on $x$. Every
$D\in\mathrm{SU}(2)$ satisfies $\sigma_y\bar D\sigma_y=D$, i.e.\ $TD=DT$; since $R_X(x)\in\mathrm{SU}(2)$, so is $D_j(x)$, and
$h_j=\langle TD_j\phi_j|D_j\Omega_j\phi_j\rangle=\langle D_jT\phi_j|D_j\Omega_j\phi_j\rangle=\langle T\phi_j|\Omega_j\phi_j\rangle$.
\emph{(iii), degree.} If $\ell$ encoding layers precede $G_j$, each amplitude of $|\phi_j\rangle$ is
$\sum_{|k|\le\ell,\,k\equiv\ell\pmod2}\xi_ke^{ikx/2}$ (induction: every entry of $R_X(x)$ is a combination of $e^{\pm ix/2}$, and the
gates do not depend on $x$), and so is each amplitude of $\Omega_j|\phi_j\rangle$. The pairing $\langle Tu|w\rangle=u_1w_0-u_0w_1$ is
bilinear (no conjugation), so $\langle T\phi_j|\Omega_j\phi_j\rangle$ has exponents $(k+k')/2$ with $k\equiv k'\pmod2$, integers with
$|k+k'|/2\le\ell$. The same argument with $\ell=L$ and $M$ in place of $\Omega_j$ applies to $\mu=\langle T\psi|M\psi\rangle$.
\end{proof}

\emph{The choice of $|\psi^\perp\rangle$ (statements after Lemma~\ref{lem:phaselock}).} The orthogonal complement of $|\psi\rangle$ in
$\mathbb C^2$ has dimension one, so its unit vectors are $e^{i\alpha}|\psi^\perp\rangle$. Let $|\chi\rangle=e^{i\alpha}|\psi^\perp\rangle$
with real $\alpha(x)$, and let $h_j'=\langle\chi|\partial_j\psi\rangle$ and $\mu'=\langle\chi|M|\psi\rangle$. Since
$\langle\chi|w\rangle=e^{-i\alpha}\langle\psi^\perp|w\rangle$ for every $w$, $h_j'=e^{-i\alpha}h_j$ and $\mu'=e^{-i\alpha}\mu$, hence
$h_j'|\chi\rangle=h_j|\psi^\perp\rangle$ and $\overline{h_j'}\mu'=\overline{h_j}\mu$: (i) and (ii) hold with $|\chi\rangle$ in place of
$|\psi^\perp\rangle$. Steps (i) and (ii) above use the gates nowhere, only $\langle\psi|\psi\rangle=1$ and the derivative $|\partial_j\psi\rangle$, so they hold for any
differentiable parameterisation.

Any unit vector orthogonal to $|\psi\rangle$ is $e^{i\alpha}|\psi^\perp\rangle$ for some phase $\alpha$. For (i) and (ii) the choice is
immaterial: with $e^{i\alpha(x)}|\psi^\perp\rangle$ in place of $|\psi^\perp\rangle$, both $h_j$ and $\mu$ are multiplied by
$e^{-i\alpha(x)}$, and the vector $h_j|\psi^\perp\rangle$ and the product $\overline{h_j}\mu$ are the same as before. For (iii) it
matters: the statement is about this particular $|\psi^\perp\rangle$, and an $x$-dependent phase moves the frequencies (the phase
$e^{ix}$ shifts every frequency of $h_j$ and $\mu$ by one), so the degree bounds need not hold. Statements (i) and (ii) use only that
$|\psi\rangle$ is a normalised state that depends differentiably on $\theta_j$, so they hold for any parameterisation. Statement (iii)
is proved for a parameter that enters exactly one gate, as every Lie-algebra coordinate does
(Appendix~\ref{app:experiments}). Geometrically, $\mu$ is the direction of steepest ascent of $f$ on the Bloch sphere, written in the basis
$|\psi^\perp\rangle$, and (ii) is the component of the motion $h_j$ along it; its length, $|\mu|=\sqrt{1-f^2}$
(Remark~\ref{rem:multiqubit}(ii)), measures how steep the ascent is.

Parts (i) and (ii) restate that the vectors orthogonal to $|\psi\rangle$ form one complex dimension, the tangent space of the Bloch
sphere at $|\psi\rangle$; part (iii) adds where each $h_j$ comes from and what its degree is. The results below use the lemma through
the factorisation of Eq.~\eqref{eq:factor}.

\begin{remark}[More than one qubit]
\label{rem:multiqubit}
On $N$ qubits $|\psi\rangle$ is a unit vector in $\mathbb C^d$, $d=2^N$, and the vectors orthogonal to it form a space of $d-1$ complex
dimensions (one for $N=1$, three for $N=2$). Let $|e_1\rangle,\dots,|e_{d-1}\rangle$ be an orthonormal basis of that space and $M$
Hermitian.

\textup{(i)} $(I-|\psi\rangle\langle\psi|)|\partial_j\psi\rangle=\sum_rh_{j,r}|e_r\rangle$ with
$h_{j,r}=\langle e_r|\partial_j\psi\rangle$. A tangent vector is given by $d-1$ complex numbers, no longer by one (Figure~\ref{fig:coupling}).

\textup{(ii)} $\partial_jf=2\,\mathrm{Re}\sum_r\overline{h_{j,r}}\,\mu_r$ with $\mu_r=\langle e_r|M|\psi\rangle$, the same for all $j$.
We call the $\mu_r$ envelopes; if $M^2=I$, then $\sum_r|\mu_r|^2=1-f^2$.

\textup{(iii)} How many envelopes are non-zero depends on the basis: $\mu_r=\langle e_r|(M-f)|\psi\rangle$, so in a basis in which
$(M-f)|\psi\rangle$ is a multiple of $|e_1\rangle$, $\mu_r=0$ for $r\ge2$ and the gradient at $x$ depends on $h_{j,1}(x)$ only. What
does not depend on the basis is the dimension of the real span of the tangent-coefficient functions $x\mapsto(h_{j,r}(x))_r$ over
all parameters, which Proposition~\ref{prop:tangent} counts: a change of basis multiplies all of them by one unitary matrix $V(x)$.
\end{remark}

\begin{proof} $|\psi\rangle$ and the $|e_r\rangle$ form an orthonormal basis of
$\mathbb C^d$, so $|\partial_j\psi\rangle=\lambda_j|\psi\rangle+\sum_rh_{j,r}|e_r\rangle$ with
$h_{j,r}=\langle e_r|\partial_j\psi\rangle$; applying $I-|\psi\rangle\langle\psi|$ gives (i). As in (i)--(ii) of the proof of
Lemma~\ref{lem:phaselock}, $\mathrm{Re}\,\lambda_j=0$ and $\langle\psi|M|\psi\rangle\in\mathbb R$, so
$\partial_jf=2\,\mathrm{Re}\sum_r\overline{h_{j,r}}\langle e_r|M|\psi\rangle$, the formula of (ii).
\emph{Norm.} $\sum_r|\mu_r|^2=\|M\psi\|^2-|\langle\psi|M|\psi\rangle|^2=1-f^2$ by Parseval's
identity in the basis $\{|\psi\rangle,|e_r\rangle\}$, using $M^2=I$ and $M=M^\dagger$. \emph{Adapted basis.} Since $\langle e_r|\psi\rangle=0$,
$\mu_r=\langle e_r|(M-f)|\psi\rangle$. Since also $\langle\psi|(M-f)|\psi\rangle=0$, the vector $(M-f)|\psi\rangle$ lies in
$|\psi\rangle^\perp$, so a basis with $|e_1\rangle$ along it exists (any basis if it vanishes); in such a basis it is orthogonal to $|e_r\rangle$ for $r\ge2$, and only
the term $r=1$ of (ii) remains.
\emph{Invariance.} Two orthonormal bases of $|\psi(x)\rangle^\perp$ differ by a unitary matrix $V(x)$, and $g\mapsto V(\cdot)g(\cdot)$ is an
invertible real-linear map on functions with values in $\mathbb C^{d-1}$, so it preserves the dimension of real spans.
\end{proof}

\paragraph{The factored Jacobian.} In matrix form, Lemma~\ref{lem:phaselock} says that every column of the coefficient
Jacobian is the same linear map applied to the Fourier coefficients of one $h_j$:
\begin{equation}
  J=A_\mu H,
  \label{eq:factor}
\end{equation}
where $H\in\mathbb R^{2(2L+1)\times P}$ collects the real and imaginary parts of the Fourier coefficients of the $h_j$, and
$A_\mu\in\mathbb R^{(2L+1)\times2(2L+1)}$ is the real-linear map $h\mapsto\bigl(\text{Fourier coefficients of }2\,\mathrm{Re}[\bar h\mu]
\text{ up to degree }L\bigr)$, determined by the envelope alone. Each parameter contributes one column
of $H$; all parameters share $A_\mu$.

\begin{proof}[Proof of Eq.~\eqref{eq:factor}] By Lemma~\ref{lem:phaselock}, $h_j=\sum_{|k|\le L}a_{jk}e^{ikx}$ and
$\mu=\sum_{|k|\le L}b_ke^{ikx}$; set $a_{jk}=b_k=0$ for $|k|>L$. Expanding $2\,\mathrm{Re}[\bar h_j\mu]=\bar h_j\mu+h_j\bar\mu$ gives the
Fourier coefficients of $\partial_jf$, $\partial_jc_m=\sum_k\bigl(\bar a_{jk}b_{k+m}+\bar b_ka_{j,k+m}\bigr)$, which are real-linear in
$\mathbf a_j$ and depend otherwise only on $\mathbf b$. They vanish for $|m|>L$, since $f$ has degree $\le L$, and the real
representation of $(\partial_jc_0,\dots,\partial_jc_L)$ is column $j$ of $J$. Writing $A_\mu$ for this real-linear map from
$\mathbb C^{2L+1}\cong\mathbb R^{2(2L+1)}$ to $\mathbb R^{2L+1}$ and $H$ for the matrix of stacked real and imaginary parts of the
$\mathbf a_j$ gives $J=A_\mu H$.
\end{proof}

\subsection{Tangent count and spare directions}
\label{sec:tangent}
\begin{proposition}[Tangent count, summary]
\label{prop:tangentmain}
For any number of gates per block, the serial circuit has at most $2E+2$ tangent directions, with equality for almost every choice of
gates; the ancilla circuit has at least $6E+6$, for almost every choice of gates. Where $J$ has full rank, the only tangent direction of the
serial circuit that leaves $f$ unchanged to first order is the rotation of the final state about the readout axis.
\end{proposition}
It combines Proposition~\ref{prop:tangent} and Corollary~\ref{cor:noredundancy} below: each block adds at most two real directions on one
qubit (one complex direction, at a new frequency) and at least six on two qubits, and the directions that leave $f$ unchanged are counted by
comparing with the $2E+1$ coefficients. So the serial circuit has no tangent direction to spare: of its $2E+2$ directions, $2E+1$ change the
function and one is invisible to the readout.
Phase-locking costs the serial circuit no Fourier direction: its coefficient Jacobian has full row rank almost everywhere
(Lemma~\ref{lem:rowrank}). What it limits is the number of directions. It is a statement at one input; a function is fitted over all
inputs, so the question is how many independent directions the parameters offer over all inputs together, compared with the $2E+1$
Fourier coefficients they have to control.
To count them we need, at every input, a basis of the vectors orthogonal to $|\psi(x)\rangle$ in which to write the tangent vectors
(Remark~\ref{rem:multiqubit}). We use $e_r(x)=U(x)|r\rangle$, $r=1,\dots,2^N-1$: the circuit applied to the other computational
basis states, which are orthonormal and orthogonal to $|\psi(x)\rangle=U(x)|0\rangle$ because $U(x)$ is unitary. For $N=1$ this is
$-|\psi^\perp\rangle$, the vector of Lemma~\ref{lem:phaselock} up to sign. Parameter $j$ then has the \emph{tangent-coefficient function}
$h_j(x)=(\langle e_r(x)|\partial_j\psi(x)\rangle)_r\in\mathbb C^{2^N-1}$, a vector of trigonometric polynomials of degree $\le E$. The number
of tangent directions of a circuit is the dimension of the real span of these functions over all parameters; by
Remark~\ref{rem:multiqubit}(iii) it does not depend on the choice of the basis. For the serial circuit it is the rank of the
matrix $H$ of Eq.~\eqref{eq:factor}.

\begin{proposition}[Tangent count]
\label{prop:tangent}
Let $\mathcal T$ be the real span of the tangent-coefficient functions $h_j$ over all Lie-algebra coordinates $j$. For any number of
gates per block:
\textup{(i)} For the serial circuit, $\dim\mathcal T\le2(E+1)$ for every choice of gates, with equality for almost every choice.
\textup{(ii)} For the ancilla circuit, $\dim\mathcal T\ge6(E+1)$ for almost every choice of gates.
\end{proposition}
These counts are the directions at one input, accumulated over the blocks. At one input the state can move in $2(d-1)$ real
directions: two on one qubit, the one complex direction of Lemma~\ref{lem:phaselock}, and six on two qubits
(Remark~\ref{rem:multiqubit}). Every one of the $E+1$ blocks of the serial and of the ancilla circuit contributes these
directions once, each block at a new frequency, which gives the $2(E+1)$ of (i) and the $6(E+1)$ of (ii).

A tangent direction can leave $f$ unchanged to first order: it then moves the state without changing the output. Our circuits have
such directions of one trivial kind, the \emph{readout rotations}: rotations of the final state that commute with the readout $M$ and
are therefore invisible to it; they span at most one dimension on a single qubit and at most seven on two (see the end of this subsection). We call every further dimension of
directions that leave $f$ unchanged \emph{spare}. Spare directions do not lower the loss by themselves; they are room to change how a
function is realised without changing the function.

\begin{proof}[Proof of Proposition~\ref{prop:tangent}] Let $d=2^N$ and let $V_\ell(x)=W^{(\ell)}S(x)W^{(\ell-1)}\cdots S(x)W^{(0)}$ be the circuit
up to block $\ell$, so that $U=W^{(L)}S\cdots S\,V_\ell$. A Lie-algebra coordinate of a gate in block $\ell$ moves the state by
$U(x)\,i\,V_\ell(x)^\dagger\Lambda V_\ell(x)|0\rangle$ for a traceless Hermitian $\Lambda$: if $A(x)$ is the circuit up to that gate, exclusive, the
coordinate with Pauli word $P$ moves the state by $U\,i\,A^\dagger PA|0\rangle$, and $A=C^\dagger V_\ell$ with $C$ the product of the gate
and of the gates of the block after it, so that $\Lambda=CPC^\dagger$. We take the tangent-coefficient functions in the orthonormal basis $e_r(x)=U(x)|r\rangle$, $r=1,\dots,d-1$, of $|\psi(x)\rangle^\perp$; by
Remark~\ref{rem:multiqubit}(iii) the dimension of their span does not depend on this choice. In this basis $U$ cancels,
$\langle e_r|\partial_j\psi\rangle=i\langle r|V_\ell^\dagger\Lambda V_\ell|0\rangle$, a trigonometric polynomial of degree $\le\ell$ since each encoding has
eigenvalues $e^{\pm ix/2}$. For the serial circuit
$U|1\rangle=-T|\psi\rangle=-|\psi^\perp\rangle$. In this basis the function of the coordinate is $h(x)=i\,\pi(V_\ell(x)^\dagger\Lambda V_\ell(x)|0\rangle)$
with $\pi(v)=(\langle r|v\rangle)_{r\ge1}$ for a vector $v$, and the Pauli words of the last gate of block $\ell$ give all traceless Hermitian
$\Lambda$, so that the span of the functions of block $\ell$ does not depend on the number of gates per block. Let $\mathcal T_\ell$ be the span of
the functions of blocks $0,\dots,\ell$. The functions of block $\ell$ have degree $\le\ell$.

\emph{Upper bounds.} The argument holds for any encoding layer $S(x)=\sum_ae^{-iax/2}\Pi_a$, with $\Pi_a$ the spectral
projections of the encoding generator (multiplicities $m_a$). Block $0$ gives constant functions, $\dim\mathcal T_0\le2(d-1)$. For $\ell\ge1$,
$V_\ell=W^{(\ell)}SV_{\ell-1}$; with $K=W^{(\ell)\dagger}\Lambda W^{(\ell)}$,
\[
S^\dagger KS=\sum_{a,b}e^{i(a-b)x/2}\,\Pi_aK\Pi_b .
\]
The block-diagonal part $K_0=\sum_a\Pi_aK\Pi_a$ commutes with $S$, so its contribution $i\,\pi(V_{\ell-1}^\dagger K_0V_{\ell-1}|0\rangle)$ is a function of
block $\ell-1$ ($K_0$ is Hermitian with $\mathrm{tr}\,K_0=\mathrm{tr}\,K=0$). The rest depends real-linearly on the off-diagonal blocks of a Hermitian $K$, of real dimension
$d^2-\sum_am_a^2$. Hence $\dim\mathcal T_\ell/\mathcal T_{\ell-1}\le d^2-\sum_am_a^2$: this is $2$ for the serial circuit ($m=(1,1)$), giving
$\dim\mathcal T\le2+2L=2(E+1)$. For one encoded qubit, $S(x)=e^{-ix/2}\Pi_++e^{ix/2}\Pi_-$ with $\Pi_\pm$ the projections onto the $\pm1$ eigenspaces of $X$ on that
qubit (rank $d/2$ each), and the off-diagonal part is $K_+=\Pi_+K\Pi_-$ and its adjoint.

\emph{Lower bounds (one encoded qubit, the first tensor factor: the serial and the ancilla circuit).} For $\ell\ge1$ the coefficients of
$h$ at $e^{\pm i\ell x}$ come only from the $K_+$ terms. With $B_\pm=W^{(\ell-1)}\Pi_\pm W^{(\ell-2)}\cdots\Pi_\pm W^{(0)}$ the extreme coefficients of $V_{\ell-1}$, and writing $K_+=|+\rangle\langle-|\otimes Q$
with $Q\in\mathbb C^{d/2\times d/2}$, they are
\[
i\,\pi\bigl(B_+^\dagger(|+\rangle\otimes Qu)\bigr)\ \text{at}\ e^{i\ell x},\qquad i\,\pi\bigl(B_-^\dagger(|-\rangle\otimes Q^\dagger w)\bigr)\ \text{at}\ e^{-i\ell x},
\]
with $u=(\langle-|\otimes I)B_-|0\rangle$ and $w=(\langle+|\otimes I)B_+|0\rangle$. Functions in $\mathcal T_{\ell-1}$ have degree $\le\ell-1$,
so $\dim\mathcal T_\ell/\mathcal T_{\ell-1}$ is at least the real rank of $Q\mapsto(Qu,Q^\dagger w)$ whenever $u,w\neq0$ and
$z\mapsto\pi(B_\pm^\dagger(|\pm\rangle\otimes z))$ are injective. That rank is $d^2/2-2(d/2-1)^2=2(d-1)$, its kernel being $\{Qu=0,\ Q^\dagger w=0\}$ (in bases with $u$ and $w$ first, the first column and the first row of $Q$ vanish; they share the entry $w^\dagger Qu$).
Together with $\dim\mathcal T_0=2(d-1)$, which holds at every choice of gates (block $0$ gives the constants $i\,\pi(W^{(0)\dagger}\Lambda W^{(0)}|0\rangle)$, and $\Lambda=W^{(0)}(|v\rangle\langle0|+|0\rangle\langle v|)W^{(0)\dagger}$ gives $iv$ for every $v\perp|0\rangle$), this gives $\dim\mathcal T\ge2(d-1)(L+1)$. At identity gates all conditions hold: $B_\pm=I$ for $\ell=1$ and $B_\pm=\Pi_\pm$ for $\ell\ge2$,
$u$ and $w$ are $2^{-1/2}|0\rangle$ on the $N-1$ unencoded qubits, and $|\pm\rangle\otimes z$ keeps $z$ in the coordinates $|1\rangle\otimes z$, which $\pi$ retains.
Since the $h_j$ have degree $\le E$, they are determined by their values on $2E+1$ equispaced points (the discrete Fourier transform on
these points is invertible), so $\dim\mathcal T$ is the rank of their real samples there, and
$\dim\mathcal T<k$ is the vanishing of a polynomial in the real and imaginary parts of the gate entries (the sum of squares of all $k\times k$ minors)
that is nonzero at the identity; the exception set is Haar-null as in Lemma~\ref{lem:genericity}. With $L=E$ this is (ii) for $d=4$ and, with the
upper bound, (i) for $d=2$.

\end{proof}

\begin{corollary}[No spare tangent direction in the serial circuit]
\label{cor:noredundancy}
For the serial circuit with $M=Z$ and any number of gates per block, with $h_j$ and $\mu$ taken in the same basis:

\textup{(i)} Rotating the state about the readout axis, $|\psi\rangle\mapsto e^{isZ}|\psi\rangle$, is a motion of the last gate. Its
tangent-coefficient function is $i\mu$, which lies in $\mathcal T$, and it leaves $f$ unchanged.

\textup{(ii)} At every choice of gates at which $J$ has full row rank, hence at almost every choice (Lemma~\ref{lem:rowrank}), a vector
$\dot{\boldsymbol\theta}\in\mathbb R^P$ satisfies $J\dot{\boldsymbol\theta}=0$ if and only if $\sum_j\dot\theta_jh_j=\lambda\,i\mu$ for some $\lambda\in\mathbb R$: the rotation of (i) is the
only tangent direction that leaves $f$ unchanged to first order. The ``if'' part holds for every choice of gates.

\textup{(iii)} For almost every target $f^*\in\mathcal F_E$ (Lebesgue measure on its coefficient vector $\tilde{\mathbf c}$), (ii) applies at
every choice of gates with $f_{\boldsymbol\theta}=f^*$.
\end{corollary}
\begin{proof} We use the basis $|\psi^\perp\rangle=T|\psi\rangle$ of Lemma~\ref{lem:phaselock}; in the
basis $-|\psi^\perp\rangle$ of Proposition~\ref{prop:tangent} all $h_j$ and $\mu$ change sign, and the statements are the same.
\emph{(i)} Let $G$ be the last gate. It is the leftmost factor of the circuit, so along $s\mapsto e^{isZ}G$ the state is
$e^{isZ}|\psi\rangle$, with tangent vector $iZ|\psi\rangle$ at $s=0$; hence $h=\langle T\psi|iZ\psi\rangle=i\mu$. Since $e^{isZ}$
commutes with $Z$, $f$ is the same for every $s$. Moreover $e^{isZ}G=G\exp(is\,G^\dagger ZG)$, and $G^\dagger ZG$ is Hermitian and
traceless, hence $\sum_at_aP_a$ with real $t_a$ and $P_a$ the Pauli words. The state is linear in each gate, so the
tangent-coefficient function is real-linear in the Lie-algebra direction, and $i\mu=\sum_at_ah_{(a)}$ with $h_{(a)}$ the functions of
the three coordinates of the last gate; thus $i\mu\in\mathcal T$.
\emph{(ii)} By Eq.~\eqref{eq:envelope} and linearity, $\sum_j\dot\theta_j\,\partial_jf=2\,\mathrm{Re}[\bar h\mu]$ for $h=\sum_j\dot\theta_jh_j$, and $J\dot{\boldsymbol\theta}$
holds its Fourier coefficients (Eq.~\eqref{eq:factor}). For $h=\lambda\,i\mu$ this function is $2\,\mathrm{Re}[-i\lambda|\mu|^2]=0$, for
every choice of gates. Conversely, the real-linear map $\Phi:\mathcal T\to\mathbb R^{2E+1}$ that sends $h$ to the Fourier coefficients
of $2\,\mathrm{Re}[\bar h\mu]$ has the column space of $J$ as its image. If $\mathrm{rank}\,J=2E+1$, then $\Phi$ is onto, and since
$\dim\mathcal T\le2E+2$ for every choice of gates (Proposition~\ref{prop:tangent}(i)), $\dim\ker\Phi\le1$. The kernel contains $i\mu$
by (i), and $\mu\not\equiv0$ because $J\neq0$. Hence $\ker\Phi=\mathbb R\,i\mu$ (and $\dim\mathcal T=2E+2$ at such gates).
\emph{(iii)} Let $F$ be the smooth map from the Euler angles of all gates to $\tilde{\mathbf c}$, as in the proof of
Lemma~\ref{lem:rowrank}. By Sard's theorem~\citep{sard1942measure} its critical values have measure zero. Let $f^*\in\mathcal F_E$ not be one of them, and let
gates give $f_{\boldsymbol\theta}=f^*$. The Euler angles are onto $\mathrm{SU}(2)$, so these gates have angles, at which
$\mathrm{rank}\,dF=2E+1$; since $dF$ is $J$ at these gates times a real matrix, $\mathrm{rank}\,J=2E+1$, and (ii) applies.
\end{proof}

Here $J\dot{\boldsymbol\theta}=0$ says that the direction $\dot{\boldsymbol\theta}$ changes no Fourier coefficient of $f$, hence
not $f$, to first order.

Generically the serial circuit thus has no spare tangent direction: of its $2E+2$ directions, $2E+1$ change the Fourier
coefficients and the last one is the readout rotation. The ancilla circuit generically has at least $4E-2$ spare directions: its at least $6E+6$ tangent directions, less the at most $2E+1$ that
change the Fourier coefficients and the at most seven readout rotations. In our numerical checks its bound in Proposition~\ref{prop:tangent} is attained, $\dim\mathcal T=6(E+1)$ (Table~\ref{tab:tangent}).
This count is taken in state space: motions of the state that leave $f$ unchanged. Section~\ref{sec:rigid} counts in parameter space
instead, motions of the gates that leave $f$ unchanged less the symmetries, and finds at least $6E-2$ for the ancilla circuit. The two
numbers differ because a motion of the gates need not move the state; for the serial circuit both are zero.

\paragraph{Readout rotations and spare directions.} For a traceless Hermitian $\Lambda$ that commutes with $M$, the curve
$s\mapsto e^{is\Lambda}G$ of the last gate $G$ rotates the final state by $e^{is\Lambda}$ and leaves $f$ unchanged, by the argument for part (i) of
Corollary~\ref{cor:noredundancy}. For $M=Z$ on one qubit these $\Lambda$ form a real space of dimension $1$, for $M=Z$ on one of two qubits
of dimension $2^2+2^2-1=7$; the tangent-coefficient function depends real-linearly on $\Lambda$, so the tangent directions of this kind span
at most one and at most seven dimensions. The directions that leave $f$ unchanged to first order form a space of dimension
$\dim\mathcal T-\mathrm{rank}\,J$, and $\mathrm{rank}\,J\le2E+1$. The number of spare directions, the dimension of this space beyond
the readout rotations, is therefore zero for the serial circuit wherever $J$ has full row rank (Corollary~\ref{cor:noredundancy}(ii)), at
least $(6E+6)-(2E+1)-7=4E-2$ for the ancilla circuit wherever its part (ii) holds.

\subsection{Rigidity}
\label{app:pf-rigid}
\par\smallskip\noindent\textbf{Theorem~\ref{thm:rigid} (restated).} \emph{Let the gates $\mathbf W$ have $\mathrm{rank}\,J=2E+1$, and let not both the input state
$W_0|0\rangle$ and the readout axis lie on the encoding axis. Then all gates in a neighbourhood of $\mathbf W$ that realise the same function are
gauge rotations of $\mathbf W$, and a gauge rotation multiplies $J$ from the right by an orthogonal matrix. Hence, near $\mathbf W$, the matrix $JJ^\top$
and the singular values of $J$ depend on the gates only through the function they realise.}\par\smallskip
The torus $G=\mathrm U(1)^{E+2}$ acts by $W_{k-1}\mapsto e^{is_kX}W_{k-1}$, $W_k\mapsto W_ke^{-is_kX}$ ($k=1,\dots,E$),
$W_0\mapsto W_0e^{it_0Z}$, $W_E\mapsto e^{it_EZ}W_E$. The gauge rotations $e^{is_kX}$ commute with $R_X(x)$, the input rotation $e^{it_0Z}$ changes
$U(x)|0\rangle$ only by a global phase, and the readout rotation $e^{it_EZ}$ commutes with the readout, so $f$ is invariant.

\begin{lemma}[Covariance]\label{lem:cov}
For $g\in G$, $J(g\cdot W)=J(W)\,O_g$ with $O_g=\mathrm{blockdiag}(O_0,\dots,O_E)$, $O_l\in\mathrm{SO}(3)$. Hence $JJ^\top$, the singular
values of $J$ and $\pf$ are $G$-invariant.
\end{lemma}
\begin{proof}
Write $(g\cdot W)_l=L_lW_lR_l^\dagger$. Then $(g\cdot W)_le^{i\theta\cdot\sigma}=L_lW_le^{i(O_l\theta)\cdot\sigma}R_l^\dagger$ with
$O_l=\mathrm{Ad}_{R_l^\dagger}$, so $f(g\cdot W,\theta)=f(W,O\theta)$; differentiate at $\theta=0$. Right multiplication by an orthogonal matrix
leaves $JJ^\top$ unchanged and rotates all rows alike, so the norms of the projections of the top rows onto the span of the lower rows
are unchanged.
\end{proof}

\begin{lemma}[The orbit has dimension $E+2$]\label{lem:orbit}
The differential of $g\mapsto g\cdot W$ at the identity is injective unless both the input state $W_0|0\rangle$ and the readout axis
$W_E^\dagger ZW_E$ lie on the encoding axis.
\end{lemma}
\begin{proof}
The generators are $u_0=Z$ at block $0$; $v_k=(W_{k-1}^\dagger XW_{k-1}$ at block $k-1$, $-X$ at block $k)$; $u_E=W_E^\dagger ZW_E$ at block $E$.
If $bu_0+\sum_ka_kv_k+cu_E=0$ and $W_0^\dagger XW_0\neq\pm Z$, block $0$ gives $a_1=b=0$, block $k$ gives $a_{k+1}W_k^\dagger XW_k=a_kX$, so all
$a_k=0$ by induction (both $X$ and $W_k^\dagger XW_k$ are nonzero, so $a_{k+1}=0$ iff $a_k=0$), and block $E$ gives $c=0$. Otherwise run the induction from block $E$ downwards.
\end{proof}

\begin{proof}[Proof of Theorem~\ref{thm:rigid}]
At a point of full rank the level set $S$ of $\mathbf c$ is, near $W$, an embedded submanifold of dimension $3(E+1)-(2E+1)=E+2$. By
Lemma~\ref{lem:orbit} and equivariance the orbit map is an immersion near the identity, from an $(E+2)$-dimensional neighbourhood of the
identity into $S$; as the dimensions agree, it is a local diffeomorphism onto an open subset of $S$. So a neighbourhood of $W$ in $S$ lies in
$G\cdot W$, which is the first claim; the second is Lemma~\ref{lem:cov}. Full rank and the condition on the input state and the readout axis
are open, so both claims hold at every point near $W$, and $JJ^\top$ is locally constant on every level set of $\mathbf c$ near $W$. In
implicit-function coordinates, in which $\mathbf c$ is a coordinate projection, these level sets are images of balls, hence connected, so
$JJ^\top$ is constant on each of them: two gate tuples near $W$ that realise the same function have the same $JJ^\top$, and hence the same
singular values of $J$.
\end{proof}
\begin{corollary}[Private fraction and coefficient flow]\label{cor:rigidflow}
Under the hypotheses of Theorem~\ref{thm:rigid}, near $\mathbf W$ the private fraction $\pf$ depends on the gates only through the function
they realise, and $JJ^\top=\Phi(\mathbf c)$ for a function $\Phi$ of the coefficients. Along gradient flow the coefficients therefore obey the
closed equation $\dot{\mathbf c}=-\tfrac2V\Phi(\mathbf c)(\mathbf c-\mathbf c^*)$ (here $\mathbf c$, $r$ and $J=\partial\mathbf c/\partial\boldsymbol\theta$ are
Parseval-weighted, as in Section~\ref{sec:structure}).
\end{corollary}
\begin{proof}
$\pf$ is $G$-invariant (Lemma~\ref{lem:cov}), and the proof of Theorem~\ref{thm:rigid} uses only the $G$-invariance of $JJ^\top$; so it
applies to $\pf$ as well. Setting $\Phi(\mathbf c)=JJ^\top$ at any gate tuple near $\mathbf W$ with coefficients $\mathbf c$ is well defined
by the theorem. Finally $\dot{\mathbf c}=J\dot\theta=-J\nabla L=-\tfrac2VJJ^\top r$.
\end{proof}
The proof uses only $\dim S=\dim G$; infinitesimally this is Corollary~\ref{cor:noredundancy}(ii) (no spare direction). For the ancilla
control the level sets have dimension $15(E+1)-(2E+1)=13E+14$, while its symmetry group (gauge in the commutant of the encoding, $7E$;
stabiliser of $|00\rangle$, $9$; commutant of the readout, $7$) has dimension at most $7E+16$ (the three subgroups overlap, for instance in $I\otimes Z$, so this bounds the dimension of the orbits from
above); so at least $6E-2$ fibre directions are not
symmetries (Lean, for the tangent space of the fibre: \texttt{AncillaCount.ancilla\_count}). (A numerical walk along both fibres is in Appendix~\ref{app:numcomp}.) The theorem is local and needs full rank; branches can meet
where $J$ loses rank, so we do not claim that a run cannot change its realisation.

\subsection{The private fraction, the single path and the no-flip point}
\label{app:new}
Lemma~\ref{lem:private} is proved in Section~\ref{sec:branch}.
\begin{lemma}[Single path]
\label{lem:singlepath}
$c_E=\phi_0\,\phi_E\prod_{l=1}^{E-1}\overline{\alpha_l}^{\,2}$ up to a unit phase, with end factors $|\phi_0|,|\phi_E|\le1$ determined by $W_0$
and $W_E$; hence $|c_E|\le\Pi$.
\end{lemma}
\begin{proof}[Sketch]
Each encoding gate is $\mathrm{diag}(z^{-1},z)$; a path through the basis states collects $z^{\pm1}$ at every encoding gate, so $z^{\pm E}$
comes only from the path that never flips, and every middle gate contributes $\alpha_l$ or $\bar\alpha_l$. The output pairs this path with its
partner. Lean: \texttt{SinglePath.single\_path\_RX}, from \texttt{PathStructure.serial\_top\_product}. Proposition~\ref{thm:identity} refines the
bound $|c_E|\le\Pi$ into an identity.
\end{proof}

\begin{remark}[Spread of the top coefficient at Haar starts]
\label{rem:clt}
For Haar-random $W_l\in\mathrm{SU}(2)$, $|\alpha_l|^2$ is uniform on $[0,1]$, so $\ln|\alpha_l|^2$ has mean $-1$ and variance $1$, and
$\ln\Pi$ has mean $-(E-1)$ and standard deviation $\sqrt{E-1}$; the strong law gives $\frac1{E-1}\ln\Pi\to-1$ almost surely (Lean, with the
uniform law as hypothesis: \texttt{InitProduct.slln\_log}). The exact law, $-\ln\Pi\sim\mathrm{Gamma}(E-1,1)$, follows because $-\ln(1-q_l)$ is
standard exponential for uniform $q_l$ (the flip probability is uniform as well, since $|\beta_l|^2=1-|\alpha_l|^2$); it is checked in
Appendix~\ref{app:tchecks}.
\end{remark}

We write the gates in the encoding basis, in which the encoding is diagonal; there the input $|0\rangle$ reads $|{+}\rangle$ and the
readout $Z$ reads $X$. Let $\psi=W_0|{+}\rangle$ (the input state $W_0|0\rangle$ of the main text) and $A=W_E^\dagger XW_E$, and $\alpha,\beta$ the angles of their Bloch vectors to the encoding axis.
\begin{proposition}[No-flip point]
\label{prop:noflip}
If every middle gate is diagonal in the encoding basis, then
\textup{(i)} $f$ has only the frequencies $0$ and $E$, and $|c_E|=\tfrac12\sin\alpha\sin\beta$;
\textup{(ii)} to first order, a flip at middle gate $l$ changes only the frequencies $l$ and $E-l$, a phase change at a middle gate only $c_E$,
and an end gate only $c_0$ and $c_E$; hence the top rows are orthogonal to the rows of $c_1,\dots,c_{E-1}$;
\textup{(iii)} for a flip at a middle gate $l$ with $2l\neq E$, along one flip coordinate ($Y$ or $Z$ of $W_l$), the two entries of $J$ in the rows of $\mathrm{Re}\,c_{E-l}$ and $\mathrm{Im}\,c_{E-l}$ form a vector of norm
$c_{\rm lo}=\sin\beta|\cos\alpha|$, and those in the rows of $c_l$ one of norm $c_{\rm hi}=\sin\alpha|\cos\beta|$; these lower-band controls satisfy
$4|c_E|^2+c_{\rm lo}^2=\sin^2\beta$ and $4|c_E|^2+c_{\rm hi}^2=\sin^2\alpha$.
\end{proposition}
\begin{proof}[Proof of Proposition~\ref{prop:noflip} (sketch)] Lean: (i) and (ii) for the circuit are \texttt{NoFlipA16}, (iii) is
\texttt{NoFlipControl.control\_lo} and \texttt{control\_hi}; the lemmas named below are their steps. With diagonal middle gates the middle block is
$\mathrm{diag}(m_1z^{-E},m_2z^{E})$ with $|m_1|=|m_2|=1$, so the output contains only $z^0$ and $z^{\pm2E}$, i.e.\ frequencies $0$ and $E$,
and $|c_E|=|\psi_0||\psi_1||A_{01}|$ (Lean \texttt{top\_coef\_norm}). A first-order flip $g$ at middle gate $l$ creates exactly two new paths,
which flip once at $l$; they carry $z^{\pm(E-2l)}$ in the state and produce frequencies $l$ and $E-l$ in the output, with coefficients
$c_{E-l}\propto A_{01}(|\psi_0|^2-|\psi_1|^2)\bar g$ and $c_l\propto\bar\psi_0\psi_1(A_{00}-A_{11})g$ after the unit phases cancel (Lean
\texttt{low\_control\_eq}, \texttt{high\_control\_eq}). A diagonal change keeps the path set, so it changes only $c_E$; end gates change
$\psi$ and $A$, hence only $c_0$ and $c_E$. The supports are disjoint, so the top rows are orthogonal to the rows of $c_1,\dots,c_{E-1}$
(\texttt{orth\_of\_disjoint}, \texttt{private\_eq\_self}). With $\psi_r$, $A_{rs}$ the entries in the encoding basis $(e_0,e_1)$, $\cos\alpha=|\psi_0|^2-|\psi_1|^2$ and $\cos\beta=(A_{00}-A_{11})/2$ ($A$ is traceless with $A^2=I$, so $A_{11}=-A_{00}$ and $A_{00}^2+|A_{01}|^2=1$),
$|c_E|=\tfrac12\sin\alpha\sin\beta$ and (iii) are \texttt{tradeoff\_low} and \texttt{tradeoff\_high}. The frequency statements are formalised on the path sums: with diagonal middle gates only $0$ and $\pm E$ occur (\texttt{NoFlipBook.coeff\_eq\_zero\_diag}), and the exact expansion in a flip $\varepsilon G$ at gate $l$ has a first-order part supported on $\pm l$ and $\pm(E-l)$ (\texttt{coeff\_expand}, \texttt{dcoeff\_eq\_zero}). The columns of $J$ are the first-order path sums of the perturbed gate (\texttt{NoFlipJ.gateJ\_eq\_dcoeff}), which gives (ii) for the entries of $J$; a phase change multiplies every path amplitude by $\pm i$, so its two first-order terms cancel at $c_0$ (\texttt{NoFlipA16}; numerical check: Appendix~\ref{app:numcomp}). For (iii), the first-order path sum at
$E-l$ (and at $l$) consists of two pairs per output state, a constant path against a path that flips once at $l$; the factors of all other gates
have modulus one and agree between the two pairs, which gives the entries of $J$ above (\texttt{NoFlipControl.control\_lo}, \texttt{control\_hi}).
For $2l=E$ the two frequencies coincide and the column carries both contributions.
\end{proof}

\subsection{The no-flip identity and the QSP branches}
\label{app:branch}
Several statements of this subsection are general-gate forms of results of \citet{wang2022energy} for symmetric QSP: the root-choice
construction of the realisations (their Theorem~4), Corollary~\ref{cor:extreme}(i) (their Lemma~18, the ``maximal solution''), and
Lemma~\ref{lem:strip} (their Theorem~1 and Corollary~3). Their proofs do not cover general gates, so we give ours.
\par\smallskip\noindent\textbf{Proposition~\ref{thm:identity} (restated).} \emph{Let $c_E\neq0$, $\gamma=|c_E|/2$, $t_a=|a_E|/|a_0|$, $t_b=|b_E|/|b_0|$, and let $\alpha$ and $\beta$ be the angles, on the Bloch
sphere, of the input state $W_0|0\rangle$ and of the readout axis $W_E^\dagger ZW_E$ to the encoding axis. Then
$|c_E|=\tfrac12\,\Pi\,\sin\alpha\,\sin\beta$, $\Pi=\gamma\,(t_a+t_a^{-1}+t_b+t_b^{-1})$,
$\log t_a=\sum_j\varepsilon^a_j\lambda^a_j$ and $\log t_b=\sum_j\varepsilon^b_j\lambda^b_j$.
For every circuit, also with $c_E=0$, $\Pi=|a_0|^2+|a_E|^2+|b_0|^2+|b_E|^2$.}\par\smallskip
\paragraph{Setting.} In the encoding frame $D(z)=\mathrm{diag}(z^{-1},z)$, $z=e^{ix/2}$, $w=z^2$; middle gates
$W_l=\bigl(\begin{smallmatrix}\alpha_l&-\bar\beta_l\\\beta_l&\bar\alpha_l\end{smallmatrix}\bigr)$, $q_l=|\beta_l|^2$;
$\varphi(z)=W_EDW_{E-1}\cdots DW_0|{+}\rangle$, $(u,v)=H\varphi$, $u=z^{-E}a(w)$, $v=z^{-E}b(w)$ with polynomials $a,b$ of degree $\le E$,
$|a|^2=g_+=(1+f)/2$ and $|b|^2=g_-=(1-f)/2$ on $|w|=1$.

\begin{proof}[Proof of Proposition~\ref{thm:identity}]
The coefficient of $z^E$ in $\varphi$ needs $z^{+1}$ from every $D$, i.e.\ the second basis component before every encoding gate:
$\varphi_E=W_Ee_1\langle e_1|W_0|{+}\rangle\prod_l\bar\alpha_l$; likewise $\varphi_{-E}=W_Ee_0\langle e_0|W_0|{+}\rangle\prod_l\alpha_l$. As
$W_E$ is unitary, $\|\varphi_E\|^2+\|\varphi_{-E}\|^2=\Pi$, and $H$ maps $\varphi_{\pm E}$ to $(a_E,b_E)$ and $(a_0,b_0)$. The $w^E$
coefficient of $|a|^2+|b|^2\equiv1$ gives $a_E\bar a_0+b_E\bar b_0=0$, and that of $f=|a|^2-|b|^2$ gives $c_E=a_E\bar a_0-b_E\bar b_0=2a_E\bar a_0$.
Hence $|a_E||a_0|=|b_E||b_0|=\gamma$ and $|a_0|^2+|a_E|^2=\gamma(t_a+t_a^{-1})$, likewise for $b$.
\end{proof}

\paragraph{End-gate form.} Let $c_E\neq0$. In the proof, $\|\varphi_E\|^2=\Pi\,|\langle e_1|W_0|{+}\rangle|^2$ and $\|\varphi_{-E}\|^2=\Pi\,|\langle e_0|W_0|{+}\rangle|^2$;
with $\psi=W_0|{+}\rangle$, $t_at_b=|\psi_1|^2/|\psi_0|^2$, which is $\tan^2(\alpha/2)$ or its inverse, $\alpha$ the angle of the Bloch vector of
$\psi$ to the encoding axis. The same computation with the rows of $HW_E$ gives $t_a/t_b=\tan^{\pm2}(\beta/2)$ for the readout axis. With
$u+u^{-1}=2/\sin\alpha$ for $u=\tan(\alpha/2)$, $t_a+t_a^{-1}+t_b+t_b^{-1}=(u+u^{-1})(v+v^{-1})=4/(\sin\alpha\sin\beta)$, so
$|c_E|=\frac12\Pi\sin\alpha\sin\beta$. (Numerical check: Appendix~\ref{app:numcomp}.)

\paragraph{Branches.} Write the roots of $w^Eg_+(w)$ as pairs $(r_j,1/\bar r_j)$, $|r_j|\le1$, $\lambda^a_j=\log(1/|r_j|)$. A realisation
chooses one root of each pair for $a$; with $\varepsilon_j=+1$ for the inner root, $a=a_E\prod_j(w-\zeta_j)$ gives
$\log t_a=\sum_j\varepsilon^a_j\lambda^a_j$. Likewise for $b$ with $g_-$. The realisation with all $\varepsilon_j=+1$ for both factors is the \emph{minimum-phase} one, with all $\varepsilon_j=-1$ the
\emph{maximum-phase} one; the names come from filter theory and describe where the roots lie, not a quantum phase.

\par\smallskip\noindent\textbf{Corollary~\ref{cor:extreme} (restated).} \emph{\textup{(i)} If $c_E\neq0$, every realisation of $f$ has $\Pi\le\Pi_{\max}=G_++G_-+\gamma^2(G_+^{-1}+G_-^{-1})$ with $G_\pm=\exp\int\log\tfrac{1\pm f}{2}\tfrac{dx}{2\pi}$,
with equality exactly when $a$ and $b$ each have all roots in the closed unit disc or all roots outside the open disc (for both in the disc: the
\emph{minimum-phase} realisation, a term from signal processing). \textup{(ii)} If $|f|<1$, such a realisation has total flip
$\sum_l\log\frac1{1-q_l}\le\frac12\int\log\frac1{1-f^2}\frac{dx}{2\pi}\le\frac{\langle f^2\rangle}{2(1-\|f\|_\infty^2)}$, independent of $E$, with $\langle f^2\rangle=\int f^2\frac{dx}{2\pi}$.}\par\smallskip
\begin{proof}[Proof of Corollary~\ref{cor:extreme}]
(i) Let $P_{\rm in}=\prod_j\min(1,|\zeta_j|)$ and $P_{\rm out}=\prod_j\max(1,|\zeta_j|)$ over the roots of $a$. Jensen's formula, in the
form of the Mahler measure, holds on every branch: $\exp\int\log|a|\tfrac{dx}{2\pi}=|a_E|P_{\rm out}$, so $G_+=|a_E|^2P_{\rm out}^2$ by
$|a|^2=(1+f)/2$; and $|a_0|=|a_E|\prod_j|\zeta_j|=|a_E|P_{\rm in}P_{\rm out}$. With $\gamma=|a_E||a_0|$,
\[G_++\frac{\gamma^2}{G_+}-|a_0|^2-|a_E|^2=|a_E|^2\,(1-P_{\rm in}^2)(P_{\rm out}^2-1)\ge0,\]
since $P_{\rm in}\le1\le P_{\rm out}$, with equality iff $P_{\rm in}=1$ or $P_{\rm out}=1$, i.e.\ iff no root of $a$ lies strictly inside or
none strictly outside. Likewise for $b$; add, and use $\Pi=|a_0|^2+|a_E|^2+|b_0|^2+|b_E|^2$ (Proposition~\ref{thm:identity}). The bound is
attained for every $f$: a root flip $\zeta\mapsto1/\bar\zeta$, rescaled by $|\zeta|$, keeps $|a|$ on the unit circle, so every root choice
keeps $|a|^2+|b|^2=1$ and is realised by a circuit with the same $f$ by the GQSP theorem~\citep{motlagh2024generalized}. (ii) $\Pi_{\max}\ge G_++G_-\ge2\sqrt{G_+G_-}=\exp\tfrac12\int\log(1-f^2)$, and
$-\log(1-y)\le y/(1-y)$.
\end{proof}

\par\smallskip\noindent\textbf{Corollary~\ref{cor:random} (restated).} \emph{Every realisation has $\Pi\ge2|c_E|$; if
$c_E\neq0$, with equality iff $t_a=t_b=1$.}\par\smallskip
\begin{proof}[Proof of Corollary~\ref{cor:random}]
$t+t^{-1}\ge2$, with equality iff $t=1$; for $c_E=0$, $\Pi\ge0$. Lean: \texttt{pi\_ge\_floor\_circuit}, \texttt{pi\_eq\_floor\_iff}.
\end{proof}
\paragraph{Branch mean.} If $c_E\neq0$, then over branches with independent fair signs
$\mathbb E\,\Pi=2\gamma(\prod_j\cosh\lambda^a_j+\prod_j\cosh\lambda^b_j)$. The mean is large: rare, nearly extreme sign choices
dominate it, so it says nothing about a typical branch. \emph{Proof.} With independent fair signs, $\mathbb E(t_a+t_a^{-1})=\mathbb E\,2\cosh\sum_j\varepsilon^a_j\lambda^a_j=2\prod_j\cosh\lambda^a_j$; the
bound $\prod\cosh\lambda_j\le\cosh\sum\lambda_j$ is elementary. That the branches are exactly the sign vectors (every realisation
takes one root of each pair, and every sign vector is realised with the same $f$ and the same $c_E$) is proved in Lean, not assumed:
\texttt{Cor54.cor54\_i}. $\square$

\par\smallskip\noindent\textbf{Proposition~\ref{prop:branchcount} (restated).} \emph{Along a continuous path of gates on which $f>-1$ (resp.\ $f<1$) everywhere, the number $k_a$ (resp.\ $k_b$) of zeros of $a$ (resp.\ $b$) in
the open unit disc $|w|<1$ is constant.}\par\smallskip
\begin{proof}[Proof of Proposition~\ref{prop:branchcount}]
The coefficients of $a_t$ are polynomials in the gate entries, so continuous in $t$; on $|w|=1$, $|a_t|^2=(1+f_t)/2>0$, so $a_t$ has no zero on
the circle. There $a_t'/a_t=\sum_j(w-\zeta_j)^{-1}$ over the zeros of $a_t$, and $\oint_{|w|=1}(w-\zeta)^{-1}dw$ is $2\pi i$ for
$|\zeta|<1$ and $0$ for $|\zeta|>1$, so $k_a=(2\pi i)^{-1}\oint a_t'/a_t$; zeros that move to infinity as the degree drops do not
contribute. The integral is continuous in $t$ and integer-valued, hence constant. Likewise for $b$ with $1-f_t$. Lean:
\texttt{BranchCount.branch\_count\_a}, \texttt{branch\_count\_b}.
\end{proof}

\begin{lemma}[Realisations of a branch]
\label{lem:strip}
Let two serial circuits with gates in $\mathrm{SU}(2)$ have the same function $f\not\equiv\pm1$, and let their polynomials $a$ have the
same roots, with multiplicity and including roots at $0$, and likewise their polynomials $b$. Let the first have no middle gate with $q_l=1$. Then they are related by the torus $G$.
\end{lemma}
\begin{proof}[Sketch]
With the same roots, and with $|a|^2=(1+f)/2$ and $|b|^2=(1-f)/2$ on the unit circle fixing their Mahler measures, $a$ and $b$ agree up
to unit factors $u_a,u_b$. Since $(u,v)=H\varphi$ are the amplitudes in the eigenbasis of the readout, these factors are a rotation
$\mathrm{diag}(u_a,u_b)$ about the readout axis after the last gate, i.e.\ the readout rotation, absorbed into $W_E$, and we may assume equal $(a,b)$, i.e.\ equal state coefficients. Let $c^{(l)}_m$ be the coefficient vector of the state
after $l$ layers, indexed by the number $m$ of encodings passed in $|{+}\rangle$. Then $c^{(l)}_m=W_lv_m$ with
$v_m=(c^{(l-1)}_{m-1,+},c^{(l-1)}_{m,-})$, so $v_0$ and $v_l$ each have one zero entry. Their other entries are the amplitudes of the two
paths that never flip below layer $l$; their moduli are the input's weights on $|{\pm}\rangle$, not both zero, times keep amplitudes,
non-zero since $q_l<1$, so at least one is non-zero. From $W_lv_m=W'_lv'_m$ for all $m$, the unitary $V=W_l'^\dagger W_l$ maps $v_m$ to
$v'_m$, and the zero entries force one off-diagonal entry of $V$ to vanish, hence both. So $W'_l=W_l\,\mathrm{diag}(\bar g)$ with $|g_i|=1$,
and the circuits below layer $l$ have equal state coefficients once $\mathrm{diag}(\bar g)$, which commutes with the encoding, is moved into
$W'_{l-1}$. Induct down to $W_0$, where only the input rotation remains. The condition $q_l<1$ is used for one circuit only. Lean:
\texttt{StripSU.realisations\_SU}, from \texttt{LayerStripFull.realisations\_of\_branch}.
\end{proof}

\subsection{The private pull}
\label{app:pull}
The part of the loss gradient in the private directions moves $c_E$
without disturbing the lower coefficients to first order: with residuals $r$ (Parseval weights included) and $V=\mathrm{Var}(f^*)$, the
\emph{private pull} is $q=(2/V)\Pp\Jtop^\top r_{\rm top}$.
\begin{lemma}[Private pull]
\label{lem:pull}
$q=\Pp\nabla L$, so along gradient flow $\dot{\boldsymbol\theta}=-\nabla L$ the loss falls at least at the rate $\|q\|^2$:
$\frac{d}{dt}L\le-\|q\|^2$.
\end{lemma}
\begin{proof}
$\nabla L=(2/V)(\Jlow^\top r_{\rm low}+\Jtop^\top r_{\rm top})$ with the first term in $\Kl$, so $\Pp\nabla L=q$; and
$\frac{d}{dt}L=-\|\nabla L\|^2\le-\|\Pp\nabla L\|^2$.
\end{proof}

\subsection{Steering the state near the no-flip point}
\label{app:nnf}
Proposition~\ref{prop:noflip} describes the point $\Pi=1$. Here we bound how fast the gates lose control over the \emph{state} as flips appear, with no condition
on phases or end gates. Let $\psi_{\boldsymbol\theta}(x)=U(x)|0\rangle$ be the state of the serial circuit and $D$ the Jacobian of
$\boldsymbol\theta\mapsto\psi_{\boldsymbol\theta}$ as a map into the functions $[-\pi,\pi)\to\mathbb C^2$, with the inner product
$\langle u,v\rangle=\mathbb E_x\,\mathrm{Re}\langle u(x)|v(x)\rangle$, $x$ uniform. Let $D_S$ be its restriction to the coordinates
$X,Y,Z$ of $W_0$ and $Y,Z$ of $W_1,\dots,W_E$ ($2E+3$ coordinates). Since $X$ commutes with the encoding, the coordinate $X$ of $W_l$ moves the state
as a combination of the coordinates of $W_{l-1}$; by induction $D_S$ has the same image as $D$. We write $|\beta_l|=\sqrt{q_l}$ for the flip
amplitude of middle gate $l$ in the encoding basis and
\[
s=\sum_{l=1}^{E-1}|\beta_l|,\qquad s^2\le(E-1)\sum_lq_l\le(E-1)\ln(1/\Pi).
\]
\begin{theorem}[Steering near the no-flip point]
\label{thm:nnf}
For all gates, every singular value $\sigma$ of $D_S$ satisfies
\[
|\sigma^2-1|\le s^2+\tau(s),\qquad \tau(s)=\cosh2s+2e^{2s}-3-4s-6s^2=\tfrac83s^3+O(s^4).
\]
In particular $\sigma_{\min}(D_S)^2\ge1-(E-1)\ln(1/\Pi)-\tau\big(\sqrt{(E-1)\ln(1/\Pi)}\big)$, which is positive for
$(E-1)\ln(1/\Pi)<0.31$. If only one middle gate flips, $\|D_S^\top D_S-I\|=q_l$ exactly.
\end{theorem}
At $\Pi=1$, $D_S$ is an isometry. The bound needs $\Pi\ge0.9567$, $0.9796$, $0.9867$, $0.9901$ at $E=8,16,24,32$; it does not reach Haar
starts or the minimum-phase realisations of Section~\ref{sec:cures}. The single flip shows that the coefficient of $s^2$ cannot be
lowered. Our searches indicate that the factor $E-1$ cannot be removed from a bound in $\Pi$ alone: with flip shares, phases and end gates chosen adversarially at
fixed $\Pi$, our searches reach $1-\sigma_{\min}^2\approx0.4\,(E-1)\ln(1/\Pi)$ at $E=8,16$ and $\Pi=0.99$ (NB42).
The theorem also locates the singular point of check C6 ($E=8$, $\Pi=0.99$, $J$ singular to machine precision): there
$\sigma_{\min}(D_S)^2\ge0.86$ by the theorem ($0.98$ measured, NB42). Since $J$ is the derivative of the readout composed with $D$, and $D_S$ has the
image of $D$, the lost direction is lost in the readout of the state, not in the gates' control of it.
\begin{proof}
\emph{Telescoping.} The coordinate $P$ of $W_l$ moves the state by $U\,i\,Q_l^\dagger PQ_l|0\rangle$, with $Q_l=R_X(x)W_{l-1}\cdots R_X(x)W_0$ the
circuit before $W_l$ (proof of Proposition~\ref{prop:tangent}). Writing traceless Hermitian $2\times2$ matrices as vectors in
$\mathbb R^3$ and $\mathrm{Ad}(V)g=VgV^\dagger$, the inner product of the moves of $g$ at $W_l$ and $g'$ at $W_m$ is
$\mathbb E_x[(\mathrm{Ad}(Q_l)^{-1}g)\cdot(\mathrm{Ad}(Q_m)^{-1}g')]$, since $\mathrm{Re}\langle0|AB|0\rangle=a\cdot b$ for $A=a\cdot\boldsymbol\sigma$, $B=b\cdot\boldsymbol\sigma$.
For $m<l$, $Q_l=P_{ml}Q_m$ with $P_{ml}=R_X(x)W_{l-1}\cdots R_X(x)W_m$, so the $(m,l)$ block of $G=D_S^\top D_S$ is $\mathbb E_x[\mathrm{Ad}(P_{ml})^{-1}]$,
and the diagonal blocks are $I$. The matrix $W_E$ does not enter (no $Q_l$ contains it; its coordinates do), and $W_0$ enters only as the factor $\mathrm{Ad}(W_0)^{-1}$ on the left of the blocks
$(0,l)$; as all three coordinates of $W_0$ belong to $S$, this is an orthogonal change of coordinates and can be dropped.

\emph{Charge walk.} $\mathrm{Ad}(R_X(x))$ has the eigenvalues $e^{-icx}$, $c\in\{1,0,-1\}$ (the \emph{charge}); $c=0$ is the axis $X$, and
$c=\pm1$ span the plane of $Y,Z$, i.e.\ the coordinates in $S$ at $l\ge1$; an orthonormal charge basis is $X$ and $(Y\pm iZ)/\sqrt2$. Expanding
$\mathrm{Ad}(P_{ml})^{-1}=A_m\Delta A_{m+1}\Delta\cdots A_{l-1}\Delta$, with $A_j=\mathrm{Ad}(W_j)^{-1}$ and $\Delta=\mathrm{diag}(e^{icx})$ in the charge basis,
writes an entry of the block as a sum over charge sequences $c_l,\dots,c_m$ (gate $j$ maps $c_{j+1}$ to $c_j$) with the factor
$e^{ix\sum_{i=m+1}^lc_i}$, so $\mathbb E_x$ keeps the \emph{zero-sum} paths, $\sum_{i=m+1}^lc_i=0$. For a middle gate the matrix elements of $A_j$ have
modulus $\le1$ if the charge is kept, $\le\sqrt2|\beta_j|$ if it changes by one (a \emph{single}) and $|\beta_j|^2$ if it changes by two
(a \emph{double}); the \emph{order} of a path is its number of singles plus twice its number of doubles. A path without changes has the constant charge $\pm1$ and a nonzero sum, and so does a path with one single, so
$G-I$ has no terms of order $0$ or $1$ in the flip amplitudes.

\emph{Schur test.} The diagonal blocks are $I$, so $\theta^\top G\theta-\|\theta\|^2=\sum_{m\neq l}\theta_m^\top G_{ml}\theta_l$. In charge coordinates
let $K_{ml}(c,c')$ be the total weight of the zero-sum paths of the entry $(c,c')$, which bounds its modulus. By $2|ab|\le a^2+b^2$,
$|\theta^\top G\theta-\|\theta\|^2|\le M\|\theta\|^2$, with $M$ the largest row sum of $K$ over the coordinates in $S$. A row is read along the
walk away from its gate: upwards through its own gate and those above it, downwards through those below; $a$ and $a'$ denote the sums of $|\beta_j|$ over
the middle gates passed, so $a+a'\le s$.

\emph{Charging a path to its last change.} Call a state \emph{approaching} if its charge is $+1$ and its running sum is negative, or $-1$
and positive. A keep moves an approaching state one step towards zero sum, and a state at zero sum leaves it at the next keep. Hence the
weight that is approaching or at zero sum with charge $\pm1$ after a gate is at most the weight approaching before it plus the inflow of the
changes at that gate: singles from the axis at nonzero sum (weight $\sqrt2|\beta_j|$; at most one of the two directions approaches) and
doubles from charge $\pm1$ (weight $|\beta_j|^2$). Summed over the gates, the weight at zero sum with charge $\pm1$ is at most the approaching
weight at the start, which is $0$ for every row, plus the inflows.

\emph{Weights by order.} Let $\sigma$ be the sum of $|\beta_j|$ over the gates passed so far. By induction over the gates, the total weight at
charge $\pm1$ and $\sqrt2$ times the total weight on the axis (over all running sums), split into the orders $0,1,2,\ge3$, are at most $(1,0,2\sigma^2,h(\sigma))$ and
$(0,2\sigma,0,g(\sigma))$ for a walk started at charge $\pm1$, and $\sqrt2$ times the same two vectors exchanged for a walk started on the
axis, where $h(\sigma)=\cosh2\sigma-1-2\sigma^2$ and $g(\sigma)=\sinh2\sigma-2\sigma$; one gate takes $\sigma$ to $\sigma+|\beta_j|$, and each step
follows from the addition formulas of $\cosh$ and $\sinh$. The zeros say that paths between charges $\pm1$ have an even number of
singles.

\emph{Rows.} Inserting these weights into the inflows and telescoping over the gates gives the row sums. Upwards from charge $\pm1$ they are
at most $\tfrac12(\cosh2a-1)=a^2+\tfrac12h(a)$, and upwards from the axis of $W_0$ at most $\tfrac1{\sqrt2}g(a)$ (this walk starts at sum $0$, so its order-$0$ weight on the axis never
approaches; only singles from the axis at nonzero sum and doubles feed the inflows). Downwards from charge $\pm1$,
the entry at gate $l$ is the zero-sum state before $W_l$ followed by $W_l$ itself, which weighs $1+|\beta_l|^2$ from charge $\pm1$ and
$2\sqrt2|\beta_l|$ from the axis; paths of order $0$ and $1$ never reach zero sum and paths of order $2$ end at charge $\pm1$ (by parity), so the axis enters only
from order $3$; at the end of the walk, $W_0$, all three charges count with no gate factor, which adds the axis term. The row sum is
at most $\tfrac12(\cosh2a'-1)+h(a')+\tfrac1{\sqrt2}g(a')$. Since $h$ and $g$ are superadditive and increasing, $a^2+a'^2\le s^2$ and
$\tau=3h+2g$, every row sum is at most $s^2+\tau(s)$, and $\|G-I\|\le s^2+\tau(s)$. Finally $\sum_l\sqrt{q_l}\le\sqrt{(E-1)\sum_lq_l}$ and
$q_l\le-\ln(1-q_l)$.

\emph{One flip.} If only gate $j$ flips, every inflow vanishes except at $W_j$, paths have order at most $2$, and the row sums are at most
$|\beta_j|^2=q_j$, so $\|G-I\|\le q_j$. For equality take $\theta$ with the coordinate $Z$ at $W_{j-1}$ (for $j=1$ after the change of
coordinates $\mathrm{Ad}(W_0)$) and a unit vector in the plane of $Y,Z$ at $W_{j+1}$. The entry of $G_{j-1,j+1}$ between the charges $+1$ and
$-1$ comes from the double at $W_j$ alone, $k=(A_{j-1})_{++}(A_j)_{+-}$ with $|k|=|\beta_j|^2$, since $W_{j-1}$ does not flip (for $j=1$,
$k=(A_1)_{+-}$); the entry between $-1$ and $+1$ is $\bar k$, and those between equal charges vanish, because $W_{j-1}$ does not flip
(for $j=1$, because their sum is not zero). Choosing the direction at $W_{j+1}$ in phase with $k$ gives
$\theta^\top G\theta-\|\theta\|^2=2|k|=q_j\|\theta\|^2$.
\end{proof}
The expansion is checked numerically (NB42, Appendix~\ref{app:d15}): built as a charge walk, $G$ reproduces the squared singular values of $D_S$
to $1.0\times10^{-14}$, the parts of order $0$ and $1$ are $I$ and $0$ to $10^{-15}$, and over 36 gate sets ($E=5,7,9,12$; random, aligned and single flips)
the row sums of the order-two part $G_2$ of $G-I$ reach at most $s^2$ (and $s^2$ to $10^{-5}$ at single flips) and those of the remainder at most $0.16\,\tau(s)$.

\subsection{Corners in several dimensions}
\label{app:corners}
Let one qubit encode $d$ inputs, each $E$ times, by encoding gates $R_X(x_{i_l})$ at positions $l=1,\dots,dE$ in any order, with gates
$W_0,\dots,W_{dE}\in\mathrm{SU}(2)$, input $|0\rangle$ and readout $Z$; $q_l$ is the flip probability of the middle gate $W_l$, and $\alpha$, $\beta$ are
as in Proposition~\ref{thm:identity}.
\begin{corollary}[Corner identity]
\label{cor:corner}
For $s\in\{\pm1\}^d$ the coefficient of $f$ at the corner $k=(s_1E,\dots,s_dE)$ satisfies $|c_k|=\tfrac12\Pi_s\sin\alpha\sin\beta$, where
$\Pi_s=\prod_{l=1}^{dE-1}p_l$ with $p_l=1-q_l$ if $s_{i_l}=s_{i_{l+1}}$ and $p_l=q_l$ otherwise.
\end{corollary}
\begin{proof}
By Theorem~\ref{thm:fourier} in each input (the others fixed), $f$ has degree $\le E$ in every variable. Let $g(x)=f(s_1x,\dots,s_dx)$. A coefficient $c_{k'}$ of $f$ contributes to frequency $\sum_rs_rk'_r\le dE$ of $g$, with equality only at
$k'=k$, so $c_k$ is the top coefficient of $g$. Since $R_X(-x)=ZR_X(x)Z$, $g$ is computed by a serial circuit with $dE$ encoding gates
$R_X(x)$ and the unitary gates $W'_l=Z^{a_{l+1}}W_lZ^{a_l}$, where $a_l=1$ if $s_{i_l}=-1$ and $a_l=0$ otherwise ($a_0=a_{dE+1}=0$).
In the encoding basis $Z$ exchanges the two eigenstates. A middle gate between encoding gates of equal sign is therefore
conjugated or left alone and keeps its flip probability $q_l$; one between opposite signs is multiplied by $Z$ on one side only, which
exchanges keeping and flipping, so its flip probability becomes $1-q_l$. The no-flip product of the new circuit is $\Pi_s$. Exchanging the
eigenstates maps the angle of the input state or of the readout axis to the encoding axis from $\alpha$ to $\pi-\alpha$, so $\sin\alpha$ and
$\sin\beta$ are unchanged. The end-gate form of Proposition~\ref{thm:identity}, which holds for every circuit with unitary gates (\texttt{BranchFacts.end\_gate\_uncond}),
gives the claim.
\end{proof}
Numerically, the corner coefficient agrees with the top coefficient of the restricted circuit to $7.9\times10^{-16}$ and the corner identity
holds to $7.1\times10^{-15}$ at 1{,}680 corners, $d=2,3$ (Appendix~\ref{app:nb35}).

\subsection{The readout factorisation}
\label{app:readout}
Write the output state of the serial circuit in the readout basis as $z^{-E}\big(a(w),b(w)\big)$ (Section~\ref{sec:branch}), and collect
the real and imaginary parts of the coefficients of $a$ and $b$ in $p=(\mathrm{Re}\,a,\mathrm{Im}\,a,\mathrm{Re}\,b,\mathrm{Im}\,b)\in\mathbb R^{4E+4}$.
Let $D=\partial p/\partial\boldsymbol\theta\in\mathbb R^{(4E+4)\times P}$. Since the readout basis change is unitary and the functions
$z^{-E}w^k$ are orthonormal on the circle, the Euclidean inner product of moves of $p$ equals the inner product $\mathbb E_x\mathrm{Re}\langle u|v\rangle$
of moves of the state used in Appendices~\ref{app:nnf} and~\ref{app:gate-control} (Parseval), so $D$ there and here have the same Gram matrix and
singular values. Let $R$ be the real-linear map that sends
$(\delta a,\delta b)$ to the coefficient vector $\tilde{\mathbf c}$ (Definition~\ref{def:jacobian}) of the real trigonometric polynomial
$4\,\mathrm{Re}\big(\bar a\,\delta a\big)$ on $|w|=1$; $J$ is the unweighted Jacobian of this appendix. With the Parseval weights of
Section~\ref{sec:structure}, $J$ and $R$ are both multiplied from the left by $\mathrm{diag}(1,\sqrt2,\dots,\sqrt2)$, which leaves every
private fraction below unchanged. The \emph{allowed space} is
$\mathcal A(a,b)=\{(\delta a,\delta b):\mathrm{Re}(\bar a\,\delta a+\bar b\,\delta b)\equiv0\text{ on }|w|=1\}$. For a matrix $M$ with
rows indexed like $\tilde{\mathbf c}$, $\pf(M)$ denotes the private fraction of Definition~\ref{def:private} computed from the rows of
$M$, so that $\pf=\pf(J)$.

\begin{proposition}[Readout factorisation]
\label{prop:readout}
Let $r=\operatorname{rank}D$, $\kappa_D=\sigma_r(D)/\sigma_1(D)$, let the columns of $Q$ be an orthonormal basis of $\operatorname{range}D$,
and let $\pf_R=\pf(RQ)$, the \emph{readout privacy}; it does not depend on the choice of $Q$. Then
\begin{enumerate}
\item[(i)] $J=RD$;
\item[(ii)] $\kappa_D\,\pf_R\le\pf\le\pf_R/\kappa_D$;
\item[(iii)] $\operatorname{range}D\subseteq\mathcal A(a,b)$. If $\dim\mathcal A(a,b)=r$, then $\operatorname{range}D=\mathcal A(a,b)$, and
$\pf_R$ depends on the gates only through $(a,b)$ and is unchanged by $a\mapsto u_a a$, $b\mapsto u_b b$ with $|u_a|=|u_b|=1$; by the
first step of the proof of Lemma~\ref{lem:strip} it is then a function of the realised function and of the root choices of $a$ and $b$.
\end{enumerate}
(i)--(iii) hold in any smooth coordinates on the manifold of unitary gates, with $J$ and $D$ taken in the same coordinates.
\end{proposition}
\begin{proof}
(i) Every circuit has $|a|^2+|b|^2=1$ and $f=|a|^2-|b|^2$ on $|w|=1$, so $f=2|a|^2-1$ there, and differentiating along any coordinate
$\theta_j$ gives $\partial_jf=4\,\mathrm{Re}(\bar a\,\partial_ja)$. Taking coefficients, a linear operation, gives
$\partial_j\tilde{\mathbf c}=R\,\partial_jp$.
(ii) $D=QG$ with $G=Q^\top D\in\mathbb R^{r\times P}$ of full row rank, whose singular values are the non-zero singular values
$\sigma_1\ge\dots\ge\sigma_r>0$ of $D$; so $J=MG$ with $M=RQ$, and every row of $J$ is $G^\top$ applied to the corresponding row of $M$.
For $u\in\mathbb R^r$, $\sigma_r\|u\|\le\|G^\top u\|\le\sigma_1\|u\|$. Let $K_M$ be the span of the lower rows of $M$; then $G^\top K_M$ is the
span $K$ of the lower rows of $J$. For each top row $m$ of $M$, $\|P_{K^\perp}G^\top m\|=\min_{k\in K_M}\|G^\top(m-k)\|$ lies between
$\sigma_r$ and $\sigma_1$ times $\min_{k\in K_M}\|m-k\|=\|P_{K_M^\perp}m\|$. Summing squares over the two top rows bounds the numerator of
$\pf(J)$ between $\sigma_r$ and $\sigma_1$ times that of $\pf(M)$, and likewise
$\sigma_r\|M_{\mathrm{top}}^\top\|_F\le\|J_{\mathrm{top}}^\top\|_F\le\sigma_1\|M_{\mathrm{top}}^\top\|_F$. The ratio gives (ii). Replacing $Q$ by
$QO$ with $O$ orthogonal replaces $M$ by $MO$, which leaves $\pf(M)$ unchanged.
(iii) Differentiating $|a|^2+|b|^2\equiv1$ gives $\mathrm{Re}(\bar a\,\partial_ja+\bar b\,\partial_jb)\equiv0$, so every column of $D$ lies in
$\mathcal A(a,b)$; equal dimensions give equality, and then $\pf_R=\pf(R\,Q_{\mathcal A})$ for an orthonormal basis $Q_{\mathcal A}$ of
$\mathcal A(a,b)$, which is determined by $(a,b)$, as is $R$. The map $(\delta a,\delta b)\mapsto(u_a\delta a,u_b\delta b)$ is an isometry
from $\mathcal A(a,b)$ onto $\mathcal A(u_aa,u_bb)$, and $\mathrm{Re}\big(\overline{u_aa}\,u_a\delta a\big)=\mathrm{Re}(\bar a\,\delta a)$, so
$\pf_R$ is unchanged. By the first step of the proof of Lemma~\ref{lem:strip}, two realisations with the same function and the same roots
of $a$ and of $b$ (with multiplicity, including roots at $0$) have the same $(a,b)$ up to such unit factors.
\end{proof}

\paragraph{Coordinates.} $\pf$ and $\kappa_D$ depend on the coordinates of the gates; $\pf_R$ does not. In the Lie-algebra coordinates
of Section~\ref{sec:structure} the coordinate $X$ of each middle gate repeats a combination of coordinates of the gate before it
(Appendix~\ref{app:nnf}), which inflates $\sigma_1(D)$; we measure medians $\kappa_D\sqrt{E+1}=0.91$--$0.95$ at minimum-phase
realisations for every $E$ from $8$ to $128$ (Appendices~\ref{app:readoutexp} and~\ref{app:d13}). In the coordinates $S$ of Theorem~\ref{thm:nnf} ($X,Y,Z$ of $W_0$; $Y,Z$ of
$W_1,\dots,W_E$) we write $\pf_S$ and $\kappa_{D_S}$; Proposition~\ref{prop:readout} holds there with the same $\pf_R$, since $D_S$ has the
range of $D$. Theorem~\ref{thm:nnf} bounds $\kappa_{D_S}$ below for $(E-1)\ln(1/\Pi)<0.31$.

\paragraph{Hypotheses.} In exact arithmetic $\operatorname{rank}D\le2E+3$: the $2E+2$ tangent directions of
Proposition~\ref{prop:tangent} and the global phase. At every point of Appendices~\ref{app:readoutexp}--\ref{app:d14} we use $r=2E+3$. On the
extreme branches $\dim\mathcal A(a,b)=2E+3$ is proved in Theorem~\ref{thm:privacy-extreme} (Step~0), so (iii) applies wherever
$\operatorname{rank}D=2E+3$; elsewhere we check it numerically where float64 resolves it (Appendix~\ref{app:readoutexp}, Part D).

\paragraph{What remains.} Proposition~\ref{prop:readout} turns the question of Section~\ref{sec:branch}, whether $\Pi$ controls $\pf$, into
three inequalities: (a) $\pf_R$ bounded below on the extreme branches by a function of $f$ alone, which
Theorem~\ref{thm:privacy-extreme} proves; (b) $\kappa_{D_S}$ bounded below there independently of $E$, which we measure
(Appendix~\ref{app:d13}) and reduce to two frame constants (Proposition~\ref{prop:gate-control}, Corollary~\ref{cor:frames}); (c) $\pf_R$
small on the balanced branches where Haar runs stall, which we measure (Appendix~\ref{app:readoutexp}). (a) and (b) together bound $\pf_S$
below on the extreme branches, not $\pf$ itself. (c) bounds $\pf$ above through (ii) only where $\kappa_D$ is not small; at the stalled points
of Appendix~\ref{app:readoutexp} $\kappa_D$ is itself small, so there the measured smallness of $\pf$ is not a consequence of (ii).
Theorem~\ref{thm:nnf} gives (b) only for $(E-1)\ln(1/\Pi)<0.31$.

\subsection{Readout privacy on the extreme branches}\label{app:theorem-a}

Throughout, $\langle u\rangle=\int_{-\pi}^{\pi}u(x)\,\frac{dx}{2\pi}$, $\mathrm{Var}\,u=\langle u^2\rangle-\langle u\rangle^2$, $f_{\max}=\max f$, $f_{\min}=\min f$,
$F=\|f\|_\infty$, and $\rho_R$ is computed on $\mathcal A(a,b)$, which equals $\operatorname{range}D$ under the hypothesis of
Proposition~\ref{prop:readout}(iii). An \emph{extreme branch} is one on which $a$ has all its roots on one side of the unit circle, and so has $b$
(Corollary~\ref{cor:extreme}).

\begin{theorem}[Readout privacy on extreme branches]\label{thm:privacy-extreme}
Let $E\ge1$, $c_E\ne0$ and $F<1$, and let $(a,b)$ lie on an extreme branch. For $\theta\in\mathbb R$ let
\begin{equation*}
C(\theta)=\Big\langle\frac{\cos^2(Ex+\theta)}{1-f^2}\Big\rangle+\frac{1+f_{\max}}2\,\mathrm{Var}\frac{\cos(Ex+\theta)}{1+f}
+\frac{1-f_{\min}}2\,\mathrm{Var}\frac{\cos(Ex+\theta)}{1-f},
\end{equation*}
$G_\pm=\exp\langle\log\frac{1\pm f}2\rangle$, $\gamma=|c_E|/2$ and $P_\pm=G_\pm+\gamma^2/G_\pm$. Then
\begin{enumerate}
\item[(i)] $\displaystyle\rho_R\;\ge\;\max_\theta\frac{\sqrt{1/C(\theta)+1/C(\theta+\pi/2)}}{\sqrt{8\min(P_+,P_-)}}$;
\item[(ii)] in particular $\displaystyle\rho_R\ge(1-F)\sqrt{\frac{1+F}{(1-F)+(1+F)^2}}$.
\end{enumerate}
Neither bound depends on $E$ other than through $f$.
\end{theorem}

\begin{proof}
\emph{Step 0: the readout is onto.} Write $p^\#(w)=w^E\overline{p(1/\bar w)}$ for a polynomial $p$ of degree $\le E$; on $|w|=1$,
$\bar p=w^{-E}p^\#$, so $S_a\delta a:=a^\#\delta a+a\,\delta a^\#=2w^E\,\mathrm{Re}(\bar a\,\delta a)$ there. By Proposition~\ref{thm:identity},
$c_E\ne0$ gives $a_0,a_E\ne0$, so $\deg a=E$; since $F<1$, $a$ has no root on the circle, so on an extreme branch its roots lie strictly on one side. If
$\mathrm{Re}(\bar a\,\delta a)\equiv0$ on the circle, then $S_a\delta a=0$ as a polynomial, so $a$ divides $a^\#\delta a$. A common root $z$ of
$a$ and $a^\#$ would be nonzero ($a^\#(0)=\bar a_E$) and make $1/\bar z$, which lies on the other side of the circle, a root of $a$ as well; so
$a$ and $a^\#$ are coprime, $a$ divides $\delta a$, and $\delta a=\mu a$ with $\mathrm{Re}\,\mu\,|a|^2\equiv0$, i.e.\ $\mu\in i\mathbb R$. The
real-linear map $\delta a\mapsto\mathrm{Re}(\bar a\,\delta a)$, from the $2E+2$ real dimensions of $\delta a$ to the $2E+1$ of real trigonometric
polynomials of degree $\le E$, therefore has a one-dimensional kernel and is onto. Hence the constraint
$(\delta a,\delta b)\mapsto\mathrm{Re}(\bar a\delta a+\bar b\delta b)$ is onto and $\dim\mathcal A(a,b)=2E+3$.

\emph{Step 1: reduction to roots outside the disc.} For all $\delta$ of degree $\le E$, $S_{a^\#}(\delta^\#)=S_a\delta$ as polynomials,
$\|\delta^\#\|=\|\delta\|$, and $|a^\#|=|a|$ on the circle. If the roots of $a$ lie inside the disc, then $a^\#$ (with $a^\#_E=\bar a_0\ne0$) has
no zeros in the closed disc, and a move $\delta$ for $a^\#$ with $\mathrm{Re}(\overline{a^\#}\delta)=\psi$ and the bound of Step~3 gives the move
$\delta^\#$ for $a$ with $\mathrm{Re}(\bar a\,\delta^\#)=\psi$ and the same bound. Likewise for $b$. In Steps~2 and~3 we may therefore assume
that $a$, and $b$, have no zeros in the closed unit disc.

\emph{Step 2: a private move.} Let $\psi_\theta=\frac12\cos(Ex+\theta)$. By Step~0 there is a polynomial $\delta a_p$ of degree $\le E$ with
$\mathrm{Re}(\bar a\,\delta a_p)=\psi_\theta$ on the circle. Let $r=\mathrm{Im}\big(\delta a_p(0)/a(0)\big)$ and $\delta a_\theta=\delta a_p-ira$. Since
$\mathrm{Re}(\bar a\cdot ia)=0$, still $\mathrm{Re}(\bar a\,\delta a_\theta)=\psi_\theta$, and $h=\delta a_\theta/a$ has $h(0)$ real. Likewise
$\delta b_\theta$ with $\mathrm{Re}(\bar b\,\delta b_\theta)=-\psi_\theta$. Then $v_\theta=(\delta a_\theta,\delta b_\theta)\in\mathcal A(a,b)$ and
$4\,\mathrm{Re}(\bar a\,\delta a_\theta)=2\cos(Ex+\theta)$: the move changes no coefficient below $E$, and $c_E$ by
$(\mathrm{Re}\,c_E,\mathrm{Im}\,c_E)=(\cos\theta,\sin\theta)$.

\emph{Step 3: its cost.} Since $a$ has no zeros in the closed disc, $h$ and $h^2$ are holomorphic on a neighbourhood of it, and the mean-value
property gives $\langle h\rangle=h(0)$ and $\langle h^2\rangle=h(0)^2$. With $h(0)$ real, taking real parts gives $\langle\mathrm{Re}\,h\rangle=h(0)$
and $\langle(\mathrm{Re}\,h)^2-(\mathrm{Im}\,h)^2\rangle=h(0)^2$, so $\langle(\mathrm{Im}\,h)^2\rangle=\mathrm{Var}\,\mathrm{Re}\,h$. On the circle
$\mathrm{Re}\,h=\mathrm{Re}(\bar a\,\delta a_\theta)/|a|^2=u_a:=\psi_\theta/|a|^2$. By Parseval, $\|\delta a_\theta\|^2=\langle|\delta a_\theta|^2\rangle
=\langle|a|^2(\mathrm{Re}\,h)^2\rangle+\langle|a|^2(\mathrm{Im}\,h)^2\rangle\le\langle\psi_\theta^2/|a|^2\rangle+\max|a|^2\,\mathrm{Var}\,u_a$, and
likewise for $b$ with $u_b=-\psi_\theta/|b|^2$. With $|a|^2=\frac{1+f}2$, $|b|^2=\frac{1-f}2$ and $\frac1{|a|^2}+\frac1{|b|^2}=\frac4{1-f^2}$ this
gives $\|v_\theta\|^2\le C(\theta)$.

\emph{Step 4: privacy.} Let $t_1,t_2\in\mathcal A(a,b)$ represent $v\mapsto\mathrm{Re}\,c_E$ and $v\mapsto\mathrm{Im}\,c_E$ of $Rv$, and $K$ the span
of the representers of the lower coefficients; then $\rho_R=\|P_{K^\perp}[t_1,t_2]\|_F/\|[t_1,t_2]\|_F$, which does not change if
$(t_1,t_2)$ is rotated to $u_\theta=\cos\theta\,t_1+\sin\theta\,t_2$, $u'_\theta=-\sin\theta\,t_1+\cos\theta\,t_2$. By Step~2, $v_\theta$ and
$v_{\theta+\pi/2}$ lie in $K^\perp$, and $\langle u_\theta,v_\theta\rangle=\langle u'_\theta,v_{\theta+\pi/2}\rangle=1$. Hence
$\|P_{K^\perp}u_\theta\|\ge1/\|v_\theta\|$, $\|P_{K^\perp}u'_\theta\|\ge1/\|v_{\theta+\pi/2}\|$, and the numerator of $\rho_R$ is at least
$\sqrt{1/C(\theta)+1/C(\theta+\pi/2)}$.

\emph{Step 5: the denominator.} The top coefficient of $4\,\mathrm{Re}(\bar a\,\delta a)$ is $c_E=2(\bar a_0\,\delta a_E+a_E\overline{\delta a_0})$,
a real-linear map of $(\delta a_0,\delta a_E)$ with squared Frobenius norm $8(|a_0|^2+|a_E|^2)$; restricting to $\mathcal A(a,b)$ does not
increase it. On $\mathcal A(a,b)$ the same coefficient equals that of $-4\,\mathrm{Re}(\bar b\,\delta b)$, which gives $8(|b_0|^2+|b_E|^2)$.
On an extreme branch, Jensen's formula as in the proof of Corollary~\ref{cor:extreme} gives $\{|a_0|^2,|a_E|^2\}=\{G_+,\gamma^2/G_+\}$ and
$\{|b_0|^2,|b_E|^2\}=\{G_-,\gamma^2/G_-\}$, so $\|[t_1,t_2]\|_F^2\le8\min(P_+,P_-)$. With Step~4 this proves (i).

(ii) Since $\cos^2(Ex+\theta)+\cos^2(Ex+\theta+\frac\pi2)=1$ and $\mathrm{Var}\,u\le\langle u^2\rangle$,
$C(\theta)+C(\theta+\frac\pi2)\le\frac1{1-F^2}+\frac{1+F}{(1-F)^2}$; and $1/C+1/C'\ge4/(C+C')$. Moreover, by
Step~5 and Parseval ($E\ge1$), $P_++P_-=|a_0|^2+|a_E|^2+|b_0|^2+|b_E|^2\le\sum_k(|a_k|^2+|b_k|^2)=\langle|a|^2+|b|^2\rangle=1$, so
$8\min(P_+,P_-)\le4$. Inserting both into (i) gives (ii).
\end{proof}

\paragraph{Remarks.} Step~3 is where the branch enters: on a balanced branch $a$ has zeros on both sides of the circle, so $\delta a/a$ has
poles in the disc whether $a$ or $a^\#$ is used, the mean-value step fails, and the bound of (i) need not hold (on balanced branches it exceeds $\rho_R$ at 16 of 24 realisations; Appendix~\ref{app:d14}). The bound is not sharp: at amplitude $0.9$ it is about half the measured $\rho_R$; when $f$ comes
close to $+1$ at one point and to $-1$ at another, $\rho_R$ itself tends to zero, so some dependence on $F$ is unavoidable (Appendix~\ref{app:d14}).

\subsection{Gate control}\label{app:gate-control}

Let $G=D_S^\top D_S$ be the Gram matrix of $D$ in the coordinates $S$ of Theorem~\ref{thm:nnf}, ordered as ($X,Y,Z$ of $W_0$; $Y,Z$ of $W_1$; \dots;
$Y,Z$ of $W_E$), so that $\kappa_{D_S}^2=\lambda_{\min}(G)/\lambda_{\max}(G)$. Write $G_{00}$ for its $3\times3$ block of $W_0$, $G_{\rm mid}$ for the block
of $W_1,\dots,W_E$, and $C$ for the coupling between them. Let $|\pm\rangle$ be the eigenvectors of the encoding generator $X$, and for
$k=1,\dots,E$ let $P_k(x)=R_X(x)W_{k-1}R_X(x)\cdots W_1R_X(x)$ be the middle circuit before $W_k$ ($P_1=R_X$), with stay and flip amplitudes
$s_k=\langle+|P_k|+\rangle$, $f_k=\langle+|P_k|-\rangle$. For $c\in\mathbb C^E$ let $s_c=\sum_kc_ks_k^2$, $f_c=\sum_kc_kf_k^2$, $p_c=\sum_kc_ks_kf_k$.

\begin{proposition}[Gate control]\label{prop:gate-control}
$G_{00}=I$, and $G_{\rm mid}$ has the same eigenvalues as the real quadratic form
\begin{equation*}
Q(c)=\big\langle\,|s_c+\overline{f_c}|^2\big\rangle+4\big\langle(\mathrm{Im}\,p_c)^2\big\rangle\qquad\text{on }\mathbb C^E\cong\mathbb R^{2E}.
\end{equation*}
In particular $\lambda_{\min}(G_{\rm mid})=\min_{\|c\|=1}Q(c)$.
\end{proposition}

\begin{proof}
\emph{The Gram matrix as an average.} Identify traceless Hermitian $2\times2$ matrices with $\mathbb R^3$. By the telescoping step of the proof of
Theorem~\ref{thm:nnf}, the inner product of the moves of the coordinate $g$ of $W_l$ and the coordinate $g'$ of $W_m$ is
$\mathbb E_x[(\mathrm{Ad}(Q_l)^{-1}g)\cdot(\mathrm{Ad}(Q_m)^{-1}g')]$, where $Q_l$ is the circuit before $W_l$ (and $Q_0=I$). So the block of $W_0$ is $\mathbb E_x[g\cdot g']=\delta_{gg'}$, i.e. $G_{00}=I$. For
$l\ge1$, $Q_l=P_lW_0$, and the constant rotation $\mathrm{Ad}(W_0)^{-1}$ cancels in the dot products, so
$(G_{\rm mid})_{(l,g),(m,g')}=\mathbb E_x[(\mathrm{Ad}(P_l)^{-1}g)\cdot(\mathrm{Ad}(P_m)^{-1}g')]$.

\emph{One gate.} For $P=\begin{psmallmatrix}p&-\bar q\\q&\bar p\end{psmallmatrix}\in\mathrm{SU}(2)$ (entries in the computational basis) let $s=\langle+|P|+\rangle=(p+\bar p+q-\bar q)/2$ and
$f=\langle+|P|-\rangle=(p-\bar p+q+\bar q)/2$. For real $a,b$ and $\gamma=a+ib$, in the coordinates $(X,Y,Z)$,
\begin{equation*}
\mathrm{Ad}(P)^{-1}(aY+bZ)=\big(-2\,\mathrm{Im}(\gamma sf),\ \mathrm{Re}(\gamma s^2+\overline{\gamma f^2}),\ \mathrm{Im}(\gamma s^2+\overline{\gamma f^2})\big).
\end{equation*}
This is a polynomial identity in $p,q,\bar p,\bar q,a,b$, checked by expanding both sides ($\mathrm{Ad}(P)^{-1}g$ has the coordinates of $P^\dagger gP$).

\emph{All gates.} Write a real coordinate vector of $G_{\rm mid}$ as $\theta=(\theta_{l,Y},\theta_{l,Z})_{l=1}^E$ and let
$\gamma_l=\theta_{l,Y}+i\theta_{l,Z}=\sqrt2\,c_l$, so that $\|\theta\|^2=\|\gamma\|^2=2\|c\|^2$. By the first step,
$\theta^\top G_{\rm mid}\theta=\mathbb E_x\|\sum_l\mathrm{Ad}(P_l)^{-1}(\theta_{l,Y}Y+\theta_{l,Z}Z)\|^2$, and by the identity with $s=s_l$, $f=f_l$ the sum is
$\big(-2\,\mathrm{Im}\,p_\gamma,\ \mathrm{Re}(s_\gamma+\overline{f_\gamma}),\ \mathrm{Im}(s_\gamma+\overline{f_\gamma})\big)$. Hence
\begin{equation*}
\theta^\top G_{\rm mid}\,\theta=\big\langle|s_\gamma+\overline{f_\gamma}|^2\big\rangle+4\big\langle(\mathrm{Im}\,p_\gamma)^2\big\rangle=Q(\sqrt2\,c)=2Q(c),
\end{equation*}
and $\theta\mapsto\sqrt2\,c$ is an orthogonal map of $\mathbb R^{2E}$, so $G_{\rm mid}$ and $Q$ have the same eigenvalues.
\end{proof}

\begin{corollary}[A frame-bound reduction]\label{cor:frames}
Let $A_S^2$ be the smallest eigenvalue of the Gram matrix $(\langle\bar s_k^2s_l^2\rangle)_{kl}$ of the stay family $\{s_k^2\}$ in $L^2$ of the circle,
and $\sigma_F^2$ the largest eigenvalue of that of the flip family $\{f_k^2\}$. Then
\begin{equation*}
\lambda_{\min}(G_{\rm mid})\ge(A_S-\sigma_F)_+^2\qquad\text{and}\qquad\lambda_{\min}(G)\ge\min\big(1,\lambda_{\min}(G_{\rm mid})\big)-\|C\| .
\end{equation*}
\end{corollary}

\begin{proof}
Writing $\|g\|^2=\langle|g|^2\rangle$ for functions, $Q(c)\ge\|s_c+\overline{f_c}\|^2\ge(\|s_c\|-\|f_c\|)^2$, with $\|s_c\|\ge A_S\|c\|$ and $\|f_c\|\le\sigma_F\|c\|$. For the second inequality, $G$ is the
block-diagonal matrix with blocks $G_{00}=I$ and $G_{\rm mid}$ plus the off-diagonal blocks $C$, $C^\top$, whose norm is $\|C\|$ (Weyl).
\end{proof}

\paragraph{Remarks.} The proposition expresses gate control through the partial products $P_k$ of the circuit, which are the partial
nonlinear Fourier transforms of its gates \citep{alexis2024qsp,ni2025inlfft}. At the no-flip point $f_k\equiv0$ and $|s_k|=1$; then
$Q(c)=\|s_c\|^2$, and $\{s_k^2\}$ is orthonormal (its members are unimodular multiples of distinct powers of $e^{ix}$), so $G=I$, as in
Theorem~\ref{thm:nnf}. A lower bound on $\kappa_{D_S}$ that does not depend on $E$ would therefore follow from bounds on $A_S$ from below and on
$\sigma_F$ from above that are uniform in the length of the circuit, together with bounds on $\|C\|$ and on $\lambda_{\max}(G)$. We measure such bounds on the extreme branches at 60 minimum-phase realisations of amplitude-$0.9$ targets
(Appendix~\ref{app:d14}: $A_S\ge0.76$, $\sigma_F\le0.24$ for $E=16$--$128$) but do not prove them. \citet{ni2025inlfft} prove bounds independent of the length for a related
matrix of the full transform (their Lemma~5.4); their error bound for layer stripping, which inverts the transform, grows as the cube of the length. For coherent flip
phases, the case in which, at fixed $\Pi$, the loss of gate control grows with $E$ (as the factor $E-1$ of Theorem~\ref{thm:nnf} allows), $A_S$ falls and $\sigma_F$ grows with $E$ (Appendix~\ref{app:d14}).

\section{Lean formalisation}
\label{app:lean}
The proofs are those of Appendix~\ref{app:proofs}. We checked them in Lean~4, except those marked ``by hand'' in Table~\ref{tab:status}, and
include the Lean files for readers who want to retrace them. Where a Lean statement is narrower than the paper's (one qubit, tangent directions, cited results as hypotheses), we say so
below and in Table~\ref{tab:status}. Every lemma, proposition, remark and corollary of Appendices~\ref{sec:background}--\ref{sec:tangent} is stated
and proved in Lean~4, two of them for the cases the paper uses: Lemma~\ref{lem:genericity}, whose polynomiality hypothesis is discharged
for the serial and the ancilla circuit with one gate per block, and Lemma~\ref{lem:stack}, whose Pauli-word instances are proved for one and two qubits. For Lemma~\ref{lem:bridge} this includes $\mathbf J_{\rm data}=BJ$ for the serial and the ancilla circuit, the orthogonality of
$BD^{-1}/\sqrt n$ on the grid and the exact factor $\sqrt n$ between the extreme singular values, for both circuits (\texttt{BridgeData}, \texttt{BridgeAncilla}). We use Lean~4 (v4.28.0) with Mathlib (commit pinned in
\texttt{lake-manifest.json}); the project is part of the repository and is built with \texttt{lake build}. The proofs were developed with the help of AI tools (the automated
prover Aristotle and a language model) and are checked by the Lean kernel. The project contains no \texttt{sorry}; \texttt{\#print axioms} reports only \texttt{propext},
\texttt{Classical.choice} and \texttt{Quot.sound} (output for every theorem in \texttt{axioms\_output.txt}). Circuits are formalised at the level of gates, as in the paper, and $J$ is the
derivative of $\tilde{\mathbf c}$ in Lie-algebra coordinates at the current gates. The README of the project maps each result to its
Lean theorems and states the conventions of the file.

\paragraph{Cited results.} They enter as explicit hypotheses of the theorems that use them, never as axioms: the Fejér--Riesz
factorisation~\citep{fejer1916trigonometrische,riesz1916problem}; Theorem~3 of~\citet{motlagh2024generalized}; Sard's
theorem~\citep{sard1942measure} for smooth maps $\mathbb R^n\to\mathbb R^m$; and that the zero set on $\mathrm{SU}(d)^m$ of a polynomial in
the real and imaginary parts of the entries that is nonzero at one point is null. For Haar measure the last one follows
from~\citet{mityagin2020zero}, which covers connected open domains of $\mathbb R^D$, applied in charts, together with the identity
theorem for real-analytic functions and the smooth density of Haar measure in charts. In Lean this property is a predicate on a measure
$\mu$, and the almost-everywhere statements are proved for every $\mu$ that satisfies it; the zero measure satisfies it trivially, so these
statements concern the Haar measure only through the cited fact that it satisfies the predicate. Theorem~\ref{thm:fourier} is cited and not
formalised; its serial and ancilla cases are proved.

\paragraph{Statements made in running text.} Formalised are: that parts (i) and (ii) of Lemma~\ref{lem:phaselock} do not depend on the
phase of $|\psi^\perp\rangle$; for the generality of Definition~\ref{def:vqc}, the equality of the outputs, that the state is $V_M$
times that of the circuit with blocks $\tilde W$, that the gates stay in $\mathrm{SU}(d)$, and that the map of the gates is a bijection; and the counts of readout rotations and of spare directions (paragraph \emph{Readout rotations and spare
directions} of Appendix~\ref{sec:tangent}). For one qubit, that an encoding generator and an observable with eigenvalues $\pm1$ and any unit
initial state are carried to $R_X$, $Z$ and $|0\rangle$ by fixed gates in $\mathrm{SU}(2)$, with the encoding written as the matrix exponential
$e^{-ixH/2}$ and the map of the gates a bijection of $\mathrm{SU}(2)^{E+1}$ (\texttt{Wlog}). For $t$ unitary gates per block, that $JJ^\top$ is $t$
times $JJ^\top$ of the circuit whose gates are the block products, so that rank, left singular vectors and condition number agree and the
singular values differ by $\sqrt t$ (\texttt{MultiGate}). That the hypotheses of Theorem~\ref{thm:rigid} hold outside a null set of gate tuples
(\texttt{AxisGeneric}), and that two realisations on the same branch with no middle gate a full flip have the same $JJ^\top$
(\texttt{BranchFacts}). Not formalised are: that the degree bounds of Lemma~\ref{lem:phaselock}(iii) can fail for
another phase; on several qubits, that an encoding generator or an observable with the same eigenvalues differs from ours by a unitary; that translations
preserve Haar measure; that $J$ and the span of the tangent vectors carry over under the change of frame, and the translated forms of
Corollary~\ref{cor:noredundancy}(i) and Lemma~\ref{lem:phaselock}(iii); the correspondence of the optimiser runs; and, on two qubits, that a readout rotation leaves $f$ unchanged
beyond first order. The
Pauli-word instances of Lemma~\ref{lem:stack} are proved for one and two qubits, the cases used in this paper. Lemma~\ref{lem:genericity}
is proved with the polynomiality of $\det(JJ^\top)$ as a hypothesis, which the paper derives from Lemma~\ref{lem:poly}; the hypothesis
is discharged for the serial and the ancilla circuit with one gate per block, the circuits to which the lemma is applied; for several
gates per block the paper's argument is by hand. Theorem~3 of~\citet{motlagh2024generalized}
enters in its authors' convention, and the conversion of the proof of Lemma~\ref{lem:rowrank}(i) is proved.

\paragraph{Rigidity, private fraction, path structure and branches.} The statements of Appendices~\ref{app:pf-rigid}--\ref{app:branch} are
formalised in separate files of the same project, with the exceptions stated in this paragraph. Theorem~\ref{thm:rigid} is formalised for the circuit in full: near
$\mathbf W$ the gates that realise the same function are gauge rotations of $\mathbf W$ (\texttt{RigidMain}); a gauge rotation multiplies $J$
from the right by a block-diagonal matrix with blocks in $\mathrm{SO}(3)$ (\texttt{RigidCov}); and near $\mathbf W$ any two tuples of gates in
$\mathrm{SU}(2)$ that realise the same function have the same $JJ^\top$ and the same characteristic polynomial of $JJ^\top$, hence the same
singular values of $J$ (\texttt{RigidUniform}). The coefficient vector of the Lean project omits the Parseval factors $\sqrt2$; they rescale
fixed rows of $J$, which changes none of these statements. The private fraction $\pf$ of Definition~\ref{def:private} is formalised for the circuit, with the top rows
identified as those of $\mathrm{Re}\,c_E$ and $\mathrm{Im}\,c_E$ and $\|\Pp t\|^2$ written as the least-squares distance of $t$ from the span
of the lower rows; the Parseval factors leave $\pf$ unchanged. Its gauge invariance (Lemma~\ref{lem:cov}) and Corollary~\ref{cor:rigidflow},
including the closed equation for $\dot{\mathbf c}$, are formalised (\texttt{RigidPrivate}, \texttt{RigidUniform}, \texttt{PrivatePull}).
Lemmas~\ref{lem:private} and~\ref{lem:pull} are formalised for the circuit (\texttt{PrivatePull}), with the loss of Section~\ref{sec:structure} on the $2E+1$
equispaced points, the Parseval-weighted $J$ and $r$, and the gradient in the Lie-algebra coordinates of the gates: $\nabla L=\tfrac2VJ^\top r$,
$q=\Pp\nabla L$, $\dot L=-\|\nabla L\|^2\le-\|q\|^2$ along every gradient flow (Lemma~\ref{lem:pull}), and Lemma~\ref{lem:private} with the two top rows and $\sigma_{\min}(J)$. Lemma~\ref{lem:singlepath} is formalised for the circuit (\texttt{SinglePath}): the product form, exactly with $\overline{\alpha_l}^{\,2}$ for
gates in $\mathrm{SU}(2)$ and with factors of modulus $|\alpha_l|^2$ for unitary gates, the bounds $|\phi_0|,|\phi_E|\le1$, and $|c_E|\le\Pi$. Proposition~\ref{prop:branchcount} is formalised through the argument principle for polynomials,
$k_a=(2\pi i)^{-1}\oint_{|w|=1}a'(w)/a(w)\,dw$, which needs no winding numbers (\texttt{BranchCount}). Of
Proposition~\ref{prop:noflip}, all three parts are formalised for the circuit (\texttt{NoFlipJ}, \texttt{NoFlipA16}): the columns of $J$
are the first-order path sums of the perturbed gate, so part~(ii) holds for the entries of $J$ itself, including that a phase change leaves
$c_0$ unchanged; $\sin\alpha=2|\psi_0||\psi_1|$, $\cos\alpha=|\psi_0|^2-|\psi_1|^2$ for the input column of the first gate and
$\sin\beta=2|M_{00}||M_{01}|$, $\cos\beta=|M_{00}|^2-|M_{01}|^2$ for the readout row of the last gate, both in the encoding basis,
which are the Bloch angles of the proposition. The count of Section~\ref{sec:rigid} for the ancilla control is formalised for tangent directions
(\texttt{AncillaSym}, \texttt{AncillaCount}): the kernel of $J$ has dimension at least $13E+14$, with equality at full rank; the tangent vectors
of the symmetries span at most $7E+16$ dimensions (commutant of the encoding, $7$ per encoding; stabiliser of $|00\rangle$, $9$; commutant of
the readout, $7$); so at least $6E-2$ dimensions of the kernel are not symmetry directions. Of Remark~\ref{rem:clt} only the law of large numbers is, for abstract i.i.d.\ variables; that these are the
$1-q_l$ of Haar gates and that $\Pi$ is their product are by hand. The end-gate form of Proposition~\ref{thm:identity}, with $\sin\alpha=2|\psi_0||\psi_1|$ and $\sin\beta$ the same expression for the
readout row of the last gate, and the factorisation of $a_0$ by the roots of $a$ with the sign form of
$\log t_a$ (its third equality, used in Corollary~\ref{cor:extreme}), are formalised as well. Corollary~\ref{cor:corner} is formalised for one qubit with $d$ inputs encoded in any order and gates in $\mathrm{SU}(2)$
(\texttt{CornerIdentity}): for every family $c$ with $f(\mathbf x)=\sum_{\mathbf k\in[-E,E]^d}c_{\mathbf k}e^{i\mathbf k\cdot\mathbf x}$ the corner coefficient satisfies
$|c_{s E}|=\tfrac12\Pi_s\sin\alpha\sin\beta$, with $\sin\alpha$, $\sin\beta$ as above, and such a family exists (\texttt{CornerExpansion}). Lemma~\ref{lem:strip} is formalised as stated (\texttt{StripSU}): for gates in $\mathrm{SU}(2)$ the two circuits are related by
the action of the torus $G$ used in Theorem~\ref{thm:rigid}, and $f\not\equiv\pm1$ makes $a$ and $b$ nonzero. Proposition~\ref{thm:identity} is formalised for the circuit itself, with no hypothesis beyond unitary gates and, for the
equalities that need it, $c_E\neq0$, which makes the four extreme coefficients nonzero (\texttt{Prop52}): the state's extreme coefficients come from the two no-flip paths, $\Pi=|a_0|^2+|a_E|^2+|b_0|^2+|b_E|^2$, and the
$w^E$-coefficients of $|a|^2+|b|^2\equiv1$ and $f=|a|^2-|b|^2$ give $|c_E|=2|a_E||a_0|=2|b_E||b_0|$. Corollary~\ref{cor:extreme}(ii) is formalised for the circuit on the
minimum-phase branch, with no hypothesis beyond unitary gates, $c_E\neq0$ and $|f|<1$: Jensen's formula (via the Mahler measure),
$|a|^2=(1+f)/2$ and $|b|^2=(1-f)/2$ on the unit circle and the integrability of $\log\frac{1\pm f}2$ are proved. Part~(i) is
formalised for the circuit on every branch, the bound and the equality case; that every root choice is realised by a circuit with the
same $f$, so that the bound is attained, follows from the cited GQSP theorem by root flips $r\mapsto1/\bar r$ (\texttt{Realise}), with the realising gates in
$\mathrm{U}(2)$; a unit phase on every gate puts them in $\mathrm{SU}(2)$ and changes neither $f$, nor the roots of $a$ and $b$, nor $\Pi$
(\texttt{RealiseSU}). Part~(ii) is formalised on every extreme branch, through $\Pi=\Pi_{\max}\ge G_++G_-$, including its last inequality, where $\|f\|_\infty^2$ is attained and below one (\texttt{RealiseSU}). Corollary~\ref{cor:random} is formalised for the circuit (\texttt{Cor54}): a circuit with the same
$f$ has the same $c_E$ (Fourier uniqueness) and takes one root of each pair $\{\zeta_j,1/\bar\zeta_j\}$, so the realisations of $f$ are
exactly the sign vectors, each realised by the cited GQSP theorem (in $\mathrm{SU}(2)$, as above); on each, $\Pi=\gamma\,(2\cosh\sum_j\varepsilon^a_j\lambda^a_j+2\cosh\sum_j\varepsilon^b_j\lambda^b_j)$,
and the mean over all $2^{2E}$ sign vectors is the branch mean of Appendix~\ref{app:branch} (\texttt{cor54\_i}); Corollary~\ref{cor:random} is proved with
its equality case. Roots on the unit circle ($\lambda_j=0$) belong to both choices of their pair, so there the sign vectors count a branch more than once. The circuit-level statements of Appendix~\ref{app:branch} are proved for the circuit with diagonal encodings, which equals, as a
matrix, the circuit of Section~\ref{sec:structure} (with $R_X$ encodings, input $|0\rangle$ and readout $Z$) with gates $W_EH$, $HW_lH$,
$HW_0$ (\texttt{circRX\_eq\_circ}). Proposition~\ref{thm:identity} with its end-gate form, Corollaries~\ref{cor:extreme} and~\ref{cor:random},
Proposition~\ref{prop:branchcount} and Lemma~\ref{lem:strip} are also stated for the circuit of Section~\ref{sec:structure} itself, so that $f$, $\Pi=\prod_l|\langle{+}|W_l|{+}\rangle|^2$ and the $q_l$ enter
in the paper's terms, and $a$, $b$ and their roots in the encoding eigenbasis, as in Section~\ref{sec:branch} (\texttt{FrameLink}, \texttt{FrameLink2},
\texttt{Cor54}, \texttt{Prop52}, \texttt{RealiseSU}). Hypothesis taken from outside: the uniform law of Haar flip probabilities (Remark~\ref{rem:clt}). Several of these statements are general-gate forms of results of~\citet{wang2022energy} for symmetric QSP
(Appendix~\ref{app:branch}); the formalised proofs are ours, and none of their results enters as a hypothesis.

\paragraph{Steering near the no-flip point.} Theorem~\ref{thm:nnf} is formalised for the circuit with gates in $\mathrm{SU}(2)$, with $\mathbb E_x$
the exact average over an equispaced grid of $2E+1$ points, in Rayleigh form (\texttt{NnfGate}, \texttt{NnfWalk}, \texttt{NnfMass}, \texttt{NnfHarvest}, \texttt{NnfRows}, \texttt{NnfProd},
\texttt{NnfLink}, \texttt{NnfMain}, \texttt{NnfCor}): for $G$ the Gram matrix of the moves of all gate coordinates and every $\theta$ supported on the
coordinates $S$, $|\theta^\top G\theta-\|\theta\|^2|\le(s^2+\tau(s))\|\theta\|^2$; with $s^2\le(E-1)\ln(1/\Pi)$, $\theta^\top G\theta\ge(1-u-\tau(\sqrt u))\|\theta\|^2$
for $u=(E-1)\ln(1/\Pi)$, and $1-u-\tau(\sqrt u)>0$ for $u<0.31$. If only middle gate $l$ flips, $|\theta^\top G\theta-\|\theta\|^2|\le q_l\|\theta\|^2$ for every
such $\theta$, with equality for some $\theta\neq0$ (\texttt{NnfSingle}, \texttt{NnfSingleMain}, \texttt{NnfSingleAttain}). The steps from this Rayleigh form to the
singular values of $D_S$ and to $\|D_S^\top D_S-I\|$, the equality of the grid average with $\int\mathrm dx/2\pi$, that $D_S$ has the image of $D$, and the expansion $\tau(s)=\tfrac83s^3+O(s^4)$ are by hand.

\paragraph{Readout factorisation and gate control.} Proposition~\ref{prop:readout}(ii) and the independence of $\pf_R$ from the
choice of $Q$ are formalised for any matrices $R$, $Q$, $G$ with $\sigma_{\min}(G)>0$, with $\kappa_D$ taken as
$\sigma_{\min}(G)/\sigma_{\max}(G)$ for $D=QG$; that these are $\sigma_r(D)$ and $\sigma_1(D)$ when $Q$ has orthonormal columns is by hand.
The private fraction is formalised for any split of the rows into top and lower rows, and for the circuit it is that of
Definition~\ref{def:private} by definition (\texttt{ReadoutFactor}). Parts (i) and (iii) are by hand. Proposition~\ref{prop:gate-control} is formalised for the circuit with gates in $\mathrm{SU}(2)$
(\texttt{StateGram}, \texttt{GramAverage}, \texttt{GateControl}): the columns of $D$ are the derivatives of the state along the coordinate curves
$s\mapsto W_l\exp(isg)$, $\mathbb E_x$ is the exact average over an equispaced grid, and the proposition is proved as $G_{00}=I$,
$\theta^\top G_{\rm mid}\theta=Q(\sqrt2\,c)=2Q(c)$ with $\|\theta\|=\|\sqrt2\,c\|$, and $\lambda_{\min}(G_{\rm mid})=\inf_{\|c\|=1}Q(c)$, with $\lambda_{\min}$ the
infimum of the Rayleigh quotient; the step from this Rayleigh form to equal spectra is by hand. Corollary~\ref{cor:frames} is formalised with
$\langle\cdot\rangle$ the exact average over an equispaced grid, $A_S^2$ and $\sigma_F^2$ as the extremes of the Rayleigh quotients
$\langle|s_c|^2\rangle$ and $\langle|f_c|^2\rangle$, and $\|C\|$ as any bound on $|a^\top Cw|/(\|a\|\|w\|)$ (\texttt{GateBounds}); part (i) is also
formalised for the circuit, $\lambda_{\min}(G_{\rm mid})\ge(A_S-\sigma_F)_+^2$ with $s_k=\langle+|P_k|+\rangle$, $f_k=\langle+|P_k|-\rangle$
(\texttt{GateControl}). Theorem~\ref{thm:privacy-extreme} is formalised, (i) and (ii), for any polynomials $a,b$ of degree $\le E$, $E\ge1$, with
$|a|^2+|b|^2=1$ on $|w|=1$ (\texttt{ExtremeOnto}, \texttt{CircleTools}, \texttt{PrivateMove}, \texttt{PrivacyBound}, \texttt{ReadoutRows},
\texttt{ExtremeFacts}, \texttt{PrivacyMoves}, \texttt{PrivacyExtreme}): $\pf_R$ is $\pf(RQ)$ for every $Q$ with orthonormal columns spanning
$\mathcal A(a,b)$, $\langle\cdot\rangle$ is the average over the unit circle, $c_E$ is the top Fourier coefficient of $f=|a|^2-|b|^2$, and the
maximum over $\theta$ is stated as a bound for every $\theta$. Step~0, including $\dim\mathcal A(a,b)=2E+3$, is formalised as well.

\paragraph{Status of every statement.} Table~\ref{tab:status} lists how every formal statement of the paper is proved, and how the
statements of the main text, and those of the appendix proved by hand only, are checked numerically. Statements of the appendix that are
formalised are not checked separately. The README of the project maps each statement to its Lean theorems.

{\small\setlength{\tabcolsep}{4pt}
\begin{longtable}{>{\raggedright\arraybackslash}p{0.3\linewidth}>{\raggedright\arraybackslash}p{0.2\linewidth}>{\raggedright\arraybackslash}p{0.44\linewidth}}
\caption{Proof and numerical check of every statement. Every proved statement has a written proof in Appendix~\ref{app:proofs}.
Lean: the written proof was also checked in Lean, possibly with the hypotheses listed above; by hand: written proof only. Checks in
Appendix~\ref{app:valid} unless noted.}
\label{tab:status}\label{tab:leanmap}\\
\toprule
statement & proof & numerical check: result \\
\midrule
\endfirsthead
\toprule
statement & proof & numerical check: result \\
\midrule
\endhead
\midrule\multicolumn{3}{r}{\emph{continued}}\\
\endfoot
\bottomrule
\endlastfoot
\multicolumn{3}{l}{\emph{Main text}}\\
Thm.~\ref{thm:fourier} (Fourier) & cited; serial and ancilla cases Lean & --- \\
Lem.~\ref{lem:rowrank} (model class) & Lean & (i) exact realisations reproduce $f$ to $10^{-8}$ at 2{,}160 branches (T3); (ii) full rank in float64 at 90/90 structured starts ($E=24$, S1, App.~\ref{app:starts}); at Haar starts in extended precision at 1000/1000 ($E=8$) and 20/20 ($E=24$, mostly below float64 resolution) (C2, App.~\ref{app:nb37}) \\
Thm.~\ref{thm:rigid} (rigidity) & Lean & fibre walk: serial $JJ^\top$ changes by $10^{-14}$, ancilla by $6\times10^{-2}$ (App.~\ref{app:numcomp}) \\
Count of Sec.~\ref{sec:rigid} (ancilla fibre) & Lean (tangent directions); manifold statement by hand & measured part (conditioning changes along the fibre): walk along a non-symmetry fibre direction at $E=24$, $JJ^\top$ changes by $6\times10^{-2}$ (App.~\ref{app:numcomp}) \\
Lem.~\ref{lem:private} (private fraction) & Lean & 900 points: 900/900 (T5) \\
Lem.~\ref{lem:pull} (private pull) & Lean & 553 resolved points and 150 checkpoints of two runs: 553/553, 150/150 (T6) \\
Prop.~\ref{thm:identity} (no-flip identity) & Lean & 80 points at 50 digits, max error $2\times10^{-47}$ (T2); end-gate form at 100 points at 50 digits, $3\times10^{-37}$ (C5); sign form of $\log t_a$ at 2{,}000 branches, $3\times10^{-14}$ (T3) \\
Cor.~\ref{cor:extreme} (extreme branches) & Lean & 160 targets, $E=8$--$32$: formula to $8\times10^{-16}$, bound 160/160 (T3) \\
Cor.~\ref{cor:random} (floor) & Lean & 2{,}000 branches: 2000/2000 (T3) \\
branch mean of $\Pi$ (App.~\ref{app:branch}) & Lean & 10/10 functions (T3) \\
Prop.~\ref{prop:branchcount} (branch count) & Lean & 40 gradient-descent paths: every change at $|f|\ge1-10^{-2}$, 32/32 (T4) \\
\midrule
\multicolumn{3}{l}{\emph{Running text of the main text}}\\
$H=X$, $M=Z$, $|0\rangle$ (Sec.~\ref{sec:structure}) & Lean, one qubit; several qubits by hand & 20 instances, $1\times10^{-14}$ (C1, App.~\ref{app:nb37}) \\
hypotheses of Thm.~\ref{thm:rigid} hold almost everywhere (Sec.~\ref{sec:final}) & Lean & full rank and off-axis input at 1000/1000 Haar starts ($E=8$, 100 digits) and full rank at 20/20 ($E=24$, 120 digits) (C2) \\
same branch, same $JJ^\top$ (Sec.~\ref{sec:branch}) & Lean & 20 targets: same branch $3\times10^{-15}$, other branch $\ge0.2$ (C3) \\
Gamma law of $-\ln\Pi$ at Haar gates (Sec.~\ref{sec:cost}) & by hand (standard) & KS $p=0.31$--$0.73$ (T1) \\
corner identity in several dimensions (Sec.~\ref{sec:final}; Cor.~\ref{cor:corner}) & Lean (\texttt{CornerIdentity}, \texttt{CornerExpansion}) & 1{,}680 corners, $7\times10^{-15}$ (NB35 C2) \\
$\pf$ at $\Pi=1$ with Haar end gates (Sec.~\ref{sec:branch}) & measured & $\pf=0.99$ at $E=24$ (App.~\ref{app:numcomp}) \\
several gates per block (Sec.~\ref{sec:cures}; after Lem.~\ref{lem:stack}) & Lean & 40 draws, $t=5$, $4\times10^{-15}$ (C4) \\
\midrule
\multicolumn{3}{l}{\emph{Appendix}}\\
Prop.~\ref{prop:tangentmain} (tangent count) & Lean & exact numerical rank at random gates (Table~\ref{tab:tangent}) \\
Lem.~\ref{lem:bridge} (from $J$ to data) & Lean & \\
Lem.~\ref{lem:genericity}--\ref{lem:phaselock}; Rem.~\ref{rem:multiqubit}; Prop.~\ref{prop:tangent}; Cor.~\ref{cor:noredundancy}; Lem.~\ref{lem:cov}, \ref{lem:orbit} & Lean & \\
Lem.~\ref{lem:singlepath} (single path) & Lean & \\
Rem.~\ref{rem:clt} (Haar product) & law of large numbers Lean (abstract variables); link to Haar gates and Gamma law by hand & Gamma law, KS $p=0.31$--$0.73$ at $E=8$--$32$ (T1) \\
Prop.~\ref{prop:noflip} (no-flip point) & Lean & bookkeeping to $10^{-12}$ at $E=10,24$ (App.~\ref{app:numcomp}) \\
Thm.~\ref{thm:nnf} (steering near the no-flip point) & Lean for gates in $\mathrm{SU}(2)$, with $\mathbb E_x$ the average over an equispaced grid, in Rayleigh form on the coordinates $S$, incl.\ the bound in $\Pi$, the threshold $0.31$ and the single flip (bound $q_l$, attained); singular values and $\|D_S^\top D_S-I\|$ from the Rayleigh form, grid average $=\int\mathrm dx/2\pi$, image of $D_S$ by hand & NB42 (App.~\ref{app:d15}): charge walk $10^{-14}$; row sums at 36 gate sets; C6; searches \\
Prop.~\ref{prop:readout} (readout factorisation) & (ii) and choice of $Q$: Lean, with $\kappa_D$ as $\sigma_{\min}/\sigma_{\max}$ of $G=Q^\top D$; (i), (iii) by hand & 1{,}054 points (NB32 points, NB39 Q0, NB40 P0): $\|J-RD\|/\|J\|\le1.3\times10^{-12}$, bounds hold at every point (App.~\ref{app:readoutexp}, \ref{app:d13}) \\
Thm.~\ref{thm:privacy-extreme} (readout privacy, extreme branches) & Lean, for any $(a,b)$ with $|a|^2+|b|^2=1$ on $|w|=1$ & bound below $\pf_R$ at 240 (NB40 P0) and 172 realisations incl.\ stress families (NB41); fails on balanced branches at 16/24 (App.~\ref{app:d14}) \\
Prop.~\ref{prop:gate-control} (gate control, exact) & Lean for gates in $\mathrm{SU}(2)$, with $\mathbb E_x$ the average over an equispaced grid, in Rayleigh form ($G_{00}=I$, $\theta^\top G_{\rm mid}\theta=Q(\sqrt2\,c)$, $\lambda_{\min}(G_{\rm mid})=\inf_{\|c\|=1}Q$); equal spectra by hand & spectra of $G_{\rm mid}$ and $Q$ agree to $5.3\times10^{-14}$, 60 realisations (NB41) \\
Cor.~\ref{cor:frames} (frame-bound reduction) & Lean, with $\langle\cdot\rangle$ the average over an equispaced grid (\texttt{GateBounds}); (i) also for the circuit (\texttt{GateControl}) & both inequalities at 60 realisations (NB41) \\
Lem.~\ref{lem:strip} (realisations of a branch) & Lean & \\
Cor.~\ref{cor:rigidflow} (private fraction and flow) & Lean & \\
\end{longtable}}

\section{Experimental details}
\label{app:experiments}
\subsection{Circuits, optimisers and data}
\paragraph{Code.} The repository \repourl{} contains the library \texttt{phaselock\_core.py} (circuits, targets, optimisers, statistics) and the
helper modules of the notebooks; two notebooks that run the experiments of the main text and write their results to CSV files,
\texttt{nb1\_geometry} for the exact loss (Section~\ref{sec:exact}: the three optimisers, conditioning along gradient descent, end points of
Levenberg--Marquardt) and \texttt{nb2\_practice} for the data (Section~\ref{sec:data}: learning-rate tuning, evaluation,
Levenberg--Marquardt on the same runs, and the checks of learning rate and step budget); the notebooks of Appendices~\ref{app:nbprot}
and~\ref{app:valid}; the script \texttt{screen\_realdata.py} for the screen of real-world series; the result files; and the Lean project
of Appendix~\ref{app:lean}. Saved gates and loss curves are attached to its release as a separate archive. The scripts \texttt{tables\_exact.py},
\texttt{tables\_data.py} and \texttt{figures.py} make the tables and data figures from the result files alone; they run no circuit.
Simulation is exact statevector simulation in float64 with JAX~\citep{jax2018github}; the README of the repository gives the hardware and
JAX version of every notebook.

\paragraph{Architectures.} All circuits have the form $W^{(L)}S(x)W^{(L-1)}\cdots S(x)W^{(0)}|0\rangle$ with
$S(x)=\bigotimes_w R_X(x)^{e_w}$, $e_w\in\{0,1\}$, and readout $Z$; on two qubits, numbered 0 and 1, the readout is on qubit~1.
\emph{Serial}: $N=1$, $L=E$. \emph{Ancilla}: $N=2$, encoding on qubit~0 only ($e=(1,0)$), $L=E$. Each block is a product of one or more general $\mathrm{SU}(2^N)$ gates. At initialisation every gate is drawn
independently from the Haar measure, the uniform distribution on $\mathrm{SU}(2^N)$, by the QR method of~\citet{mezzadri2007generate}
followed by division by a $2^N$-th root of the determinant.

\paragraph{Parameterisation.} Each trainable gate is stored as a matrix $G\in\mathrm{SU}(2^N)$, not as a product of rotations with
angles, and can be compiled into native gates after training (three rotations for $\mathrm{SU}(2)$; at most three CNOT gates and
single-qubit rotations for $\mathrm{SU}(4)$~\citep{vatan2004optimal}). Its coordinates are those of Section~\ref{sec:background}: the derivative along $G\exp(i\boldsymbol\theta\cdot\mathbf P)$, with $\mathbf P$ the
$4^N-1$ non-identity Pauli words, is computed at $\boldsymbol\theta=0$ by automatic differentiation, and an update multiplies the gate by
$\exp(i\boldsymbol\theta\cdot\mathbf P)$, evaluated by eigendecomposition of the Hermitian generator. Gradient descent in these coordinates
is Riemannian steepest descent on $\mathrm{SU}(2^N)$ with its bi-invariant metric~\citep{absil2008optimization,wiersema2023optimizing}.
On hardware, the gradient in these coordinates can be estimated with the parameter-shift rule, since every Pauli word has eigenvalues
$\pm1$~\citep{wiersema2023optimizing}; the gate is updated classically and compiled again after each step.
The choice does not change what a circuit can represent: angles that cover $\mathrm{SU}(2^N)$ and these coordinates reach the same gates,
and where such angles are regular they span the same tangent directions, so that the counts of Appendix~\ref{sec:tangent} hold for them as
well. The two differ where angles are singular, as Euler angles are where two of them act alike: there a direction is lost, and nearby
directions are distorted. Our coordinates offer every direction at every gate at the same scale, so that a difference between circuits
cannot stem from such points. Optimiser trajectories do depend on the coordinates, and training with angles was not studied.

\paragraph{Exact loss (Section~\ref{sec:exact}).} The loss is $\mathcal L=\|f-f^*\|^2_{L^2(S^1)}/\mathrm{Var}_{S^1}(f^*)$, evaluated on the
$2E+1$ equispaced points of $[-\pi,\pi)$; since $(f-f^*)^2$ has degree $\le2E$, this grid average equals the integral over the
circle exactly. Targets: \emph{flat} ($c_0,a_k,b_k\sim\mathcal N(0,1)$) and \emph{$1/k$} ($a_k,b_k\sim\mathcal N(0,1)/k$) random real
trigonometric polynomials of degree $E$, shifted and scaled so that $\sup|f^*|=0.9$ on a grid of 8192 points; they lie strictly inside
the serial and ancilla model classes. $E\in\{8,12,16,24,32\}$; 100 runs per circuit and configuration, each with its own target and
initialisation; the targets are the same for both circuits.

\paragraph{Optimisers (exact loss).} \emph{Gradient descent}: Riemannian steepest descent~\citep[Ch.~4]{absil2008optimization} with Armijo
backtracking~\citep{armijo1966minimization} (constant
$10^{-4}$; the first trial step doubles the previous accepted step; halving on rejection; initial step $1/\lambda_{\max}$ of the
Gauss--Newton matrix at initialisation); budget 50{,}000 steps.
\emph{Levenberg--Marquardt}~\citep{levenberg1944method,marquardt1963algorithm}: steps
$-\mathbf V\,\mathrm{diag}\big(s_i/(s_i^2+\lambda s_1^2)\big)\mathbf U^\top\mathbf r$ from the singular value decomposition
$\mathbf J_{\rm data}=\mathbf U\,\mathrm{diag}(s)\mathbf V^\top$, $\lambda$ multiplied by $0.3$ on
acceptance and by $5$ on rejection (initial value $10^{-3}$, clipped to $[10^{-15},10^{8}]$), 2{,}000 iterations. \emph{Adam}~\citep{kingma2014adam}: $\beta_1=0.9$, $\beta_2=0.999$,
$\epsilon=10^{-12}$, cosine schedule from the peak rate to $1/100$ of it, budget 50{,}000 steps; peak rate from
$\{10^{-3},3\cdot10^{-3},10^{-2},3\cdot10^{-2},10^{-1}\}$, selected on 5 separate targets by the \emph{tuning score}, the 75th
percentile of the final $\log_{10}$ loss: the selected rate is the centre of the \emph{near-best} rates, those whose score is within
0.25 of the best. Gradient descent and Levenberg--Marquardt stop updating a run once
its loss is below $10^{-14}$. Adam always runs the full schedule, and every run is scored at its end, in tuning and in evaluation.

\paragraph{Reported quantities (exact loss).} \emph{Convergence}: a run converges if its final loss is at most $10^{-6}$. We report
the share of converged runs per circuit with a 95\% Wilson interval~\citep{wilson1927probable}, and the difference of two shares with
Newcombe's hybrid score interval~\citep{newcombe1998interval}; a difference is \emph{significant} if this interval excludes zero.
Table~\ref{tab:rates_exact} gives the counts at $10^{-4}$, $10^{-6}$, $10^{-8}$ and $10^{-10}$. \emph{Budget}: for gradient descent, whose
loss decreases monotonically under the line search, the number of runs with loss $\le10^{-6}$ after 7{,}500, 14{,}500, 27{,}000 and
50{,}000 steps (Table~\ref{tab:gd_budget}).

\emph{Conditioning}: for the loss of Section~\ref{sec:exact}, with $n=2E+1$ grid points, residual $\mathbf r$ and $V=\mathrm{Var}(f^*)$,
$\nabla\mathcal L=\frac{2}{nV}\mathbf J_{\rm data}^\top\mathbf r$ and $\|\mathbf r\|^2=nV\mathcal L$, so that $\|\nabla\mathcal L\|^2\ge\frac{4}{nV}\sigma_{\min}(\mathbf J_{\rm data})^2\mathcal L
\ge\frac4V\sigma_{2E+1}(J)^2\mathcal L$ by Lemma~\ref{lem:bridge}; this is the inequality of \citet[Theorem~1]{liu2022loss} for our loss, and under
gradient flow it bounds the rate of decrease of the loss from below. We record $\kappa_J$ at initialisation and at 14 checkpoints along
gradient descent, spaced geometrically from step 500 to step 50{,}000, and the loss at every step. Computed singular values carry
absolute errors of order $\varepsilon\sigma_1$, $\varepsilon\approx1.1\times10^{-16}$~\citep[Sec.~8.6]{golub2013matrix}, so that values of $\kappa_J$
below about $10^{-15}$ are not resolved.

\emph{Adam.} Converged runs of Adam end not at machine precision but at a level that depends on the learning rate
(Appendix~\ref{app:additional}). The threshold $10^{-8}$, which we had fixed before the runs, touches the upper end of these levels for
the ancilla circuit; we therefore report $10^{-6}$. Two checks of the learning rate apply to every circuit alike. \emph{Bracketing}: if the near-best rates include the smallest rate of the grid, the
rates $3\cdot10^{-4}$ and $10^{-4}$ are added (if they include the largest, $0.3$), the rate is selected again on the same tuning targets,
and if it changes, the 100 runs are repeated with the new rate. \emph{Common rate}: all configurations at the rate $10^{-3}$.

\paragraph{Training on data (Section~\ref{sec:data}).} 300 equispaced points of $[-\pi,\pi)$ with one fixed random 200/100 train/test split.
Targets: \emph{flat} as above ($\sup|f^*|=0.9$; reported in Appendix~\ref{app:additional}) and \emph{Poisson-kernel}, a sum of three Poisson kernels with random centres, signs and weights
and Fourier coefficients decaying as $r^{|k|}$ with $r^E=0.05$ ($\sup|f^*|=0.8$), which is not representable at degree $E$;
$E\in\{8,12,16,24,32\}$. Real data: monthly mean air temperatures at Nottingham, 1920--1939 (the \texttt{nottem} data set distributed
with R; Series~R of \citealp{anderson1976time}), the whole series of 240 values mapped to $[-\pi,\pi)$, one fixed random
180/60 split, centred on the training mean and scaled so that the degree-$E$ least-squares fit has $\sup=0.9$; $E\in\{24,32\}$.
Screened series (\texttt{gasprice}, weekly US gasoline prices from 1990 to 2003, from the R package \texttt{quantreg} \citep{koenker2026quantreg}; \texttt{ibmclose}, daily IBM
closing prices, Series~B of \citet{box1970time}, taken from the R package \texttt{timeSeriesDataSets} \citep{rossi2024timeseriesdatasets}): the last 300 values with an affine trend removed, mapped to $[-\pi,\pi)$, one fixed
random 200/100 split, centred and scaled as for the temperatures; $E\in\{24,32\}$; learning-rate grid extended downwards by $3\cdot10^{-4}$.
Circuits at matched parameter counts: serial with five gates per block (15 parameters per block), ancilla with one
(15). Full-batch Adam as above with 20{,}000 steps, peak rate
tuned per circuit and configuration on 5 separate targets (for the real-world series: 5 separate initialisations) by the 75th percentile
of the $\log_{10}$ excess training loss, with the selection rule of the exact loss; 100 runs per circuit and configuration (real-world
series: 100 initialisations). Levenberg--Marquardt as on the exact loss (2{,}000 iterations), from the same data and initial gates, in every configuration.

\paragraph{Screening of real-world series.} The screen (\texttt{screen\_realdata.py} in the repository) was fixed before any circuit was run on
a new series. Candidates: every univariate, regularly sampled time series in Rdatasets \citep{arelbundock2026rdatasets} with at least 240
values. At the commit of the archive that we cite (3{,}648 data sets) these are 30 series; three have a window identical to that of another
candidate and two have missing values in the window, which leaves 25. For each, the last $\min(n,300)$ values, affine trend removed,
scaled to $\max|y|=0.8$, on the grid $x_k=-\pi+2\pi k/n$; least-squares fits of degree $E\in\{4,8,12,16,24,32\}$ on a random two thirds
of the points, held-out $R^2$ on the rest, median over five splits. A series passes if $R^2_{32}-R^2_{16}\ge0.02$, $R^2_{32}\ge R^2_{24}-0.005$,
$R^2_{32}\ge0.95$, and the degree-32 fit has supremum at most 1, so that it lies in the serial and ancilla model classes; circuits are run on every
passing series (at most three, by the gain $R^2_{32}-R^2_{16}$). The threshold $0.95$ was set after seeing the temperature series, which was part of
the design before the screen; no other series influenced the criteria. Table~\ref{tab:screen} lists all screened series. Three series have
$R^2_{32}\ge0.95$: the two that pass and \texttt{boot/wool}, which fails on the gain ($0.017$). The temperatures fail on two criteria
($R^2_{32}=0.868$, below $0.95$ and below $R^2_{24}-0.005=0.876$).

\paragraph{Measures (data).} The reference is the least-squares fit over real trigonometric polynomials of degree $\le E$ on the
training points, with training loss $L_{\rm LS}$. For every target its supremum is below 1, so it is the best fit within the serial and
ancilla model classes. Excess training loss:
$(L_{\rm train}-L_{\rm LS})/\mathrm{Var}(y_{\rm train})$, on a $\log_{10}$ scale floored at $-16$. A run \emph{reaches the optimum} if
its excess training loss is at most $10^{-6}$. This threshold was set after the step-budget runs, in place of the $10^{-15}$ used at
first: the level at which converged runs end depends on the optimiser and its schedule (Appendix~\ref{app:additional}), and a threshold
near double precision counts some converged runs as failures. Shares with Wilson intervals, differences with Newcombe's interval. Test
$R^2$ on the held-out points. Held-out error: $1-R^2$, compared as the ratio serial over ancilla for runs with the same run index (the
same target; for real-world series the same data, so that the pairing is arbitrary there and the share of pairs in which the serial
run is worse estimates the probability that a serial run is worse than an independent ancilla run).

\paragraph{Learning-rate bracketing (data).} This check was added after the main experiment; its rule was fixed before the additional
runs and is the same for every circuit. It tests whether the selected rate is limited by the edge of the grid. If the near-best rates
include the smallest rate of the grid, the next two smaller rates of the sequence $10^{-k}$, $3\cdot10^{-k}$ are added (below $10^{-3}$:
$3\cdot10^{-4}$ and $10^{-4}$); if they include the largest, the next larger rate, $0.3$, is added. The rate is then selected again on the same
tuning runs, and if the selection changes, the 100 evaluation runs are repeated with the new rate, from the same targets and initial
gates. A configuration whose best tuning score is at the floor of $-16$ is not extended, since no rate can score better.

\paragraph{Step budget (data).} This check, too, was added after the main experiment. It tests whether the shortfall of the serial
circuit is a matter of the step budget. It covers seven configurations, serial and ancilla circuits on flat and Poisson-kernel targets at
$E\in\{12,24\}$ and on the three real-world series at $E=24$, and the first 30 evaluation runs of each, with the same data and initial
gates as in the main experiment. Each run takes 80{,}000 steps instead of 20{,}000, at the peak rate selected in the main experiment,
which is not tuned again. Under the \emph{held} schedule, which Section~\ref{sec:data} reports, the rate stays at its peak for 60{,}000
steps and then follows the cosine decay of the main experiment over the last 20{,}000, so that only the time at the peak rate grows.
Under the \emph{stretched} schedule the same decay is spread over all 80{,}000 steps.

\paragraph{Seeds.} Every target and every initialisation has its own seed. The seed of a target depends on the target family, $E$ and the
run index; the seed of an initialisation depends on these and on the circuit. Neither depends on the optimiser, so that all circuits are
trained on the same targets and all optimisers start from the same initial gates. Tuning uses run indices that do not occur in
evaluation, so that no tuning target or initialisation is evaluated.

\paragraph{Uniform rotation angles.} The common practical initialisation draws the three angles of a \texttt{qml.Rot} gate uniformly, which
is not uniform on $\mathrm{SU}(2)$. In NB21 ($E=24$, flat targets, gradient descent) it behaves like the Haar start: 4 of 40 runs converge
within 25{,}000 steps and 21 of 40 within 200{,}000, against 3 and 45 of 80 from Haar starts; the per-run prediction by the private
fraction holds for both (AUC 0.89 and 0.78). Under the protocol of the starts table, on the targets and seeds of its Haar row, see Appendix~\ref{app:nb35}.

\subsection{Experimental protocols of the mechanism notebooks}
\label{app:nbprot}
All runs: serial circuit (and the ancilla where stated), $E=24$, exact loss on $2E+1$ points, gradient descent with Armijo steps in
Lie coordinates (first trial twice the last accepted step, halving), float64, converged at $L\le10^{-6}$.
\begin{center}\small\setlength{\tabcolsep}{3pt}
\begin{tabular}{llll}
\toprule
notebook & question & runs & status \\
\midrule
NB21 & two phases; private fraction predicts & 80+40+\dots & pre-registered, fresh seeds \\
NB22--24 & starts and recipe & 30 per arm & exploratory (shared seeds) \\
NB26 & Hessian at stall points; kicks & 60 & pre-registered, fresh seeds \\
NB27 & target vs start; scaled top coefficient & 144+192+192 & pre-registered, fresh seeds \\
NB28 & QSP oracle; low-pass start; basins & 30 per arm & pre-registered; oracle on NB21/22 targets, basins fresh \\
NB29 & sample density; spare depth & 30 per arm & pre-registered, fresh seeds \\
NB30 & designed end gates & 30 per arm & pre-registered; NB22 seeds (paired) \\
NB31 & optimisers, precision, boosts, fresh seeds, steps & App.~\ref{app:nb31} & pre-registered \\
\bottomrule
\end{tabular}
\end{center}
Pre-registered: predictions written into the notebook before the run. Fresh seeds: targets and starts not used in any earlier notebook;
NB22 seeds: the same targets and starts as NB22, so that each arm is paired with an NB22 arm; the oracle of NB28 needs no training.
NB31--NB42: Appendix~\ref{app:valid}.

\subsection{Baselines for the per-run prediction}
\label{app:baselines}
Runs not converged at the snapshot; outcome: converges by 200k steps or not. Cross-validated AUC (logistic regression, 5 folds, 20
repeats), NB21 serial Haar and uniform pooled.
\begin{center}\small
\begin{tabular}{lcc}
\toprule
predictor (log scale) & 25k ($n=113$) & 50k ($n=91$) \\
\midrule
current loss & 0.41 & 0.48 \\
gradient norm & 0.60 & 0.77 \\
loss slope over the last 5k steps & 0.58 & 0.72 \\
$\kappa_J$ & 0.77 & 0.82 \\
gradient norm, slope, $\kappa_J$, stiffness & 0.77 & 0.90 \\
private pull $\|q\|$ & 0.78 & 0.83 \\
private fraction $\pf$ & 0.80 & 0.84 \\
no-flip product $\Pi$ & 0.57 & 0.67 \\
all baselines $+\,\pf$ & 0.79 & 0.90 \\
$\pf$ in Euler-angle coordinates & 0.80 & 0.83 \\
\bottomrule
\end{tabular}
\end{center}
Stiffness is $\lambda_{\max}(H)/|\lambda_{\min}(H)|$. Rank correlations at 25k: $\pf$ with $\sigma_{\min}(J)$ and $\kappa_J$ 0.98--0.99; with the
gradient norm 0.00. The smallest left singular vector of $J$ has a median weight of 0.86 on the two top rows. $\Pi$ is recomputed from the gate snapshots of NB21 with the same procedure (which reproduces $0.81$ and $0.84$ for $\pf$ and
$0.77$ and $0.82$ for $\kappa_J$); its rank correlation with $\pf$ is $0.73$ at 25k and $0.77$ at 50k. The current loss is computed in the same way. With folds grouped by run index,
so that a Haar and a uniform-angle run with the same index (and possibly the same target) never fall in different folds, the AUC is
$0.80$ and $0.83$ for $\pf$, $0.76$ and $0.80$ for $\kappa_J$, $0.49$ and $0.58$ for $\Pi$, and $0.37$ and $0.42$ for the current loss.
Separately, $\pf$ gives $0.77$ and $0.81$ on the Haar runs and $0.86$ and $0.92$ on the uniform-angle runs.

\section{Validation and confirmatory experiments}
\label{app:valid}
All notebooks in this appendix were run with predictions written into the notebook before the run and on seeds not used before
(offsets in the repository). Serial circuit, exact loss on $2E+1$ points relative to the target variance, converged at $L\le10^{-6}$, gradient
descent with Armijo steps in Lie coordinates, float64, unless stated.

\subsection{Theory checks (NB33, T1--T6)}
\label{app:tchecks}
T1--T6 test the statements of Section~\ref{sec:branch} and Remark~\ref{rem:clt} with predictions fixed before the run; their results are
collected, with those of every other statement, in Table~\ref{tab:status}. T1: $-\ln\Pi\sim\mathrm{Gamma}(E-1,1)$ at 2{,}000 Haar starts per
$E$. T2: the equalities $\Pi=|a_0|^2+|a_E|^2+|b_0|^2+|b_E|^2$ and $\Pi=\gamma(t_a+t_a^{-1}+t_b+t_b^{-1})$ of Proposition~\ref{thm:identity} at 50 digits, gates made exactly unitary, at 40 Haar starts and 40 stalled points.
T3: on 10 flat targets at $E=24$ with 200 random branches each (independent fair signs, 60-digit oracle), the sign form
$\log t_a=\sum_j\varepsilon^a_j\lambda^a_j$ of Proposition~\ref{thm:identity} and its analogue for $t_b$ (T3a), the branch mean of $\Pi$
(Appendix~\ref{app:branch}; T3b) and the floor $\Pi\ge2|c_E|$ of Corollary~\ref{cor:random} (T3c); on 20 targets per $E\in\{8,16,24,32\}$ and family (flat, $1/k$), Corollary~\ref{cor:extreme}: the value $\Pi_{\max}$ on the
minimum-phase branch (T3d), its equality on all four extreme combinations (T3e, first five targets per cell), and the two bounds of part (ii)
on the total flip (T3f). T4: branch counts every 100 steps along 40 gradient-descent paths, with $\min f$ and $\max f$ along every step. T5:
Lemma~\ref{lem:private} at the 900 points of S1. T6: Lemma~\ref{lem:pull} at the same 900 points and every 250 steps along two
gradient-descent runs of 20k steps (one Haar, one no-flip start). The gradient is computed by five-point finite differences of the loss
($h=10^{-4}$), without the Jacobian, and compared with $(2/V)J^\top r$; $q$ is compared with the projection of that gradient, built
independently; and the rate $\dot L$ along $-\nabla L$, also by finite differences, is compared with $-\|\nabla L\|^2$ and $-\|q\|^2$. At the 553
points and 150 checkpoints with $\|\nabla L\|\ge10^{-4}$, all three agree to $7\times10^{-7}\|\nabla L\|$ and $-\dot L-\|q\|^2\ge0.13\,\|\nabla L\|^2$. Below
$\|\nabla L\|=10^{-4}$, finite differences lose resolution; over all 900 points the gradient and $q$ still agree to $2\times10^{-6}\|\nabla L\|$, the rate to $3\times10^{-5}\|\nabla L\|^2$, and $-\dot L>\|q\|^2$ holds everywhere.
Where $\|q\|\ge10^{-3}\|\nabla L\|$ (231 points), $q$ matches the projected gradient to a median relative error of $4\times10^{-11}$; at Haar and all-flip
points $\|q\|$ is often below $10^{-10}\|\nabla L\|$, where the check is not informative.
\emph{Results.} T1: mean of $-\ln\Pi$ $7.05$, $14.9$, $23.0$, $30.9$ against $E-1$ at $E=8,16,24,32$; KS $p=0.73$, $0.31$, $0.64$, $0.32$. T2:
maximum relative error $2.1\times10^{-47}$ over the 80 points (median $\Pi$ at the Haar starts $3\times10^{-11}$). T3a: 2{,}000 of 2{,}000
branches, maximum deviation $3.0\times10^{-14}$. T3b: 10 of 10 functions within 3 standard errors (largest $|z|=1.9$). T3c: 2{,}000 of 2{,}000,
median $\Pi/(2|c_E|)=1.26$. T3d: maximum relative error $8.3\times10^{-16}$ over 160 targets; T3e: $5.1\times10^{-16}$ over 40; T3f: 160 of 160.
T4: 14 changes of $k_a$ and 18 of $k_b$, all in windows with $\min f\le-0.9994$ or $\max f\ge0.9997$ respectively; 200 of 214 windows that come
within $10^{-2}$ of $-1$, and 190 of 208 that come within $10^{-2}$ of $+1$, have no change; $(k_a,k_b)$ at 12.5k steps equals that at 25k in
$95\%$ of 38 unconverged runs. T5: 900 of 900. In float64, the no-flip identity of T2 fails at Haar starts by up to a factor 46.
The float64 failure of the identity at Haar starts is a resolution effect ($|c_E|\approx10^{-10}$--$10^{-15}$ is computed with an absolute
error near $10^{-16}$), not a violation; it is why the check is done at 50 digits. The grid of $32(2E+1)$ points in the branch-count check misses
an extremum of $f$ by at most $E^2h^2/8\approx1.2\times10^{-3}$ (Bernstein), below the threshold $10^{-2}$.

\subsection{Checks of running-text statements (NB37, C1--C6)}
\label{app:nb37}
Predictions fixed before the run; $J$ by the adjoint method (checked against central differences to $4\times10^{-10}$), in the Lie-algebra
coordinates of the gates on the $2E+1$ grid points. C1 ($H=X$, $M=Z$, $|0\rangle$; Section~\ref{sec:structure}): for 20 random generators and
observables with eigenvalues $\pm1$ and random unit states at $E=8$, the model with encoding $e^{-ixH/2}$ equals the reference model at the
re-framed gates to $1.1\times10^{-14}$ (prediction $\le10^{-12}$). C2 (hypotheses of Theorem~\ref{thm:rigid} at Haar gates): at $E=8$, $J$ has full
rank at 1000 of 1000 Haar starts in $\mathrm{SU}(2)$ at 100 digits (smallest $\sigma_{\min}/\sigma_{\max}=10^{-15.0}$) and the input state is off
the encoding axis at all of them ($\min(1-\langle X\rangle^2)=8.9\times10^{-4}$); at $E=24$, at 20 of 20 Haar starts at 120 digits, confirmed at
160 digits to relative $4\times10^{-57}$, with $\sigma_{\min}/\sigma_{\max}$ from $10^{-33.3}$ to $10^{-11.1}$, below float64 resolution at most
starts. C3 (same branch, same $JJ^\top$; Section~\ref{sec:branch}): for 20 targets at $E=8$, two realisations with the same roots of $a$ and $b$
but a different relative phase of $b$ (gates at distance $\ge0.39$ beyond a phase per gate) have the same $JJ^\top$ to relative
$3.3\times10^{-15}$, while a realisation with one root of $a$ flipped differs by at least $0.22$; all three reproduce $f$ to
$3.5\times10^{-14}$. C4 (several gates per block): with five Haar gates per block at $E=8$ and $24$ (20 draws each),
$J_{\rm gates}J_{\rm gates}^\top=5\,J_{\rm blocks}J_{\rm blocks}^\top$ to relative $4.0\times10^{-15}$. C5 (end-gate form
$|c_E|=\frac12\Pi\sin\alpha\sin\beta$; Section~\ref{sec:branch}): at 50 digits with gates made exactly unitary and $|c_E|$ from the $2E+1$
grid, the form holds to relative $2.6\times10^{-37}$ at 40 Haar starts each at $E=8$ and $24$ and at 20 starts near the no-flip point
($E=24$, flip amplitude $0.1$); in float64 the error is $10^{-14}$ near the no-flip point and up to $9\times10^{-10}$ at $E=8$, but up to
$9\times10^{-2}$ at Haar starts at $E=24$, where $|c_E|\approx4\times10^{-11}$. C6 ($\Pi$ does not determine $\pf$; Section~\ref{sec:branch}): at $E=24$, with Haar end gates and middle gates of equal flip
probability $q$ and uniform random phases (40 draws per $q$), $\pf$ has median $0.99$ and minimum $0.93$ at $q=0$, minimum $0.17$ at $q=10^{-6}$
and $0.08$ at $q=10^{-4}$ ($\Pi=0.998$), and median $0.12$ at $\Pi\approx0.8$ and $0.5$; minimising $\sigma_{\min}(J)$ at $E=8$ over gates with
$\Pi$ fixed and $\sin\alpha,\sin\beta\ge0.9$ gives $\sigma_{\min}/\sigma_{\max}\le1.4\times10^{-15}$ at $\Pi=0.99$, $0.9$ and $0.5$, with
$\pf=0.77$ at the singular point for $\Pi=0.99$.

\subsection{Starts (NB33, S1)}
\label{app:starts}
\begin{figure*}[t]
  \centering
  \includegraphics[width=\textwidth]{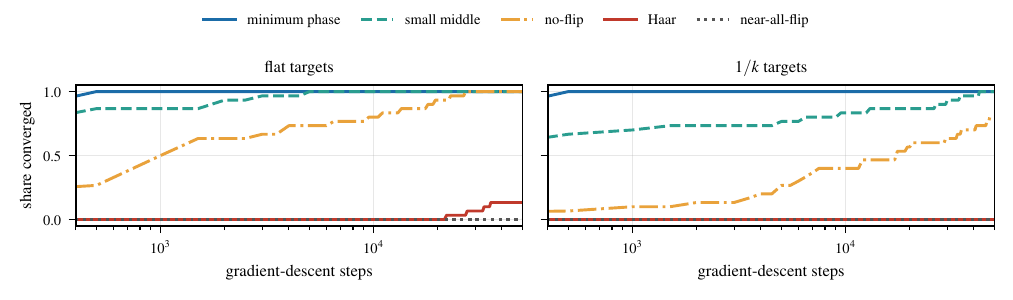}
  \caption{Share of runs converged against gradient-descent steps for the starts of Section~\ref{sec:cures} and the near-all-flip start
  ($E=24$, 30 runs each). Starts on or near an extreme QSP branch converge almost at once; the near-all-flip start never does.}
  \label{fig:starts}
\end{figure*}
$E=24$, 30 targets per family (flat and $1/k$, $\sup|f^*|=0.9$), the same 30 start seeds in every arm, 50k steps. No-flip: middle gates diagonal
in the encoding frame up to a flip amplitude $0.1$, random phases, Haar ends. Near-all-flip: the same with flip probability $0.99$.
Minimum-phase: the exact realisation (60-digit oracle, error $\le10^{-8}$) of an independent random flat function of amplitude $0.9$, not of the target.
Small middle angles: \texttt{Rot} angles $N(0,\kappa/E)$ with $\kappa=0.03$ on middle gates, Haar ends. Branch counts are the numbers of zeros
of $a$ and $b$ in the open disc (median over runs).
\begin{center}\small\setlength{\tabcolsep}{3pt}
\begin{tabular}{lcccccccc}
\toprule
 & \multicolumn{2}{c}{converged /30} & \multicolumn{2}{c}{median steps} & \multicolumn{3}{c}{at the start} & final $\Pi$ \\
start & flat & $1/k$ & flat & $1/k$ & $\Pi$ & $\mathrm{rank}\,J$ (float64) & $(k_a,k_b)$ & (conv.) \\
\midrule
Haar & 4 & 0 & 30{,}250 & -- & $3\times10^{-10}$ & 44 & (12,12) & 0.13 \\
near-all-flip & 0 & 0 & -- & -- & $10^{-46}$ & 15 & (12,12) & -- \\
no-flip & 30 & 24 & 1{,}250 & 9{,}750 & 0.79 & 49 & (16,7) & 0.46 / 0.31 \\
small middle, $\kappa=0.03$ & 30 & 30 & $\le$500 & $\le$500 & 0.98 & 49 & (20,0) & 0.92 / 0.83 \\
minimum-phase & 30 & 30 & $\le$500 & $\le$500 & 0.93 & 49 & (24,24) & 0.93 / 0.87 \\
\bottomrule
\end{tabular}
\end{center}
Predictions: S1a ($\kappa\ge27$, no-flip and minimum-phase $\ge25$ per family) fails by one run for no-flip on $1/k$; S1b (Haar $\le12$), S1c
(near-all-flip no better than Haar, strictly worse pooled) and S1d (near-all-flip median rank $\le E+1$; the other structured starts full rank)
hold. S1e (converged runs from no-flip, $\kappa$ and minimum-phase starts end with $\Pi\ge0.5$ in $\ge90\%$) was not testable as worded (four
Haar runs converged, five required) and its first half fails: $72\%$ pooled, because the no-flip arm ends at $\Pi\approx0.3$--$0.5$. Minimum-phase
runs never change their branch counts; no-flip runs change them in 60--73\% of runs; Haar runs move by $\pm1$--$2$ within the first
$\sim$100 steps and stay in $[10,14]$ (frozen after 25k steps in $97\%$).

\subsection{The branch picture across $E$ (NB32 and NB33 D2)}
\label{app:branchE}
Flat targets, Haar starts; NB32 at $E=24$ (60 runs), D2 at $E=16,32$ (40 runs each). Points: gates of runs not converged at 25k and 50k steps
(28+17, 59+49 and 40+39 runs; a run that stalls throughout counts twice). At $E=24$ the realisations of one point (run 21 at 50k) miss the
$10^{-8}$ reconstruction check (error $7\times10^{-7}$); the rows on realisations and their restarts use the other 107 points.
Restarts: 5k GD steps from the run's gates, from the minimum-phase realisation of the current function, or from one random-branch realisation.
\begin{center}\small\setlength{\tabcolsep}{4pt}
\begin{tabular}{lccc}
\toprule
 & $E=16$ & $E=24$ & $E=32$ \\
\midrule
unconverged points & 45 & 108 & 79 \\
run's own $\pf$, median & $2.8\times10^{-3}$ & $5.9\times10^{-4}$ & $1.2\times10^{-6}$ \\
minimum-phase realisation $\pf\ge0.5$ & $100\%$ & $100\%$ & $100\%$ \\
restart converges: minimum phase & $100\%$ & $100\%$ & $100\%$ \\
\phantom{restart converges:} random branch & $38\%$ & $12\%$ & $4\%$ \\
\phantom{restart converges:} continue run & $13\%$ & $3\%$ & $1\%$ \\
\bottomrule
\end{tabular}
\end{center}
At $E=24$ and $32$ every prediction holds. At $E=16$ two fail: the run's median $\pf$ is above the predicted $10^{-3}$, and random-branch
restarts converge 25 points more often than continuing the run (predicted within 10). We read this as the stall weakening at small $E$,
not as a different mechanism; minimum-phase restarts work at every $E$. A pre-registered branch-reset arm (minimum-phase realisation of
the current function every 5k steps) converged in every run; at $E=16$ and $32$, 79 of 80 runs converged before the second reset, so it amounts to a
minimum-phase start and is not reported further.

\subsection{Optimisers, precision and boosts (NB31)}
\label{app:nb31}
80 Haar starts, $E=24$, flat targets. Converged runs: gradient descent 17, Adam 48, damped Gauss--Newton 72, LM 74, L-BFGS 75, natural gradient
(input-averaged Fubini--Study metric) 80. Median cost to converge, in forward-pass equivalents: L-BFGS $7.6\times10^3$; gradient descent
$2.4\times10^5$ on its 15 easiest runs. From the starts of the 35 NB21 runs in which gradient descent never converged: LM 31, Gauss--Newton 33,
L-BFGS 32, natural gradient 35, Adam 17. LM restarted from gradient descent's 25k points converges in all 42 late runs but only 23 of 35
never-runs (prediction $\ge80\%$, fails). Precision: extended precision reproduces double precision run by run (trajectories agree to
$10^{-13}$); on the 40 runs of the precision part, double and extended precision converge in 20 and single precision in 8. At 29 of 184 stall points the 50-digit private fraction is
$10^{-28}$--$10^{-14}$ while float64 reports about $10^{-11}$: values below $\sim10^{-10}$ are below float64 resolution and we report them as
such. Boosts from gradient descent's 25k points (10k steps): adding the private pull converges 23 runs, a random direction of equal norm 5,
nothing 4, a sham inside the lower span 13 (prediction: within $\pm5$ of nothing; fails), a top-band Gauss--Newton step 4; no arm rescues a
never-run.

\subsection{Cures on data (NB33, D1)}
\label{app:data}
Data, splits and scaling as in Appendix~\ref{app:experiments}: flat and bump targets (bump: the Poisson-kernel family of Appendix~\ref{app:experiments}, amplitude $0.8$) on a 300-point grid (200/100 split, 30 fresh targets),
Nottingham temperatures (180/60), \texttt{gasprice} and \texttt{ibmclose} (last 300 points, affine trend removed, 200/100). Serial circuit with five
gates per block, $E=24$ (375 parameters); ancilla with one gate per block. 30 runs per arm; on the real series the runs differ only in their
initial gates. lexc is $\log_{10}$ of the training loss's excess over the least-squares oracle, relative to the variance (floor $-16$);
``at the optimum'' means lexc $\le-6$. Budgets: gradient descent and L-BFGS 20k iterations, natural gradient 5k, Adam 20k (learning rate selected
on separate tuning targets or seeds by the rule of Appendix~\ref{app:experiments}; the grid and choices are in the repository). Minimum-phase gates and small-middle
starts go into the first gate of each block; the other four gates start at the identity. The minimum-phase start realises an independent
random function, never the data's fit.
\begin{center}\small\setlength{\tabcolsep}{3pt}
\begin{tabular}{lccccc}
\toprule
median lexc & flat & bump & nottem & gasprice & ibmclose \\
\midrule
Adam, Haar & $-16$ & $-2.7$ & $-3.3$ & $-2.7$ & $-2.7$ \\
GD, Haar & $-1.1$ & $-1.7$ & $-0.1$ & $-1.8$ & $-1.6$ \\
L-BFGS, Haar (50k) & $-14$ & $-13.7$ & $-3.3$ & $-14.4$ & $-14.3$ \\
natural gradient, Haar (20k) & $-14$ & $-4.4$ & $-13.8$ & $-13.2$ & $-13.1$ \\
GD, small middle & $-14$ & $-16$ & $-14.7$ & $-16$ & $-16$ \\
GD, minimum phase & $-14$ & $-16$ & $-14.8$ & $-16$ & $-16$ \\
Adam, small middle & $-16$ & $-16$ & $-15.8$ & $-16$ & $-16$ \\
Adam, minimum phase & $-16$ & $-16$ & $-15.9$ & $-16$ & $-16$ \\
ancilla Adam, Haar & $-16$ & $-5.7$ & $-15.8$ & $-16$ & $-16$ \\
\bottomrule
\end{tabular}
\end{center}
NB33 predictions: reproduction of the serial Adam baseline of Appendix~\ref{app:results} (median lexc in $[-4.5,-1.5]$ on each real series) and the ancilla
($\le-10$) holds; gradient descent from Haar starts at the optimum in $\le10/30$ holds (0 everywhere); small-middle and minimum-phase gradient
descent at the optimum in $\ge20/30$ on every data set hold (28--30), as does a branch-reset arm that we do not report further; L-BFGS and natural gradient at the optimum in
$\ge20/30$ on every data set fail at NB33's budgets (L-BFGS 9--28 at 20k, natural gradient 2--30 at 5k).

\paragraph{With logging and larger budgets (NB34 A).} Same data, splits, targets, start seeds and Adam rates as NB33; the loss is logged every
100 iterations; L-BFGS runs 50k and natural gradient 20k iterations. At NB33's budgets the counts reproduce NB33 in all 50 cells (A0). Natural
gradient at 20k: 30/30 on flat, Nottingham, gasprice and ibmclose, 9/30 on bump (from 4); pre-registered decision ``mixed'' (budget on three
data sets, not on bump). L-BFGS at 50k: Nottingham 10 (from 9), gasprice 21 (from 19): ``stall''. Gradient descent from a minimum-phase start
reaches the optimum in a median of 100--700 steps and with less work than L-BFGS on all five data sets (A2 holds; Table~\ref{tab:data}). Work counts every function evaluation, line searches included, plus 3 per gradient and $P=375$ per Jacobian. With five gates per block
natural gradient pays for five times as many Jacobian columns as the circuit of block products needs (Lemma~\ref{lem:stack}); at a fifth of
its work it would still need two orders of magnitude more than gradient descent from a minimum-phase start.
Adam's first visits to the optimum were kept to the end of its schedule in every run. Held-out $R^2$: every run at the optimum is within
$1.4\times10^{-5}$ of the oracle; Adam from Haar starts loses a median $0.0004$--$0.0034$ on the real series, gradient descent from Haar starts $0.02$--$1.2$. Relative to the
optimum's held-out error $1-R^2$ (one split; $0.100$, $0.041$ and $0.036$ for \texttt{nottem}, \texttt{gasprice}, \texttt{ibmclose}), Adam from
Haar starts has a median held-out error $0.4\%$, $8.3\%$ and $8.9\%$ higher (largest $1.8\%$, $15\%$, $17\%$; on \texttt{nottem} 13 of
its 30 runs are about $3.5\%$ below the optimum), gradient descent from Haar starts $1200\%$, $74\%$ and $62\%$.

\paragraph{Stacking under Adam (NB38).}\label{app:nb38} Lemma~\ref{lem:stack} makes the number of gates per block irrelevant for gradient
descent, not for Adam, so the Adam rows of Table~\ref{tab:data} change the branch and the stacking at once. Same data, splits, start seeds and
Adam rates as NB34 A (tuned for five gates per block, not retuned for one), 20k steps. Runs at the optimum out of 30 on flat, bump, \texttt{nottem}, \texttt{gasprice}, \texttt{ibmclose}: one Haar
gate in the first slot of each block and identities in the other four (stacked like the structured starts) 15, 1, 1, 1, 0; one gate per block,
Haar, 9, 1, 1, 0, 0; one gate per block, minimum phase, 30 on every data set (median 200--600 steps). For comparison (NB34 A), all five gates
Haar 16, 0, 2, 3, 1, and minimum phase with five gates per block 30 on every data set. All four predictions fixed before the run hold. Under
Adam, too, the outcome follows the start's branch, not the stacking.

\subsection{The no-flip start along its paths (NB34 B1)}
\label{app:noflip}
Starts as in S1 (same targets and seeds), 20k steps, metrics every 250 steps. Medians; $\Pi/(2|c_E|)=1/(\sin\alpha\sin\beta)$ at the first
converged snapshot (or the end).
\begin{center}\small\setlength{\tabcolsep}{3pt}
\begin{tabular}{lcccccc}
\toprule
 & \multicolumn{2}{c}{converged /30} & \multicolumn{2}{c}{$\Pi/(2|c_E|)$ at the end} & \multicolumn{2}{c}{$\min\pf$ along the path} \\
start & flat & $1/k$ & flat & $1/k$ & flat & $1/k$ \\
\midrule
Haar & 0 & 0 & 1.08 & 1.40 & $3\times10^{-10}$ & $4\times10^{-9}$ \\
no-flip & 28 & 17 & 6.3 & 21 & 0.26 & 0.13 \\
small middle & 30 & 26 & 10 & 42 & 0.90 & 0.94 \\
minimum phase & 30 & 30 & 13 & 57 & 0.94 & 0.95 \\
\bottomrule
\end{tabular}
\end{center}
All start with $\Pi/(2|c_E|)=1.54$ except minimum phase ($8.9$): the ratio depends on the end gates only, which the first three share.
Predictions: B1a (converged no-flip runs end with ratio $\ge2$ in $\ge90\%$) holds, $96\%$ of 45 (no Haar run converged within 20k, so its
half was not testable); B1b ($\pf\ge10^{-3}$ along converged no-flip paths in $\ge80\%$; Haar $\min\pf\le10^{-4}$ in $\ge90\%$) holds, $100\%$
and $98\%$; B1c (the ratio at 2,500 steps predicts which slow no-flip runs converge, AUC $\ge0.7$) fails, $0.62$ (22 vs 15 runs).

\subsection{Flip-amplitude sweep (NB34 B2)}
\label{app:sweep}
Middle gates diagonal in the encoding frame up to flip amplitude $\varepsilon$ (flip probability $\varepsilon^2$), random phases, Haar ends; S1 targets
and seeds, gradient descent 50k steps; $\varepsilon=0.1$ and $0.995$ are S1's no-flip and near-all-flip arms.
\begin{center}\small\setlength{\tabcolsep}{4pt}
\begin{tabular}{lccccccc}
\toprule
$\varepsilon$ & 0.1 & 0.3 & 0.5 & 0.7 & 0.9 & 0.995 & Haar \\
\midrule
flat & 30 & 25 & 19 & 11 & 3 & 0 & 4 \\
$1/k$ & 24 & 2 & 0 & 0 & 0 & 0 & 0 \\
\bottomrule
\end{tabular}
\end{center}
Predictions B2a (counts non-increasing in $\varepsilon$, Spearman $\le-0.8$: $-1.00$, $-0.85$) and B2b ($\varepsilon=0.9$ no better than Haar$+3$;
$\varepsilon=0.3$ $\ge25/30$ on flat) hold.

\subsection{Uniqueness of realisations (NB34 C)}
\label{app:unique}
At the 239 gradient-descent endpoints of NB34 B1 whose root pairs are separated
($\min_j\lambda_j\ge10^{-3}$), we computed $(a,b)$ exactly from the gates (50 digits) and stripped the layers: the middle-gate flip
probabilities agree with the run's to $4\times10^{-14}$ at all 239. The pre-registered form of the check realised the function from its float64
coefficients through root finding and the oracle; it agreed at 200 of 239 (prediction $\ge95\%$, failed). All 39 disagreements are stalled
Haar endpoints, and their size grows as $|c_E|$ falls ($|c_E|$ between $10^{-14}$ and $10^{-3}$): there a root pair of $(1\pm f)/2$ approaches
$0$ and $\infty$, and float64 coefficients do not determine its position. The disagreement is a conditioning property of the map from $f$ to
the gates, not a second realisation.

\subsection{Uniform angles, the band below the top, and corners (NB35, NB36)}
\label{app:nb35}
Predictions fixed before the run; 23 of 26 hold in NB35, and the three that fail there hold when recomputed at 80 digits (NB36,
below). Setup as NB33. Private fraction of a set $S$ of coefficient rows:
$\|P_\perp J_S^\top\|_F/\|J_S^\top\|_F$, with $P_\perp$ the projector off the span of all other rows; \emph{band privacy} of width $B$ is that of
the top $B$ frequencies against all lower ones (so $B=1$ is $\pf$).

\paragraph{Uniform rotation angles (U).} $E=24$, the 30 flat and 30 $1/k$ targets and start seeds of the Haar arm of NB33 S1; every gate a
\texttt{qml.Rot} with angles uniform in $[0,2\pi)$; the optimiser acts on the gate matrices, as for every start. At the start the median $\Pi$ is
$4.7\times10^{-11}$ and the median branch counts are $(12,12)$ (U2: $\Pi\le10^{-6}$, counts in $[9,15]$; holds). Converged: 9 of 30 (flat) and 0
of 30 ($1/k$) (U1: $\le12$; holds), against 4 and 0 from Haar starts (U3: within 6; holds). The converged runs need a median of 25{,}000 steps
(Haar: 30{,}250) and end at median $\Pi=0.11$ (Haar: 0.13). Branch counts stay at $(12,12)$ in the median from start to end, in both families.

\paragraph{The band below the top (B).} Fresh seeds; $E=16,24,32$; 30 flat and 30 $1/k$ targets with Haar starts each; gradient descent for
50k steps. Points: runs unconverged at 25k and 50k steps, at most 40 per cell (36 at $E=16$, flat). Medians over points; at $E=24$ and $32$
every measure except the residual share is the 80-digit value of NB36 (at $E=16$ float64):
\begin{center}\small\setlength{\tabcolsep}{3pt}
\begin{tabular}{lcccccc}
\toprule
 & \multicolumn{2}{c}{$E=16$} & \multicolumn{2}{c}{$E=24$} & \multicolumn{2}{c}{$E=32$} \\
 & flat & $1/k$ & flat & $1/k$ & flat & $1/k$ \\
\midrule
residual share, top 4 frequencies & 1.00 & 0.96 & 0.98 & 0.48 & 0.90 & 0.25 \\
$\pf$ of $E$ & $4.5{\times}10^{-3}$ & $8.8{\times}10^{-4}$ & $1.6{\times}10^{-4}$ & $1.9{\times}10^{-9}$ & $3.6{\times}10^{-7}$ & $8.5{\times}10^{-16}$ \\
$\pf$ of $E-1$ & $4.1{\times}10^{-3}$ & $5.4{\times}10^{-4}$ & $9.8{\times}10^{-5}$ & $9.3{\times}10^{-10}$ & $3.1{\times}10^{-7}$ & $4.7{\times}10^{-16}$ \\
$\pf$ of $E-2$ & $9.7{\times}10^{-3}$ & $4.2{\times}10^{-4}$ & $1.5{\times}10^{-4}$ & $1.7{\times}10^{-9}$ & $2.3{\times}10^{-7}$ & $2.7{\times}10^{-16}$ \\
$\pf$ of $E-3$ & $1.2{\times}10^{-2}$ & $5.8{\times}10^{-4}$ & $2.2{\times}10^{-4}$ & $2.1{\times}10^{-9}$ & $3.9{\times}10^{-7}$ & $4.2{\times}10^{-16}$ \\
$\pf$ of $E/2$ & 0.17 & $1.8{\times}10^{-2}$ & $1.7{\times}10^{-2}$ & $1.0{\times}10^{-4}$ & $4.7{\times}10^{-3}$ & $8.4{\times}10^{-10}$ \\
band privacy, $B=2$ & 0.22 & $5.7{\times}10^{-2}$ & $2.3{\times}10^{-2}$ & $5.8{\times}10^{-5}$ & $3.6{\times}10^{-4}$ & $6.1{\times}10^{-11}$ \\
band privacy, $B=4$ & 0.48 & 0.28 & 0.27 & $2.6{\times}10^{-2}$ & $8.1{\times}10^{-2}$ & $3.3{\times}10^{-5}$ \\
band privacy, $B=8$ & 0.80 & 0.69 & 0.60 & 0.13 & 0.38 & $2.2{\times}10^{-2}$ \\
smallest $B$ with band privacy $\ge0.1$ & 2 & 3 & 3 & 8 & 5 & $>8$ \\
two smallest $\sigma/\sigma_{\max}$ (order) & $10^{-4}$ & $10^{-6}$ & $8{\times}10^{-7}$ & $3{\times}10^{-15}$ & $10^{-10}$ & $8{\times}10^{-26}$ \\
third smallest $\sigma/\sigma_{\max}$ & $2{\times}10^{-2}$ & $4{\times}10^{-4}$ & $1.1{\times}10^{-3}$ & $4{\times}10^{-10}$ & $3{\times}10^{-7}$ & $2{\times}10^{-19}$ \\
minimum phase: $\pf\ge0.5$ for $E\dots E-3$ & 36/36 & 40/40 & 40/40 & 40/40 & 40/40 & 40/40 \\
\bottomrule
\end{tabular}
\end{center}
B1 (flat, $E=24,32$: top-4 share $\ge0.6$) holds. B2 (median $\pf$ of $E-1$ and $E-2$ $\le10^{-3}$, and band privacy non-decreasing in
$B=1,2,4,8$ at $\ge90\%$ of points) and B3 (two smallest singular values $\le10^{-6}\sigma_{\max}$, the third ten times larger) hold in float64 on
flat targets. On $1/k$ targets the monotonicity part of B2 fails in float64 at $E=24$ (35 of 40, 36 needed) and $E=32$ (4 of 40), and B3 at
$E=32$, where the smallest singular values sit at float64 resolution.

\paragraph{The band at 80 digits (NB36).} Every measure above recomputed exactly at the stored points of $E=24$ and $32$: gates made exactly
unitary (they drift from unitarity during training by up to $2\times10^{-13}$), the Gram matrix $JJ^\top$ and all projections in mpmath at 80
digits. Predictions fixed before the run, all hold. P0: 80 and 120 digits agree to $3\times10^{-15}$. P1 (B3 at $E=32$, $1/k$): the two smallest
singular values are $7\times10^{-26}$ and $9\times10^{-26}\,\sigma_{\max}$, the third $2\times10^{-19}$. P2 (monotonicity at $1/k$): 37 of 40 points at
both $E=24$ and $E=32$. P3 (control): wherever float64 resolves the second-smallest singular value ($\ge10^{-12}\sigma_{\max}$), float64 and 80
digits agree within a factor 2 in all 912 comparisons. P4: perturbing every gate at the size of the unitarity drift changes band privacy at
$B=1,2$ by a median factor $1+4\times10^{-12}$; at 38 of 40 points by less than a factor 2. So the NB35 failures were resolution: in float64 the
private fractions near the top at $E=32$, $1/k$, read $10^{-6}$--$10^{-9}$, while their exact values are $\approx10^{-16}$; the stall there is deeper than
float64 can show, not shallower.
B4 (minimum-phase realisations private for each of $E\dots E-3$) holds in every cell, with medians $0.91$--$0.97$
and no realisation dropped.

\paragraph{Corners in several dimensions (C).} $d=2$ with $E=2,3,4$ and $d=3$ with $E=2,3$; encoding order alternating ($x_1x_2x_1x_2\dots$) or
blocked ($x_1\dots x_1x_2\dots x_2\dots$); middle gates Haar or near no-flip with flip amplitude $\varepsilon=0.05$ or $0.3$, Haar end gates;
20 circuits per cell; every sign pattern $s$ with $s_1=+1$, corner $(s_1E,\dots,s_dE)$. C1: the corner coefficient equals the top coefficient of
the one-dimensional circuit with $x_i=s_ix$ (max deviation $7.9\times10^{-16}$ over 1{,}680 corners; holds). C2:
$|c_k|=\frac12\Pi_s\sin\alpha\sin\beta$, where $\Pi_s$ takes $1-q_l$ for a middle gate between encoding gates of equal sign in $s$ and $q_l$ for
one between opposite signs (max deviation $7.1\times10^{-15}$; holds). C3 ($d=2$, $\varepsilon=0.05$): median $|c_k|$ by order and corner
\begin{center}\small\setlength{\tabcolsep}{4pt}
\begin{tabular}{lcccc}
\toprule
 & \multicolumn{2}{c}{$E=3$} & \multicolumn{2}{c}{$E=4$} \\
 & $(E,E)$ & $(E,-E)$ & $(E,E)$ & $(E,-E)$ \\
\midrule
alternating & 0.28 & $2.7\times10^{-14}$ & 0.40 & $2.1\times10^{-16}$ \\
blocked & 0.31 & $7.9\times10^{-4}$ & 0.42 & $1.0\times10^{-3}$ \\
Haar middle gates, alternating & $2.4\times10^{-3}$ & $8.3\times10^{-3}$ & $3.6\times10^{-4}$ & $2.1\times10^{-4}$ \\
\bottomrule
\end{tabular}
\end{center}
(prediction: alternating $(E,-E)\le10^{-6}$ and $(E,E)\ge0.05$, blocked $(E,-E)\ge10^{-5}$; holds). From Haar middle gates every corner is
small, with no preference between them.

\subsection{Numerical companions to Appendix~\ref{app:proofs}}
\label{app:numcomp}
\paragraph{Rigidity (NB32 R5, $E=24$).} Along a non-symmetry fibre walk the ancilla's tangent kernel $JJ^\top$ changes by $5.8\times10^{-2}$
(relative, median point; the ancilla fibre has dimension $13E+14=326$, of which at least $6E-2=142$ tangent directions are not symmetry directions); along the serial fibre by $1.1\times10^{-14}$ (prediction: serial $\le10^{-6}$, ancilla
$\ge10^{-2}$; holds).

\paragraph{No-flip point.} With Haar end gates the private fraction at the no-flip point is $0.99$ at $E=24$; the remainder is the shared
end-gate coordinates of $c_0$. The frequency bookkeeping of Proposition~\ref{prop:noflip} is checked to $10^{-12}$ at $E=10$ and $24$. Ends
with $\alpha=\beta\neq90^\circ$ are not harmful in practice, only the identity limit is (NB30).

\paragraph{End-gate form.} $|c_E|=\frac12\Pi\sin\alpha\sin\beta$ holds to relative $3\times10^{-37}$ at 50 digits (C5,
Appendix~\ref{app:nb37}); in float64 it cannot be resolved at Haar starts for $E=24$, as the identity of T2.

\paragraph{From the no-flip product to the private fraction.} At $\Pi=1$ (no-flip point) $\pf=0.99$ with Haar end gates
(Proposition~\ref{prop:noflip}), and $\pf$ is continuous. Away from $\Pi\approx1$, $\Pi$ alone gives no bound. At the NB32 points, every
realisation with $\Pi\ge0.5$ has $\pf\ge0.5$ ($107/107$); of those with $\Pi\le10^{-2}$, 53 of 59 have $\pf\le10^{-2}$ (the six exceptions
have $\pf=0.01$--$0.13$), so the pre-registered prediction R8 of NB32 ($\ge95\%$ in both) failed. What bounds $\pf$ is
Proposition~\ref{prop:readout}, with Theorem~\ref{thm:privacy-extreme} and Proposition~\ref{prop:gate-control} for its two factors; their
checks are in Appendices~\ref{app:readoutexp}--\ref{app:d14}.

\subsection{The readout factorisation (NB32 points, NB39)}
\label{app:readoutexp}
All quantities from exact derivatives in float64: $\pf$ (Definition~\ref{def:private}, coordinates of Section~\ref{sec:structure}),
$\pf_R$, $\kappa_D$ (Proposition~\ref{prop:readout}), and $\pf_S$, $\kappa_{D_S}$ in the coordinates of Theorem~\ref{thm:nnf}.
Predictions written before each run, except Part D.

\paragraph{At the points of NB32 ($E=24$; no retraining).}
The 322 points of R8: the run's gates at the unconverged 25k and 50k snapshots (108), their minimum-phase (107) and first random-branch
(107) realisations. S0: $\pf$ recomputed matches NB32's stored value to a median relative $2\times10^{-14}$ (holds; the largest
differences, up to $0.6$, are at four points with $\pf=10^{-12}$--$10^{-9}$, at float64 resolution). S1: $\|J-RD\|/\|J\|\le1.3\times10^{-12}$
and the bounds of Proposition~\ref{prop:readout} hold at all 322 points (holds). P1: minimum-phase $\pf_R\ge0.5$ at $\ge95\%$ (holds: all
107; minimum $0.69$, median $0.95$). P2: median $\pf_R$ at the run's gates $\le10^{-2}$ (fails: $1.3\times10^{-2}$, against median
$\pf=5.9\times10^{-4}$). P3: $\pf_R$ ranks $\pf$ better than $\Pi$ (holds: Spearman $0.98$ against $0.88$; $R^2$ of $\log\pf$ $0.88$ against
$0.63$). P4: in the middle range $10^{-2}\le\Pi<0.5$ ($n=156$) Spearman with $\pf_R\ge0.8$ (holds: $0.94$; with $\Pi$: $0.46$).
$\pf_R\le10^{-2}$ implies $\pf\le10^{-2}$ at 87 of 87 points ($\Pi\le10^{-2}$: 53 of 59, R8); the six exceptions of R8 have
$\pf_R=0.015$--$0.075$. At the minimum-phase realisations $\kappa_D=0.179$--$0.194$, $\kappa_D\pf_R\ge0.12$, and $\pf$ lies within a factor
$1.14$ of $\pf_R$. At the run's gates the medians are $\log_{10}\pf=-3.23$, $\log_{10}\pf_R=-1.88$, $\log_{10}\kappa_D=-2.46$; a fit of
$\log\pf$ on $\log\pf_R$ and $\log\kappa_D$ has coefficients $0.94$ and $0.52$ ($R^2=0.90$).

\paragraph{NB39.} Fresh seeds. Minimum-phase realisations from the 60-digit oracle of NB32, accepted if they reproduce the target on the
grid to $10^{-8}$ (all 330 accepted; largest error $1.5\times10^{-14}$). Q0: $\|J-RD\|/\|J\|\le10^{-10}$ and the bounds of
Proposition~\ref{prop:readout} in both coordinate systems at every point (holds: 492 of 492 points, residual $\le1.5\times10^{-13}$).

\emph{Part A: minimum-phase realisations, flat and $1/k$ targets, amplitude $0.9$, 30 per cell} (medians; minima in brackets, rounded down):
\begin{center}\small
\begin{tabular}{llcccccc}
\toprule
family & $E$ & $\Pi$ & $\pf$ & $\pf_R$ & $\kappa_D\sqrt{E+1}$ & $\kappa_{D_S}$ & $\kappa_{D_S}\pf_R$\\
\midrule
flat & 8  & 0.89 & 0.89 & 0.89 (0.87) & 0.95 & 0.84 (0.76) & 0.75 (0.68)\\
flat & 16 & 0.92 & 0.92 & 0.93 (0.90) & 0.95 & 0.84 (0.79) & 0.78 (0.73)\\
flat & 24 & 0.93 & 0.94 & 0.94 (0.92) & 0.95 & 0.84 (0.81) & 0.80 (0.75)\\
flat & 32 & 0.93 & 0.93 & 0.95 (0.90) & 0.94 & 0.85 (0.79) & 0.81 (0.73)\\
$1/k$ & 8  & 0.83 & 0.92 & 0.92 (0.88) & 0.93 & 0.80 (0.75) & 0.73 (0.67)\\
$1/k$ & 16 & 0.86 & 0.94 & 0.94 (0.92) & 0.93 & 0.82 (0.72) & 0.77 (0.67)\\
$1/k$ & 24 & 0.87 & 0.95 & 0.95 (0.93) & 0.93 & 0.84 (0.73) & 0.79 (0.69)\\
$1/k$ & 32 & 0.88 & 0.96 & 0.96 (0.94) & 0.93 & 0.84 (0.77) & 0.80 (0.73)\\
\bottomrule
\end{tabular}
\end{center}
Q1: $\pf_R\ge0.5$ at $\ge95\%$ in every cell, medians within $0.15$ across $E$ per family (holds: all 240; spread $0.06$). Q2: median
$\kappa_D\sqrt{E+1}\in[0.8,1.2]$ in every cell, median $\kappa_{D_S}\ge0.3$, and $\kappa_{D_S}$ at $E=32$ at least half that at $E=8$
(holds). $\pf$ lies within a factor $1.11$ of $\pf_R$, and $\pf_S$ within $1.05$.

\emph{Part B: dependence on $\|f\|_\infty$} ($E=24$, flat, 30 per amplitude; medians, minima in brackets, rounded down):
\begin{center}\small
\begin{tabular}{lcccccc}
\toprule
amplitude & $\sup|f|$ on the grid & $\Pi$ & $\pf_R$ & $\kappa_{D_S}$ & $\kappa_{D_S}\pf_R$\\
\midrule
0.5  & 0.49 & 0.98 & 0.998 (0.99) & 0.96 (0.94) & 0.96 (0.94)\\
0.9  & 0.86 & 0.92 & 0.94 (0.91)   & 0.85 (0.78) & 0.79 (0.73)\\
0.99 & 0.98 & 0.90 & 0.65 (0.56)   & 0.79 (0.67) & 0.52 (0.42)\\
\bottomrule
\end{tabular}
\end{center}
Q3: median $\pf_R$ non-increasing in the amplitude and $\ge0.5$ at $0.99$ (holds). $\pf$ lies within a factor $1.25$ of $\pf_R$.

\emph{Part C: the picture of NB32 at $E=16$} (its own runs, not those of D2: 40 Haar starts, gradient descent 25k steps; 9 and 17 runs converged by 12.5k and 25k; 162
points: the run's gates, minimum-phase and random-branch realisations at the 54 unconverged points; $\pf$ matches the stored value to a
median relative $3\times10^{-15}$). Q4: $\pf_R\le10^{-2}\Rightarrow\pf\le10^{-2}$ at $\ge95\%$, Spearman with $\pf_R$ above that with $\Pi$,
minimum-phase $\pf_R\ge0.5$ at $\ge95\%$ (holds: 26 of 26, against 16 of 19 for $\Pi\le10^{-2}$; Spearman $0.98$ against $0.90$, middle
range $0.96$ against $0.66$; 54 of 54). At the run's gates the medians are $\log_{10}\pf=-1.92$, $\log_{10}\pf_R=-1.30$,
$\log_{10}\kappa_D=-1.82$; $R^2$ of $\log\pf$ on $\log\pf_R$ $0.92$, on $\log\Pi$ $0.57$, on both $\log\pf_R$ and $\log\kappa_D$ $0.97$.

\emph{Part D: hypotheses of Proposition~\ref{prop:readout}} (added after Parts A--C; a check, not a prediction). 28 circuits,
$E=8,16,24,32$, four Haar and three near-no-flip each. $\sigma_{2E+4}(D)/\sigma_1(D)\le5\times10^{-16}$ at every circuit, and
$\operatorname{range}D$ lies in $\mathcal A(a,b)$ to $2\times10^{-15}$. Where float64 resolves both $\sigma_{2E+3}(D)/\sigma_1(D)$ and the
smallest singular value of the constraint map $(\delta a,\delta b)\mapsto\mathrm{Re}(\bar a\delta a+\bar b\delta b)$ (both $\ge10^{-10}$;
18 of 28 circuits, all near-no-flip ones among them), $\pf_R$ computed on $\mathcal A(a,b)$ and on $\operatorname{range}D$ agree to a
relative $4\times10^{-7}$, so (iii) applies there. At the Haar circuits below that resolution ($E\ge16$) they differ by up to a factor $1.3$
for $E\le24$ and $8.2$ at $E=32$.

\subsection{Gate control at large $E$ (NB40)}\label{app:d13}

Predictions fixed before the run. Minimum-phase realisations of fresh targets from the 60-digit oracle, accepted if they reproduce the
target on the grid to $10^{-8}$ (all 270 accepted; largest error $6\times10^{-14}$); quantities as in Appendix~\ref{app:readoutexp}, with $q_l$ the flip
probabilities of the middle gates. Part A: flat and $1/k$ targets of amplitude 0.9, $E\in\{16,32,48,64,96,128\}$, 20 per cell. Part B: six
fixed shapes (a random part of degree 6 plus a top coefficient $0.03$, $\sup|f|=0.85$), each realised at $E\in\{12,24,48,96,128\}$, so that
$\sum_lq_l$ does not change with $E$.
\begin{center}\small
\begin{tabular}{llcccccc}
\toprule
family & $E$ & $\sum_lq_l$ & $\|G-I\|$ & $\kappa_{D_S}$ (min) & $\rho_R$ (min) & bound (i)$/\rho_R$\\
\midrule
flat & 16  & 0.090 & 0.160 & 0.853 (0.774) & 0.924 (0.898) & 0.46\\
flat & 32  & 0.071 & 0.166 & 0.849 (0.785) & 0.950 (0.935) & 0.48\\
flat & 64  & 0.065 & 0.166 & 0.850 (0.783) & 0.963 (0.958) & 0.50\\
flat & 128 & 0.053 & 0.151 & 0.861 (0.825) & 0.977 (0.970) & 0.53\\
$1/k$ & 16  & 0.157 & 0.203 & 0.816 (0.756) & 0.940 (0.926) & 0.38\\
$1/k$ & 32  & 0.125 & 0.180 & 0.835 (0.786) & 0.956 (0.938) & 0.40\\
$1/k$ & 64  & 0.133 & 0.186 & 0.830 (0.787) & 0.962 (0.944) & 0.41\\
$1/k$ & 128 & 0.108 & 0.158 & 0.854 (0.805) & 0.972 (0.964) & 0.45\\
\bottomrule
\end{tabular}
\end{center}
(medians; $E=48$ and $96$ in the repository, in line with their neighbours.) P0 (theorems): $\|J-RD\|/\|J\|\le3.3\times10^{-14}$, the bounds of
Proposition~\ref{prop:readout} in both coordinate systems and the bound (i) of Theorem~\ref{thm:privacy-extreme} below $\rho_R$ at all 240 realisations (holds). P1
($\kappa_{D_S}$ at $E=128$ at least $0.9\times$ its value at $E=16$, and $\ge0.5$ everywhere): flat $0.853\to0.861$, $1/k$ $0.816\to0.854$, minimum
$0.756$ (holds). P2 (median $\|G-I\|/\sum_lq_l\in[1,4]$ in every cell, growing by at most $1.25\times$ from $E=16$ to $128$): in range everywhere
($1.28$--$2.99$), but the ratio grows by $1.58\times$ for flat targets ($1.21\times$ for $1/k$) (fails). P3 (fixed shapes: $\kappa_{D_S}$ varies by at most
$0.01$ across $E$, $\sum_lq_l$ by at most $5\%$): at most $0.005$ and $3.1\%$ (holds). P4 ($\rho_R\ge0.5$; median $\kappa_D\sqrt{E+1}\in[0.8,1.2]$): $\rho_R\ge0.898$,
medians $0.91$--$0.95$ (holds).

\emph{Reading of P2 (after the run).} $\|G-I\|$ does not grow: its median is $0.15$--$0.18$ (flat) and $0.16$--$0.20$ ($1/k$) at every $E$. The ratio
grows because $\sum_lq_l$ falls for flat targets of higher degree ($0.090\to0.053$; the largest $q_l$ falls from $0.019$ to $0.002$), and within
the flat family $\|G-I\|$ and $\sum_lq_l$ are nearly uncorrelated ($0.25$). Gate control on the extreme branches is therefore not set by the
total flip, as the prediction assumed; Proposition~\ref{prop:gate-control} describes what sets it. Reported: $\kappa_{D_S}\cdot$bound (i) is at least $0.22$
(median $0.37$) at every realisation, a lower bound on $\rho_S$ in which only $\kappa_{D_S}$ is measured; the measured $\rho_S$ is at least $0.92$.

\subsection{Checks of Theorem~\ref{thm:privacy-extreme}, Proposition~\ref{prop:gate-control} and Corollary~\ref{cor:frames} (NB41)}\label{app:d14}

Checks of proved statements, run on the gates saved by NB40 (5 realisations per cell of its Part A) and on generated circuits; no training.

\emph{Proposition~\ref{prop:gate-control}.} The spectra of $G_{\rm mid}$ and of the form $Q$ agree to $5.3\times10^{-14}$ and $\|G_{00}-I\|\le8.3\times10^{-14}$ at all 60
realisations (holds).

\emph{Corollary~\ref{cor:frames}.} Both inequalities hold at all 60 realisations (holds). Medians over $E=16$--$128$: $A_S=0.80$--$0.84$ (smallest $0.766$),
$\sigma_F=0.15$--$0.20$ (largest $0.231$), $(A_S-\sigma_F)^2=0.36$--$0.47$ (smallest $0.286$), against $\lambda_{\min}(G_{\rm mid})=0.80$--$0.85$;
$\|C\|=0.02$--$0.09$. None of these changes systematically with $E$.

\emph{The failure mode.} Circuits whose total flip $0.1$ is spread evenly over the middle gates: with coherent flip phases,
$\lambda_{\min}(G_{\rm mid})=0.65,0.42,0.21$, $A_S=0.53,0.30,0.21$ and $\sigma_F=0.44,0.62,0.69$ at $E=16,32,64$; with random phases $\lambda_{\min}(G_{\rm mid})=0.85,0.83,0.80$,
$A_S=0.80,0.78,0.75$ and $\sigma_F=0.19,0.21,0.25$. Coherent phases, the case in which, at fixed $\Pi$, the loss of gate control grows with $E$ (as the factor $E-1$ of Theorem~\ref{thm:nnf} allows), destroy gate control through
exactly the two frame constants of Corollary~\ref{cor:frames}.

\emph{Theorem~\ref{thm:privacy-extreme}.} On minimum-phase realisations (float64 Fejér--Riesz factorisation) of flat and $1/k$ targets, $E\in\{8,17,32,49\}$, and on
three stress families ($f$ within $10^{-1}$--$10^{-4}$ of $\pm1$ at one point, of $+1$ and $-1$ at two points, or on two wide plateaus), the bound (i)
and the corollary (ii) lie below $\rho_R$ at all 172 realisations (holds). Medians of bound (i)$/\rho_R$: $0.63$, $0.48$, $0.29$, $0.22$ at amplitude
$0.5$, $0.9$, $0.99$, $0.999$ (smallest $0.37$ at $0.9$). When $f$ comes within $10^{-4}$ of $+1$ and of $-1$, $\rho_R$ itself falls to a median $0.08$,
so the dependence on $F$ in Theorem~\ref{thm:privacy-extreme} cannot be removed. On balanced branches (random root choices) the same bound exceeds $\rho_R$ at 16 of 24
realisations (smallest ratio $0.19$): the hypothesis of extreme branches is needed.

\subsection{Checks of Theorem~\ref{thm:nnf} (NB42)}\label{app:d15}
Checks of the numbers around a proved statement, on generated circuits; no training. The root of $1-u-\tau(\sqrt u)$ is $u^*=0.3100031$,
so $\Pi$ must be at least $0.956680$, $0.979545$, $0.986612$, $0.990050$ at $E=8,16,24,32$; $\tau(s)/s^3\to8/3$. The Gram matrix from
$\mathrm{Ad}$ (the proof's telescoping) and from the state moves agree to $10^{-15}$. Over 32 gate sets ($E=5$--$12$; random, aligned,
single and larger flips), $\max|\sigma^2-1|$ is at most $0.71\,(s^2+\tau(s))$, and single flips give $\|G-I\|=q$ to $10^{-14}$.
Charge walk: see the paragraph after the proof of Theorem~\ref{thm:nnf}. Check C6 (gates regenerated with NB37's seed and optimiser, $\Pi=0.990000$,
$\sigma_{\min}/\sigma_{\max}(J)=5\times10^{-15}$): $\sigma_{\min}(D_S)^2=0.985$, against $0.8692$ from $\Pi$ alone and $0.879$ from $s$. Adversarial search at
$\Pi=0.99$ (Powell, four restarts): $1-\sigma_{\min}^2=0.0306=0.435\,(E-1)\ln(1/\Pi)$ at $E=8$ and $0.0617=0.409\,(E-1)\ln(1/\Pi)$ at $E=16$;
a search bounds the worst case from below only.

\section{Additional results}
\label{app:additional}
\paragraph{Conditioning at initialisation.} The coefficient Jacobian of the serial circuit starts worse conditioned than that of the
ancilla circuit at every $E$: the median $\kappa_J$ is $10^{-4.5}$, $10^{-7.1}$ and $10^{-11.6}$ at $E=8,12,16$ (flat targets), against
$10^{-2.8}$, $10^{-4.2}$ and $10^{-5.7}$ for the ancilla circuit, and from $E=24$ on it is below the resolution of double precision. Initial
conditioning does not account for the difference: configurations of the two circuits with the same median initial $\kappa_J$ still differ
significantly in convergence (for instance serial at $E=16$ and ancilla at $E=32$, both $10^{-11.6}$: 65 and 99 of 100 runs under gradient
descent), and within a configuration the initial $\kappa_J$ of a run does not predict its outcome (area under the ROC curve $0.40$ to $0.65$ in
the eleven configurations with at least five runs of each kind, every interval containing $0.5$). This matches the main text: the start
does not decide which runs stall.

\begin{table}[h]
\centering\small
\caption{Tangent directions added by every layer after the first (exact numerical rank at random gates) for the two circuits of this
paper, against the upper bound $d^2-\sum_am_a^2$ of Appendix~\ref{app:proofs}, with $m_a$ the multiplicities of the eigenvalues of the
encoding generator. The row marked $^\ast$ is proved, the row marked $^\dagger$ is proved as a lower bound
(Proposition~\ref{prop:tangent}).}
\label{tab:tangent}
\begin{tabular}{lccccc}\toprule
circuit & qubits & encoded & $(m_a)$ & per layer & upper bound\\\midrule
serial$^\ast$ & 1 & 1 & (1,1) & 2 & 2\\
ancilla$^\dagger$ & 2 & 1 & (2,2) & 6 & 8\\
\bottomrule\end{tabular}
\end{table}
\label{app:results}
\begin{table}[h]
\centering\small
\caption{Real-world series, full-batch Adam, 100 initialisations per architecture on the same data. Runs at the optimum: excess training loss over the least-squares fit at most $10^{-6}$ of the variance. Held-out error ratio paired by run index;
last column: share of serial runs with higher held-out error than the paired ancilla run.}
\label{tab:real}
{\begin{tabular}{lccccccc}\toprule
 & & \multicolumn{2}{c}{runs at the optimum} & \multicolumn{2}{c}{median test $R^2$} & held-out error & serial runs \\
\cmidrule(lr){3-4}\cmidrule(lr){5-6}
series & $E$ & serial & ancilla & serial & ancilla & serial/ancilla [10\%, 90\%] & worse \\\midrule
\texttt{gasprice} & 24 & 0.04 & 1.00 & 0.9569 & 0.9591 & 1.05 [1.00, 1.15] & 0.92 \\
\texttt{gasprice} & 32 & 0.00 & 0.98 & 0.9682 & 0.9786 & 1.49 [1.30, 1.83] & 1.00 \\
\texttt{ibmclose} & 24 & 0.01 & 1.00 & 0.9607 & 0.9641 & 1.09 [1.04, 1.14] & 0.97 \\
\texttt{ibmclose} & 32 & 0.00 & 1.00 & 0.9629 & 0.9638 & 1.03 [0.97, 1.09] & 0.70 \\
\texttt{nottem} & 24 & 0.09 & 1.00 & 0.8995 & 0.9004 & 1.01 [0.97, 1.02] & 0.71 \\
\texttt{nottem} & 32 & 0.00 & 1.00 & 0.8879 & 0.8873 & 0.99 [0.96, 1.03] & 0.39 \\
\bottomrule
\end{tabular}
}
\end{table}
\begin{table}[h]
\centering\small
\caption{Screening of real-world series (Appendix~\ref{app:experiments}): median held-out $R^2$ of the least-squares fit of degree $E$ over five
random splits, the gain from $E=16$ to $32$, and the supremum of the degree-32 fit. Duplicates and series with missing values omitted.}
\label{tab:screen}
\begin{tabular}{lrcccccc}\toprule
series (Rdatasets) & length & $R^2_{16}$ & $R^2_{24}$ & $R^2_{32}$ & $R^2_{32}-R^2_{16}$ & $\sup|f_{32}|$ & passes \\\midrule
\texttt{AER/DJFranses} & 770 & $0.762$ & $0.822$ & $0.858$ & $+0.096$ & 0.76 &  \\
\texttt{AER/DutchSales} & 425 & $0.271$ & $0.168$ & $0.020$ & $-0.251$ & 0.46 &  \\
\texttt{AER/MarkDollar} & 518 & $-0.057$ & $-0.089$ & $-0.283$ & $-0.226$ & 0.31 &  \\
\texttt{AER/MarkPound} & 1974 & $-0.137$ & $-0.245$ & $-0.525$ & $-0.388$ & 0.15 &  \\
\texttt{CardioDataSets/cv\_mortality\_ts} & 508 & $0.600$ & $0.594$ & $0.544$ & $-0.056$ & 0.75 &  \\
\texttt{DAAG/mdbAVtJtoD} & 867 & $-0.347$ & $-0.596$ & $0.670$ & $+1.017$ & 0.51 &  \\
\texttt{boot/manaus} & 1080 & $0.377$ & $0.474$ & $0.620$ & $+0.242$ & 0.57 &  \\
\texttt{boot/wool} & 309 & $0.939$ & $0.950$ & $0.957$ & $+0.017$ & 0.74 &  \\
\texttt{datasets/nottem} & 240 & $-0.529$ & $0.881$ & $0.868$ & $+1.397$ & 0.72 &  \\
\texttt{datasets/sunspot.m2014} & 3177 & $0.874$ & $0.873$ & $0.870$ & $-0.004$ & 0.56 &  \\
\texttt{datasets/sunspot.month} & 3310 & $0.846$ & $0.844$ & $0.850$ & $+0.004$ & 0.67 &  \\
\texttt{datasets/sunspot.year} & 289 & $-0.012$ & $-0.077$ & $0.687$ & $+0.699$ & 0.71 &  \\
\texttt{datasets/sunspots} & 2820 & $0.816$ & $0.837$ & $0.833$ & $+0.017$ & 0.65 &  \\
\texttt{datasets/treering} & 7980 & $0.022$ & $0.103$ & $0.004$ & $-0.018$ & 0.33 &  \\
\texttt{forecast/gas} & 476 & $-0.079$ & $-0.226$ & $0.844$ & $+0.923$ & 0.64 &  \\
\texttt{fpp2/auscafe} & 426 & $0.590$ & $0.545$ & $0.624$ & $+0.034$ & 0.47 &  \\
\texttt{fpp2/gasoline} & 1355 & $0.402$ & $0.465$ & $0.430$ & $+0.029$ & 0.52 &  \\
\texttt{fpp2/goog} & 1000 & $0.816$ & $0.867$ & $0.878$ & $+0.062$ & 0.66 &  \\
\texttt{fpp2/hyndsight} & 365 & $-0.118$ & $-0.215$ & $-0.595$ & $-0.477$ & 0.33 &  \\
\texttt{fpp2/usmelec} & 486 & $-0.252$ & $-0.483$ & $-0.097$ & $+0.155$ & 0.48 &  \\
\texttt{quantreg/MelTemp} & 3650 & $0.456$ & $0.418$ & $0.583$ & $+0.127$ & 0.72 &  \\
\texttt{quantreg/gasprice} & 695 & $0.893$ & $0.955$ & $0.981$ & $+0.088$ & 0.75 & yes \\
\texttt{timeSeriesDataSets/co2\_ts} & 468 & $-0.245$ & $-0.526$ & $0.878$ & $+1.123$ & 0.72 &  \\
\texttt{timeSeriesDataSets/elec\_ts} & 476 & $-0.189$ & $-0.306$ & $0.553$ & $+0.742$ & 0.60 &  \\
\texttt{timeSeriesDataSets/ibmclose\_ts} & 369 & $0.920$ & $0.949$ & $0.962$ & $+0.043$ & 0.68 & yes \\
\bottomrule
\end{tabular}

\end{table}
\begin{table}[h]
\centering\small
\caption{Training on data, step budget (Appendix~\ref{app:experiments}): runs at the optimum out of 30, serial $|$ ancilla, for the main
experiment ($1\times$, 20{,}000 steps) and for 80{,}000 steps with the additional steps at the peak rate ($4\times$ held) or with the
decay spread over the whole run ($4\times$ stretched); last column: ancilla minus serial under the held schedule, with a 95\% Newcombe
interval. Same data and initialisations in all three columns. Runs are scored at the last step; a run that reached the optimum earlier and
left it under the held peak rate counts as not at the optimum. The $1\times$ column uses the first 30 runs of the main experiment, with its
targets, seeds and learning rates; Table~\ref{tab:data} uses fresh targets, start seeds and its own learning-rate selection, so its Haar rows differ.}
\label{tab:budget}
\begin{tabular}{llcccc}\toprule
 & & \multicolumn{3}{c}{runs at the optimum, serial $|$ ancilla} & ancilla$-$serial [95\%] \\
\cmidrule(lr){3-5}
target & $E$ & 1$\times$ & 4$\times$ held & 4$\times$ stretched & 4$\times$ held \\\midrule
flat & 12 & 23/30 $|$ 30/30 & 24/30 $|$ 30/30 & 21/30 $|$ 30/30 & +0.20 [+0.05, +0.37] \\
flat & 24 & 22/30 $|$ 30/30 & 15/30 $|$ 30/30 & 15/30 $|$ 30/30 & +0.50 [+0.30, +0.67] \\
Poisson kernel & 12 & 14/30 $|$ 30/30 & 20/30 $|$ 30/30 & 21/30 $|$ 30/30 & +0.33 [+0.15, +0.51] \\
Poisson kernel & 24 & 1/30 $|$ 17/30 & 0/30 $|$ 28/30 & 3/30 $|$ 30/30 & +0.93 [+0.75, +0.98] \\
\texttt{nottem} & 24 & 6/30 $|$ 30/30 & 1/30 $|$ 30/30 & 3/30 $|$ 30/30 & +0.97 [+0.79, +0.99] \\
\texttt{gasprice} & 24 & 2/30 $|$ 30/30 & 7/30 $|$ 30/30 & 7/30 $|$ 30/30 & +0.77 [+0.56, +0.88] \\
\texttt{ibmclose} & 24 & 0/30 $|$ 30/30 & 18/30 $|$ 30/30 & 17/30 $|$ 30/30 & +0.40 [+0.21, +0.58] \\
\bottomrule
\end{tabular}

\end{table}
\paragraph{Adam on the exact loss (Section~\ref{sec:exact}).} Converged runs end at a level that falls with the
learning rate: for the ancilla circuit the median final loss of the runs below $10^{-6}$ is $10^{-9.6}$ at the rate $10^{-3}$ (5\% and
95\% quantiles $10^{-11.8}$ and $10^{-8.2}$) and $10^{-11.8}$ at $10^{-4}$. A threshold of $10^{-10}$ therefore cuts through the converged
runs: at $E=8$ on flat targets, 77 serial and 54 ancilla runs lie below it, although all 200 are below $10^{-6}$. Up to 7 of 100 runs per
configuration lie between $10^{-8}$ and $10^{-6}$; under either threshold the same nine of ten differences are significant.

\paragraph{Checks on data (Section~\ref{sec:data}).} \emph{End levels and threshold.} Under the schedule of the main experiment Adam
leaves 3{,}255 of the 4{,}800 evaluation runs below $10^{-15}$; with the same decay spread over four times as many steps, every converged
ancilla run ends between $10^{-10}$ and $10^{-6}$; Levenberg--Marquardt leaves its converged runs between $10^{-16}$ and $10^{-14}$. For
every threshold from $10^{-4}$ to $10^{-15}$, the difference in the share of runs at the optimum between the ancilla and the serial circuit
is positive with a 95\% interval that excludes zero in all 16 configurations. \emph{Learning-rate bracketing.}
30 of the 48 configurations have their best tuning score at the floor and are not extended; in 26 of them the near-best rates include an
end of the grid. Of the other 18, nine were extended and four selections changed; no share of runs at the optimum changes by more than
$0.02$. In the ten configurations with $E\ge16$ in which most serial runs fail, no rate, the added ones included, brings the tuning score
of the serial circuit below $10^{-4.2}$. Five tuning runs cannot separate rates when fewer than about a quarter of the runs fail: for the
serial circuit on flat targets at $E\le16$ the tuning score is at the floor for at least three rates of the grid, while $0.05$ to $0.16$
of its evaluation runs fail at the selected rate. \emph{Step budget} (Table~\ref{tab:budget}). Under the stretched schedule converged runs
end higher, at a level that appears to rise with the peak rate, so that shares at a fixed threshold are less comparable between
configurations. Under both schedules the serial circuit reaches the optimum less often than the control in all seven configurations;
under the held schedule the difference ranges from $0.20$ to $0.97$ with every interval excluding zero. Its size depends on the budget,
and not in one direction: on \texttt{ibmclose} the serial circuit reaches the optimum in 18 of 30 runs instead of none, on flat targets at
$E=24$ and on the temperatures less often than with 20{,}000 steps (15 instead of 22, and 1 instead of 6).

\paragraph{Flat targets at matched parameter counts (Sections~\ref{sec:adam} and~\ref{sec:data}).} The flat targets of
Section~\ref{sec:exact}, trained in the setup of Section~\ref{sec:data}: 200 sampled training points, Adam with 20{,}000 steps and a tuned
learning rate, and parameter counts matched (five gates per block in the serial circuit). The serial circuit reaches the optimum in 95, 87,
84, 60 and 36 of 100 runs at $E=8,12,16,24,32$, the ancilla circuit in every run. The difference to the ancilla control is significant at every $E$; at $E=8$ the lower end of the interval is
$0.003$. The setup differs from that of Section~\ref{sec:adam} in more than the parameter counts (sampled points, a shorter budget), so
that the counts of the two sections are not to be compared with each other; what the runs show is that the difference between the
circuits is there at matched counts. Levenberg--Marquardt from the same initial gates brings the serial circuit to the optimum in 98, 100, 96, 94
and 82 runs at $E=8$ to $32$, the ancilla circuit in all. These targets are also part of the step-budget check (Table~\ref{tab:budget}).
\begin{table}[h]
\centering\small
\caption{Exact loss: runs out of 100 with final loss at or below the threshold, serial $|$ ancilla. The threshold of the main text
is $10^{-6}$. GD: gradient descent; LM: Levenberg--Marquardt. For Adam, $10^{-10}$ lies within the range at which converged runs end
(Appendix~\ref{app:experiments}).}
\label{tab:rates_exact}
\begin{tabular}{lllccccc}\toprule
optimiser & targets & threshold & $E=8$ & $E=12$ & $E=16$ & $E=24$ & $E=32$ \\\midrule
GD & flat & $10^{-4}$ & 93 $|$ 100 & 86 $|$ 100 & 65 $|$ 100 & 27 $|$ 100 & 3 $|$ 100 \\
 &  & $10^{-6}$ & 93 $|$ 100 & 85 $|$ 100 & 65 $|$ 100 & 24 $|$ 99 & 3 $|$ 99 \\
 &  & $10^{-8}$ & 92 $|$ 100 & 83 $|$ 100 & 63 $|$ 100 & 21 $|$ 99 & 1 $|$ 98 \\
 &  & $10^{-10}$ & 92 $|$ 100 & 83 $|$ 100 & 62 $|$ 100 & 19 $|$ 99 & 1 $|$ 98 \\
\cmidrule(l){2-8}
 & $1/k$ & $10^{-4}$ & 78 $|$ 100 & 43 $|$ 100 & 11 $|$ 93 & 0 $|$ 29 & 0 $|$ 0 \\
 &  & $10^{-6}$ & 71 $|$ 100 & 32 $|$ 99 & 5 $|$ 77 & 0 $|$ 16 & 0 $|$ 0 \\
 &  & $10^{-8}$ & 68 $|$ 100 & 29 $|$ 96 & 4 $|$ 72 & 0 $|$ 14 & 0 $|$ 0 \\
 &  & $10^{-10}$ & 63 $|$ 100 & 27 $|$ 94 & 3 $|$ 69 & 0 $|$ 13 & 0 $|$ 0 \\
\midrule
LM & flat & $10^{-4}$ & 100 $|$ 100 & 100 $|$ 100 & 96 $|$ 100 & 90 $|$ 100 & 83 $|$ 100 \\
 &  & $10^{-6}$ & 100 $|$ 100 & 100 $|$ 100 & 96 $|$ 100 & 90 $|$ 100 & 83 $|$ 100 \\
 &  & $10^{-8}$ & 100 $|$ 100 & 100 $|$ 100 & 96 $|$ 100 & 90 $|$ 100 & 83 $|$ 100 \\
 &  & $10^{-10}$ & 100 $|$ 100 & 100 $|$ 100 & 96 $|$ 100 & 90 $|$ 100 & 83 $|$ 100 \\
\cmidrule(l){2-8}
 & $1/k$ & $10^{-4}$ & 100 $|$ 100 & 99 $|$ 100 & 96 $|$ 100 & 91 $|$ 100 & 75 $|$ 100 \\
 &  & $10^{-6}$ & 100 $|$ 100 & 99 $|$ 100 & 96 $|$ 100 & 86 $|$ 100 & 73 $|$ 100 \\
 &  & $10^{-8}$ & 100 $|$ 100 & 99 $|$ 100 & 96 $|$ 100 & 86 $|$ 100 & 73 $|$ 100 \\
 &  & $10^{-10}$ & 100 $|$ 100 & 99 $|$ 100 & 96 $|$ 100 & 86 $|$ 100 & 73 $|$ 100 \\
\midrule
Adam & flat & $10^{-4}$ & 100 $|$ 100 & 95 $|$ 100 & 83 $|$ 100 & 66 $|$ 100 & 54 $|$ 100 \\
 &  & $10^{-6}$ & 100 $|$ 100 & 95 $|$ 100 & 83 $|$ 100 & 66 $|$ 100 & 53 $|$ 100 \\
 &  & $10^{-8}$ & 98 $|$ 100 & 95 $|$ 100 & 83 $|$ 98 & 66 $|$ 94 & 47 $|$ 93 \\
 &  & $10^{-10}$ & 77 $|$ 54 & 76 $|$ 43 & 61 $|$ 37 & 45 $|$ 18 & 22 $|$ 7 \\
\cmidrule(l){2-8}
 & $1/k$ & $10^{-4}$ & 94 $|$ 100 & 88 $|$ 100 & 65 $|$ 100 & 32 $|$ 100 & 12 $|$ 98 \\
 &  & $10^{-6}$ & 94 $|$ 100 & 88 $|$ 100 & 62 $|$ 100 & 23 $|$ 100 & 9 $|$ 97 \\
 &  & $10^{-8}$ & 94 $|$ 100 & 86 $|$ 100 & 62 $|$ 99 & 19 $|$ 97 & 5 $|$ 92 \\
 &  & $10^{-10}$ & 79 $|$ 61 & 70 $|$ 47 & 45 $|$ 44 & 6 $|$ 24 & 3 $|$ 14 \\
\bottomrule
\end{tabular}

\end{table}
\begin{table}[h]
\centering\small
\caption{Gradient descent on the exact loss: runs out of 100 with loss $\le10^{-6}$ after 7{,}500, 14{,}500, 27{,}000 and all 50{,}000
steps. These are four of the checkpoints at which the loss was recorded (Appendix~\ref{app:experiments}), each about twice the one
before.}
\label{tab:gd_budget}
\begin{tabular}{llcccccccc}\toprule
 & & \multicolumn{4}{c}{serial} & \multicolumn{4}{c}{ancilla} \\\cmidrule(lr){3-6}\cmidrule(lr){7-10}
targets & $E$ & 7{,}500 & 14{,}500 & 27{,}000 & 50{,}000 & 7{,}500 & 14{,}500 & 27{,}000 & 50{,}000 \\\midrule
flat & 8 & 74 & 86 & 90 & 93 & 100 & 100 & 100 & 100 \\
 & 12 & 51 & 74 & 80 & 85 & 100 & 100 & 100 & 100 \\
 & 16 & 16 & 34 & 55 & 65 & 97 & 100 & 100 & 100 \\
 & 24 & 0 & 3 & 6 & 24 & 51 & 92 & 98 & 99 \\
 & 32 & 0 & 0 & 0 & 3 & 1 & 46 & 94 & 99 \\
\midrule
$1/k$ & 8 & 26 & 40 & 52 & 71 & 94 & 97 & 100 & 100 \\
 & 12 & 1 & 5 & 13 & 32 & 46 & 78 & 92 & 99 \\
 & 16 & 0 & 0 & 0 & 5 & 5 & 18 & 53 & 77 \\
 & 24 & 0 & 0 & 0 & 0 & 0 & 0 & 5 & 16 \\
 & 32 & 0 & 0 & 0 & 0 & 0 & 0 & 0 & 0 \\
\bottomrule
\end{tabular}

\end{table}

\end{document}